\documentclass[12pt]{article}
\usepackage{graphicx,bm,amsmath,amssymb,latexsym,epsfig,euscript,multirow,color,url}
\usepackage{hyperref}
\allowdisplaybreaks

\long\def\symbolfootnote[#1]#2{\begingroup%
\def\thefootnote{\fnsymbol{footnote}}\footnote[#1]{#2}\endgroup}

\newcommand{\be}{\begin{equation}}
\newcommand{\ee}{\end{equation}}
\newcommand{\pd}{\partial}
\newcommand{\bea}{\begin{eqnarray}}
\newcommand{\eea}{\end{eqnarray}}
\newcommand{\mat}{\begin{pmatrix}}
\newcommand{\rix}{\end{pmatrix}}
\newcommand{\nn}{\nonumber}
\renewcommand{\l}{\left}
\renewcommand{\r}{\right}

\renewcommand{\bar}{\overline}
\renewcommand{\slash}[1]{#1\!\!\!/}

\newcommand{\vv}{\mathbf}

\newcommand{\cM}{\mathcal{M}}
\newcommand{\cL}{\mathcal{L}}

\newcommand{\go}{{\tilde g}}
\newcommand{\sq}{{\tilde q}}
\newcommand{\st}{{\tilde t}}
\newcommand{\kkg}{{g^\star}}
\newcommand{\kkq}{{q^\star}}
\newcommand{\kkB}{{B^\star}}
\newcommand{\kkqR}{{q^\star_R}}
\newcommand{\kkqL}{{q^\star_L}}
\newcommand{\kkqX}{{q^\star_\chi}}
\newcommand{\kkqXX}{{q^\star_{\slash\chi}}}
\newcommand{\kkt}{{t^\star}}

\newcommand{\kktX}{{t^\star_\chi}}

\newcommand{\q}{\quad}
\newcommand{\qq}{\qquad}
\newcommand{\qqq}{\qquad\quad}
\newcommand{\qqqq}{\qquad\qquad}
\newcommand{\qqqqq}{\qquad\qquad\quad}
\newcommand{\qqqqqq}{\qquad\qquad\qquad}
\newcommand{\ds}{\displaystyle}

\input prepictex
\input pictex
\input postpictex
\newdimen\tdim
\tdim=\unitlength
\def\stpltsmbl{\setplotsymbol ({\small .})}

\def\barrow{\arrow <8\tdim> [.3,.6]}

\begin{document}

\begin{titlepage}

\begin{flushright}
\small{RUNHETC-2011-03}\\
\end{flushright}

\vspace{0.5cm}
\begin{center}
\Large\bf
Distinguishing spins at the LHC\\using bound state signals
\end{center}

\vspace{0.2cm}
\begin{center}
{\sc Dilani Kahawala$^a$\symbolfootnote[1]{kahawala@physics.harvard.edu} and Yevgeny Kats$^{\,a,b}$\symbolfootnote[2]{kats@physics.rutgers.edu}}\\
\vspace{0.4cm}
\small\textit{$^a$Department of Physics, Harvard University, Cambridge, MA 02138, USA\\
$^b$Department of Physics and Astronomy, Rutgers University, Piscataway, NJ 08854, USA\\}
\end{center}

\vspace{0.5cm}
\begin{abstract}
\vspace{0.2cm}
\noindent
A pair of new colored particles produced near the threshold can form a bound state and then annihilate into standard model particles. In this paper we show how the bound state signals can distinguish between two scenarios that have similar particle content and interactions but different spins, such as the minimal supersymmetric standard model (MSSM) and universal extra dimensions (UED). We find, for example, that bound states of KK gluons (KK gluonia) have an order of magnitude larger cross sections than gluinonia and they may be detectable as resonances in the $b\bar b$, $t\bar t$ or $\gamma\gamma$ channels if the KK gluon is very light and sufficiently long-lived. KK gluonia can be distinguished from gluinonia by their much larger cross sections and distinct angular distributions. Similarly, KK quarkonia can be distinguished from squarkonia by the size of their diphoton cross section and by their dilepton signals. Since many of our results are largely determined by gauge interactions, they will be useful for many other new physics scenarios as well.
\end{abstract}
\vfil

\end{titlepage}

\tableofcontents

\section{Introduction}
\setcounter{equation}{0}

Studying the new particles that will be produced at the LHC will likely be a non-trivial task. In many scenarios the particles of the new sector are  charged under a new parity that makes the lightest such particle a dark matter candidate. The new particles will then predominantly be produced in pairs, giving multiple decay products some of which will be undetectable (including at least two dark matter particles). The reconstruction of such events with many objects and missing energy in the final states will be complicated and often ambiguous. It may therefore be useful to also look at signals arising from the near-threshold formation and annihilation of QCD bound states of the new particles. The bound states behave as resonances which annihilate primarily into just two particles (or jets) and no missing energy, so the analysis and the interpretation of the signal are much more straightforward than in the more conventional channels. In particular, detecting the peak in the invariant mass distribution of any of the possible annihilation channels provides a direct measurement of the particle mass. The disadvantages of the bound state signals are their relatively small cross sections and the fact that the dominant annihilation mode is into dijets (which have large QCD background), with only small branching ratios for annihilation into cleaner signals such as $\gamma\gamma$. However, whenever the bound state signals are observable, they can provide an entirely independent method for characterizing or verifying the properties of the new particles.

In this paper we study how the spins of the new particles are reflected in the properties of their bound states and the resulting signals at the LHC. As an example, we compare the bound states of the level-$1$ Kaluza-Klein (KK) modes in the universal extra dimensions model (UED) with bound states of the superpartners in the minimal supersymmetric extension of the standard model (MSSM). In both scenarios, the new particles are charged in the same way under the standard model gauge groups, and the masses of the particles are to a large extent free parameters. The fundamental difference between the two scenarios is the spin of the new particles. The KK modes in UED have the same spins as the standard model particles (which are the zero modes), while the spins of the superpartners in MSSM are different. However, the collider signatures of the decay products of a pair of level-1 KK modes of UED and a pair of analogous MSSM superpartners are unfortunately very similar and it is often challenging to determine the spin~\cite{Smillie:2005ar,Datta:2005zs,Barr:2005dz,Alves:2006df,Athanasiou:2006ef,Athanasiou:2006hv,Wang:2006hk,Kilic:2007zk,Csaki:2007xm,Burns:2008cp,Gedalia:2009ym,Cheng:2010yy,MoortgatPick:2011ix}. It is therefore interesting to study whether and when the bound state annihilation signals can give information about the spin.

While much is known about bound states of the MSSM particles (see~\cite{Kats:2009bv} and references therein), the bound states in UED models have not been explored, with the exception of bound states of KK quarks in the context of a lepton collider~\cite{Carone:2003ms,Fabiano:2008xk}. Here we will study the various possible bound states of colored particles of UED in the context of the LHC and compare them to the corresponding bound states in the MSSM. We will see as we go along that once the spin of the particle is specified many of the results are largely model-independent since they are determined by gauge interactions. Thus our comparisons between particles of different spins (in the same color representation) will hold even more generally.

We start in section~\ref{sec-setup} by discussing the UED model in the context of our study and reviewing the general formalism for bound state computations. In sections~\ref{sec-KKgluonium}--\ref{sec-diKKquarks-KKqKKg} we present the production cross sections, branching ratios and angular distributions for the various bound states and compare signals obtained from UED and MSSM. In section~\ref{sec-simulation} we simulate the bound state signals and the relevant backgrounds in the dijet, $b\bar b$, $t\bar t$, $\gamma\gamma$ and $e^+e^-$ channels and estimate the LHC reach for the various cases. We summarize our conclusions in section~\ref{sec-conclusions}.

\section{Setup\label{sec-setup}}
\setcounter{equation}{0}

\subsection{Particle spectrum of UED}

In the simplest UED scenario~\cite{Appelquist:2000nn} (for a review, see~\cite{Hooper:2007qk}), all the standard model fields are propagating in a single extra dimension of size $R \sim \mbox{TeV}^{-1}$. The right- and left-handed standard model fermions are represented by separate fields in 5 dimensions, and each has two chiralities when reduced to 4 dimensions. To restrict each fermion zero-mode to a single chirality, the extra dimension is assumed to be an $S^1/\mathbb{Z}_2$ orbifold. The zero modes of the unwanted chiralities of the fermions and the 5th components of the gauge fields are then projected out by declaring them to be odd under the $\mathbb{Z}_2$ orbifold symmetry. The KK modes of the right- and left-handed quarks will be denoted by $\kkqR$ and $\kkqL$, or collectively by $q^\star$. We will refer to $\kkqR$ and $\kkqL$ as right- and left-handed KK quarks, even though each one of them is a full Dirac fermion. (Similarly, we will refer to the MSSM partners of the right- and left-handed quarks, as right- and left-handed squarks.) All the standard model interactions are contained in the bulk Lagrangian. Since in 5 dimensions the gauge, Yukawa and quartic-Higgs couplings have negative mass dimension, this is an effective theory with a UV cutoff $\Lambda$ which is assumed to be above the compactification scale: $\Lambda > 1/R$.

In the 4-dimensional description, the theory contains towers of KK modes for each standard model particle, with the mass of the $n$-th mode given (at tree level) by
\be
m_n^2 = \frac{n^2}{R^2} + m_0^2
\ee
where $m_0$ is the mass of the zero mode (that is the standard model particle itself). As a result, all the KK modes at a particular level $n$ are approximately degenerate, with masses ordered like in the standard model. Loop corrections shift the masses~\cite{Cheng:2002ab,Cheng:2002iz}, and the resulting spectrum of the first KK level has the KK gluon as the heaviest particle, followed by the KK quarks, and then the KK excitations of the electroweak gauge bosons, the Higgs, and the leptons. The lightest KK particle (LKP), which is typically one of the neutral KK gauge bosons, is a dark matter candidate~\cite{Servant:2002aq,Cheng:2002ej}. At tree level, the LKP is stable because of momentum conservation in the extra dimension. When loops are taken into account it remains stable because the theory retains KK parity under which odd-$n$ modes are charged. The KK parity also prevents tree-level contributions to the electroweak observables, thus allowing a compactification scale $1/R$ as low as a few hundred GeV. For our purposes it is important that due to  KK parity the level-1 KK particles will be produced in pairs, and thus they can form bound states.

In general, the theory also includes boundary terms (on the orbifold fixed points) whose coefficients are free parameters. The boundary terms can be assumed to be symmetric under the interchange of the two orbifold fixed points, and then the theory still preserves KK parity. However, the mass spectrum of the KK modes can be affected dramatically. Even if boundary terms are absent at a particular scale, they are generated by the renormalization group running~\cite{Georgi:2000ks,Cheng:2002ab,Cheng:2002iz}, and the resulting corrections to the masses are of the order of the loop corrections. The spectra of~\cite{Cheng:2002ab,Cheng:2002iz} described above were obtained with the simplifying assumption that the coefficients of the boundary terms vanish at a specific cutoff scale. However, as has been shown in~\cite{Carena:2002me} for the case of the gauge kinetic term, even when the coefficient of the boundary term is not much larger than the expectation from na\"{\i}ve dimensional analysis, there is a large effect on the spectrum of the KK modes. Similarly, the effects of boundary kinetic and mass terms for a massive scalar field have been analyzed in~\cite{Flacke:2008ne}, where the results were applied to the electroweak sector and it was found that the identity of the LKP was sensitive to the boundary terms.

In both UED and MSSM, annihilation may or may not be the dominant decay mode of the various bound states, depending on the intrinsic decay rates of the constituent particles, which in turn depend on the mass spectrum of the model. Since the theoretical and experimental constraints on the possible mass spectra are a question that is decoupled from the bound state analysis, we will leave it out of the scope of this paper. We will assume that the bound states decay predominantly by annihilation, but the reader should remember that if the constituent particles are not sufficiently long-lived, the cross sections will need to be multiplied by appropriate branching ratios based on the annihilation rates that we compute here and the model-dependent single-particle decay rates. Note, however, that annihilation branching ratios close to $1$ are not implausible. In particular, in the MSSM there exist various motivated scenarios in which this happens to be true for the gluinonium~\cite{Kats:2009bv} or the stoponium~\cite{Martin:2008sv}. Or looking at this from a different perspective, the observation or non-observation of bound state signals can give us certain information about the spectrum of the model.

In UED, the spectrum of the level-1 KK modes plays a crucial role in determining which of them are sufficiently long-lived for their bound states to decay by annihilation rather than by the decays of the constituent particles. If the assumption of~\cite{Cheng:2002ab,Cheng:2002iz} that the boundary terms vanish at the UV cutoff were true, the KK gluon would have strong two-body decays into a KK quark and an antiquark, and the KK quarks could decay electroweakly into a KK electroweak gauge boson and a quark. In this case, similarly to what happens in much of the parameter space of the MSSM~\cite{Kats:2009bv}, the branching ratios for annihilation will be small. However, we believe (although without constructing explicit examples) that the presence of boundary terms can change the spectrum in ways that would make these decays kinematically forbidden for some of the particles.

For example, if the KK gluon $\kkg$ becomes lighter than the KK quarks $\kkq$, its 2-body decays will be replaced by 3-body decays into a KK electroweak gauge boson, a quark and an antiquark through diagrams involving an off-shell KK quark, a process suppressed by the electroweak coupling and the KK quark mass. This is analogous to the decay of the gluino into a neutralino, quark and antiquark in MSSM scenarios with squarks heavier than the gluino, which easily makes the gluino sufficiently stable for our purposes~\cite{Kats:2009bv}. Similarly to the case of the gluino, there is no need for an unnaturally large gap between the KK gluon and KK quark masses in order for the annihilation to dominate. For example, suppose that the dominant decay process is $\kkg \to q \bar q \kkB$, where $\kkB$ is the KK hypercharge gauge boson. Then the rate is
\be
\Gamma_\kkg \simeq \frac{11}{45\pi}\,\alpha_s\,\frac{\alpha}{\cos^2\theta_W} \l(1 - \frac{m_\kkB}{m_\kkg}\r)^5 \l(\frac{m_\kkg}{m_\kkq}\r)^4 m_\kkg
\sim 10^{-4}\, \l(1 - \frac{m_\kkB}{m_\kkg}\r)^5 \l(\frac{m_\kkg}{m_\kkq}\r)^4 m_\kkg
\label{KKg-decay}
\ee
where for simplicity we assumed $m_\kkq \gg m_\kkg \approx m_\kkB$.\footnote{The assumption $m_\kkq \gg m_\kkg$ is not essential for the validity of this estimate. We have checked that for $m_\kkq = 2m_\kkg$, for example, the exact result differs from (\ref{KKg-decay}) by less than a factor of $2$.} On the other hand, the bound state annihilation rates are
\be
\Gamma_{\rm annih} \sim \alpha_s^5\, m_\kkg \sim 10^{-5}\, m_\kkg
\ee
which can easily dominate over $2\Gamma_{\kkg}$, especially when we include the numerical prefactors that appear in the exact expressions for $\Gamma_{\rm annih}$ (see appendix~\ref{app-rates-KKgKKg}), which are as large as ${\cal O}\,(100)$ for color-singlet bound states, primarily due to multiple powers of color factors.

The collider signatures of the 3-body decays of the KK gluon or gluino have been studied in~\cite{Csaki:2007xm} in an attempt to distinguish between UED and MSSM. One of the goals of the present paper is to find out to what extent detecting the annihilation decays of KK gluonia (which are bound states of two KK gluons) compared to gluinonia (bound states of two gluinos) can provide an additional method for distinguishing between UED and MSSM and determining the properties of the underlying particles. While gluinonia have been studied extensively in the literature, our paper is the first study of KK gluonia.

Another object of our study is KK-quarkonia, which are KK quark-KK antiquark bound states. These have been studied in~\cite{Carone:2003ms,Fabiano:2008xk} in the context of their production and detection at a lepton collider. It was found that the spectrum of~\cite{Cheng:2002ab,Cheng:2002iz} does allow the KK quarks to be sufficiently stable for forming KK-quarkonia, but not for the annihilation decays to dominate over the single KK quark decays.\footnote{It was claimed in~\cite{Carone:2003ms} that KK top quarks can be very stable so that their annihilation decays dominate, but this was incorrect, as explained in~\cite{DePree:2005yv}.} Including boundary terms can probably change these results in either direction, and we are particularly interested in the situation in which some of the KK quarks are more long-lived so that the branching ratio for annihilation is enhanced.

Other than that, it is useful to note that many of our results are largely determined by gauge interactions alone and depend only on the properties of the binding particles rather than the full spectrum of the model and are therefore valid much more generally than just for MSSM or UED. It is therefore useful to study all the possible bound states in these models, even if some of the binding particles are not sufficiently stable in generic MSSM or UED scenarios. With this motivation in mind, our study covers all the possible bound states of pair-produced colored particles in MSSM and UED.

\subsection{Bound state formalism}

\begin{table}[t]
$$\begin{array}{|c|c|c|c|c|}\hline
\, \mbox{UED}                       \,&\, \mbox{MSSM}             \,&\, \mbox{binding}             \,&\, \mbox{non-binding} \,\\\hline
\, gg \to \kkg \kkg           \,&\,  gg \to \go\go          \,&\, \vv{1}, \vv{8_S}, \vv{8_A} \,&\, \vv{10}, \vv{\bar{10}}, \vv{27} \,\\
\, q\bar q \to \kkg \kkg      \,&\,  q\bar q \to \go\go     \,&\, \vv{1}, \vv{8_S}, \vv{8_A} \,&\, \,\\\hline
\, gg \to \kkq \bar \kkq      \,&\,  gg \to \sq\sq^\ast      \,&\, \vv{1}                     \,&\, \vv{8} \,\\
\, q\bar q \to \kkq \bar \kkq \,&\,  q\bar q \to \sq\sq^\ast \,&\, \vv{1}                     \,&\, \vv{8} \,\\\hline
\, qq \to \kkq \kkq           \,&\,  q q \to \sq\sq         \,&\, \vv{\bar 3}                \,&\, \vv{6} \,\\\hline
\, qg \to \kkq \kkg           \,&\,  q g \to \sq\go         \,&\, \vv{3}, \vv{\bar 6}        \,&\, \vv{15} \,\\\hline
\end{array}$$
\caption{Pair production processes and the color representations of the pair. Two color-octets can form an octet with either a symmetric ($\propto d^{abc}$, denoted $\vv{8_S}$) or antisymmetric ($\propto f^{abc}$, $\vv{8_A}$) wave function. The Clebsch-Gordan coefficients for all the color decompositions can be found in~\cite{Beneke:2009rj}. We have not explicitly listed processes obtained by replacing particles by antiparticles.}
\label{tab-processes}
$$\begin{array}{|c|c|c|c|}\hline
\,\mbox{UED}                    \,&\,\mbox{MSSM}                 \,&\, \mbox{SU(3)}        \,&\, C \,\\\hline
(\kkg\kkg)                      \,&\, (\go\go)                   \,&\, \vv{1}              \,&\, 3 \,\\
                                \,&\,                            \,&\, \vv{8}              \,&\, 3/2 \,\\\hline
(\kkq\bar\kkq)                  \,&\, (\sq\sq^\ast)               \,&\, \vv{1}              \,&\, 4/3 \,\\\hline
(\kkq\kkq), (\bar\kkq\,\bar\kkq)\,&\, (\sq\sq), (\sq^\ast\sq^\ast) \,&\, \vv{\bar 3}, \vv{3} \,&\, 2/3 \,\\\hline
(\kkq\kkg), (\bar\kkq\kkg)      \,&\, (\sq\go), (\sq^\ast\go)     \,&\, \vv{3}, \vv{\bar 3}  \,&\, 3/2 \,\\
                                \,&\,                            \,&\, \vv{\bar 6}, \vv{6}  \,&\, 1/2 \,\\\hline
\end{array}$$
\caption{Bound states in UED and MSSM, their color representations and the strength of their potential~(\ref{V(r)}). The possible spins of these bound states will be discussed later.}
\label{tab-C}
\end{table}

Assuming that the masses of the pair of particles satisfy $m_1, m_2 \gg \Lambda_{\rm QCD}$, their dynamics can be described by the single-gluon exchange potential with the Coulomb-like form
\be
V(r) = -C\frac{\bar\alpha_s}{r}
\label{V(r)}
\ee
where $\bar\alpha_s$ denotes the strong coupling constant evaluated self-consistently at the scale of the inverse Bohr radius $a_0^{-1} = C\bar\alpha_s \mu$, where $\mu \equiv m_1m_2/(m_1+m_2)$ (while the plain $\alpha_s$ will be reserved for its value at the scale of $m_1$ and $m_2$) and
\be
C = \frac{1}{2}\l(C_1 + C_2 - C_{(12)}\r)
\label{color-factor}
\ee
where $C_1$ and $C_2$ are the quadratic Casimirs of the color representations of the two particles and $C_{(12)}$ of the bound state. For bound states considered in this paper, the values of $\bar\alpha_s$ are between $0.11$ and $0.15$, so using (\ref{V(r)}) should be a good approximation, even though subleading corrections may have sizeable effects and it would be desirable to compute them, along with higher-order corrections to the production and annihilation processes, in a future work. Table~\ref{tab-processes} lists all the possible pair production processes in UED and MSSM and specifies in each case which of the color representations of the pair have an attractive potential and which do not, based on the sign of $C$. The values of $C$ for all the attractive configurations are given in table~\ref{tab-C}. In cases where the bound states are colored, they will further hadronize with ordinary quarks or gluons to become color-neutral (if they are sufficiently long-lived for this to happen). However, these processes will be happening at much larger distance scales than both the production and annihilation of the bound states and we therefore ignore their effects.

One can get the matrix elements for bound state production and annihilation by representing the bound state as a superposition of states of two free particles with momenta distributed according to the bound state wavefunction $\psi(\vv{r})$. For an $S$-wave bound state, neglecting the dependence of the short-distance process on the momenta, it can be easily shown that the matrix element between the bound state and the particles from which the pair is produced is given by~\cite{Peskin-Schroeder,Novikov:1977dq}
\be
\cM_{\rm bound} = \frac{\psi(\vv{0})}{\sqrt{2\mu}}\,\cM_0
\label{times-wavefunction}
\ee
where $\cM_0$ is the matrix element describing the production of the free constituent particles at threshold and $\mu$ is their reduced mass. The cross section will then be proportional to
\be
\l|\psi(\vv{0})\r|^2 = \frac{C^3 \bar\alpha_s^3 (2\mu)^3}{8\pi}
\label{wavefunction}
.\ee
More specifically, the bound state production cross section can be written in terms of the near-threshold production cross section of the pair of particles $\hat\sigma_0(\hat s)$ as\footnote{While (\ref{times-wavefunction})--(\ref{wavefunction}) assume the binding particles to be distinct, the relation (\ref{sigma-bound}), with $|\psi(\vv{0})|^2$ given by (\ref{wavefunction}), is valid also if they are identical. Later, in (\ref{ann-rate}), $|\psi(\vv{0})|^2$ again refers to the expression (\ref{wavefunction}) even in case of identical particles.}
\be
\hat\sigma_{\rm bound}(\hat s) = \frac{8\pi}{2\mu}\,\frac{\hat\sigma_0(\hat s)}{\beta(\hat s)}\, |\psi(\vv{0})|^2\,2\pi\,\delta(\hat s-M^2)
\label{sigma-bound}
\ee
where $M \simeq m_1 + m_2$ is the mass of the bound state and
\be
\beta(\hat s) \equiv \sqrt{\frac{2\mu}{(M/2)^2} \l(\sqrt{\hat s} - m_1 - m_2\r)}
\ee
is the factor that makes the continuum production cross section $\hat\sigma_0(\hat s)$ vanish at threshold. For $m_1 = m_2$, $\beta$ is the velocity of each particle in the center-of-mass frame.

The annihilation rate of the bound state into two mass-$m_0$ particles is
\bea
\Gamma &=& \frac{|\psi(\vv{0})|^2}{64\pi\, m_1 m_2} \sqrt{1 - \frac{m_0^2}{(M/2)^2}} \int_0^\pi d\theta\,\sin\theta\,\sum|\cM_0(\theta)|^2\nn\\
&&\q\l(\times\frac{1}{2}\;\begin{array}{c}\mbox{ for identical}\\ \mbox{ bound particles}\end{array}\r)\l(\times\frac{1}{2}\;\begin{array}{c}\mbox{ for identical}\\ \mbox{ final particles}\end{array}\r)
\label{ann-rate}
\eea
where $\cM_0(\theta)$ is the matrix element between the pair of particles that form the bound state (with a particular spin and color representation) and the annihilation products, for any polarization and color state of the bound state, and the sum is over the colors and polarizations of the products.

\section{KK gluonia vs. gluinonia\label{sec-KKgluonium}}
\setcounter{equation}{0}

In this section we will study bound states of pairs of color octets: KK gluons $\kkg$ (spin $1$) in UED and gluinos $\go$ (spin $1/2$) in MSSM. Later in this section we will also look at bound states of spin $0$ color octets.

It is useful to classify the various possible bound states in the model according to their spin $J$, color representation, parity $P$, charge conjugation $C$, and the transformation under the chiral $U(1)$ symmetry of the quarks. This allows one to immediately determine the possible production and annihilation channels and describe the corresponding processes using effective interaction vertices involving the bound state and Standard Model fields. These effective vertices can then be used to simulate the bound state processes in an event generator. The coefficients of the vertices can be determined by matching to the relevant Feynman diagrams of the free pair of particles (times the wavefunction at the origin and the other factors).

\begin{table}[t]
$$\begin{array}{c|c|c|c}
\mbox{} & \,\mbox{color} \,&\, J^{PC} \,&\, \mbox{can couple to} \,\\\hline
(\kkg\kkg)      & \vv{1}, \vv{8_S} & 0^{++} & G^{\rho\sigma}G_{\rho\sigma},\, \bar t t \\
                & \vv{8_A}         & 1^{+-} & \epsilon^{\mu\nu\rho\sigma}i\bar t \l[\gamma_\rho,\gamma_\sigma\r] t \\
                & \vv{1}, \vv{8_S} & 2^{++} & \; G^{\rho\mu}G_{\rho\nu},\, G^{\rho\sigma}D_\mu D_\nu G_{\rho\sigma},\, i\bar q \gamma^\mu D_\nu q \\\hline
(\go\go)        & \vv{1}, \vv{8_S} & 0^{-+} & G^{\rho\sigma}\tilde G_{\rho\sigma},\, i\bar t \gamma^5 t \\
                & \vv{8_A} & 1^{--} & \bar q\gamma^\mu q,\, i\bar t \l[\gamma^\mu, \gamma^\nu\r] t  \\
\end{array}$$
\caption{KK gluonia $(\kkg\kkg)$ (UED) and gluinonia $(\go\go)$ (MSSM) and their possible couplings to gluons and quarks. Here $G^{\mu\nu}$ is the gluon field strength and $\tilde G^{\mu\nu} = \frac{1}{2}\epsilon^{\mu\nu\rho\sigma}G_{\rho\sigma}$; $q$ denotes any quark flavor, while $t$ refers to the top quark whose mass we do not neglect. In part of the spin-$1$ cases, the coupling is not to the bound state operator $V_\mu$ but to its derivative $\pd_\mu V_\nu$. }
\label{tab-KKgKKg-gogo-couplings}
\end{table}

In table~\ref{tab-KKgKKg-gogo-couplings} we list the allowed KK gluonia and gluinonia (taking the spin-statistics theorem into account) and the effective vertices through which each of them can couple to gluon or quark bilinears. The possible couplings to a pair of photons (for color-singlets) are like the couplings to a pair of gluons.

To determine the charge conjugation properties of the color-octet bound states, it is useful to know that the action of charge conjugation $C$ on a non-abelian gauge field $V_\mu^a$ is~\cite{Tyutin:1982fx,Smolyakov:1980wq,Burgess-Moore}
\be
T^a V_\mu^a \to -\l(T^a\r)^{\rm T}V_\mu^a
\label{C-nonAbelian}
\ee
where $T^a$ are the generators of gauge transformations. The minus sign here can be described by saying that gluons have $C = -1$ (this is similar to the photon which transforms as $A_\mu \to -A_\mu$, thus having $C = -1$ in a more straightforward sense). Analogous transformation rules apply to KK gluons and gluinos. In the same sense, we find that $C = +1$ for $\vv{8_S}$ KK gluonia and gluinonia and $C = -1$ for $\vv{8_A}$ KK gluonia and gluinonia.\footnote{Our result for the charge conjugation of the $\vv{8_A}$ gluinonium ($(\go\go),\,\vv{8_A},\,J^{PC} = 1^{--}$), or perhaps just the convention for defining the sign of $C$ that would describe (\ref{C-nonAbelian}), differs from that of~\cite{Goldman:1984mj,Keung:1983wz,Kauth:2009ud}. Note that our result (but not theirs) allows the coupling of this gluinonium to massless $q\bar q$ pairs (via the vector current $\bar q\gamma^\mu q$), which must be possible as we know explicitly from the diagrams.}

From table~\ref{tab-KKgKKg-gogo-couplings} we see that the scalar KK gluonia and pseudoscalar gluinonia can be produced only by gluon fusion (but can annihilate also to $t\bar t$). The spin-$1$ KK gluonium cannot be produced at all from gluons or massless quarks while the vector gluinonium can be produced from quarks. The tensor KK gluonium couples to both gluons and quarks. We have checked that the explicit results that we will now present, based on diagrams in figures~\ref{fig-diagrams-KKgKKg} and~\ref{fig-diagrams-gogo}, indeed match these effective vertices. The tensor KK gluonium happens to couple to gluons only through $G^{\rho\mu}G_{\rho\nu}$ but not $G^{\rho\sigma}D_\mu D_\nu G_{\rho\sigma}$.

\begin{figure}[t]
$$
\beginpicture
\setcoordinatesystem units <0.571\tdim,0.571\tdim>
\stpltsmbl
\startrotation by 0.6 -0.6 about -100 35
\plot -84 30 -84 45 /
\plot -69 30 -69 45 /
\plot -54 30 -54 45 /
\ellipticalarc axes ratio 2:1 150 degrees from -92 31 center at -99 35
\ellipticalarc axes ratio 2:1 280 degrees from -77 31 center at -84 35
\ellipticalarc axes ratio 2:1 280 degrees from -62 31 center at -69 35
\ellipticalarc axes ratio 2:1 280 degrees from -47 31 center at -54 35
\stoprotation
\startrotation by 0.6 0.6 about -100 -35
\plot -84 -30 -84 -45 /
\plot -69 -30 -69 -45 /
\plot -54 -30 -54 -45 /
\ellipticalarc axes ratio 2:1 -150 degrees from -92 -31 center at -99 -35
\ellipticalarc axes ratio 2:1 -280 degrees from -77 -31 center at -84 -35
\ellipticalarc axes ratio 2:1 -280 degrees from -62 -31 center at -69 -35
\ellipticalarc axes ratio 2:1 -280 degrees from -47 -31 center at -54 -35
\stoprotation
\startrotation by 0.6 0.6 about -30 35
\ellipticalarc axes ratio 2:1 -150 degrees from -38 31 center at -31 35
\ellipticalarc axes ratio 2:1 -280 degrees from -53 31 center at -46 35
\ellipticalarc axes ratio 2:1 -280 degrees from -68 31 center at -61 35
\ellipticalarc axes ratio 2:1 -280 degrees from -83 31 center at -76 35
\stoprotation
\startrotation by 0.6 -0.6 about -30 -35
\ellipticalarc axes ratio 2:1 150 degrees from -38 -31 center at -31 -35
\ellipticalarc axes ratio 2:1 280 degrees from -53 -31 center at -46 -35
\ellipticalarc axes ratio 2:1 280 degrees from -68 -31 center at -61 -35
\ellipticalarc axes ratio 2:1 280 degrees from -83 -31 center at -76 -35
\stoprotation
\put {$\kkg$} at -115 30
\put {$\kkg$} at -115 -30
\put {$g$} at -20 30
\put {$g$} at -20 -30
\linethickness=0pt
\putrule from 0 -52 to 0 52
\putrule from -120 0 to 0 0
\endpicture
\beginpicture
\setcoordinatesystem units <0.571\tdim,0.571\tdim>
\stpltsmbl
\plot -54 30 -54 45 /
\plot -39 30 -39 45 /
\plot -24 30 -24 45 /
\plot -9  30  -9 45 /
\plot -54 -30 -54 -45 /
\plot -39 -30 -39 -45 /
\plot -24 -30 -24 -45 /
\plot -9  -30  -9 -45 /
\plot -5  26 10  26 /
\plot -5  11 10  11 /
\plot -5  -4 10  -4 /
\plot -5 -19 10 -19 /
\plot -5 -34 10 -34 /
\ellipticalarc axes ratio 2:1 150 degrees from -62 31 center at -69 35
\ellipticalarc axes ratio 2:1 280 degrees from -47 31 center at -54 35
\ellipticalarc axes ratio 2:1 280 degrees from -32 31 center at -39 35
\ellipticalarc axes ratio 2:1 280 degrees from -17 31 center at -24 35
\ellipticalarc axes ratio 2:1 220 degrees from -0 35 center at -10 35
\ellipticalarc axes ratio 2:1 -220 degrees from 0 35 center at 10 35
\ellipticalarc axes ratio 2:1 -280 degrees from 17 31 center at 24 35
\ellipticalarc axes ratio 2:1 -280 degrees from 32 31 center at 39 35
\ellipticalarc axes ratio 2:1 -280 degrees from 47 31 center at 54 35
\ellipticalarc axes ratio 2:1 -150 degrees from 62 31 center at 69 35
\ellipticalarc axes ratio 2:1 -150 degrees from -62 -31 center at -69 -35
\ellipticalarc axes ratio 2:1 -280 degrees from -47 -31 center at -54 -35
\ellipticalarc axes ratio 2:1 -280 degrees from -32 -31 center at -39 -35
\ellipticalarc axes ratio 2:1 -280 degrees from -17 -31 center at -24 -35
\ellipticalarc axes ratio 2:1 -220 degrees from 0 -35 center at -10 -35
\ellipticalarc axes ratio 2:1 220 degrees from 0 -35 center at 10 -35
\ellipticalarc axes ratio 2:1 280 degrees from 17 -31 center at 24 -35
\ellipticalarc axes ratio 2:1 280 degrees from 32 -31 center at 39 -35
\ellipticalarc axes ratio 2:1 280 degrees from 47 -31 center at 54 -35
\ellipticalarc axes ratio 2:1 150 degrees from 62 -31 center at 69 -35
\startrotation by 0 -1 about 0 30
\ellipticalarc axes ratio 2:1 -220 degrees from -5 29 center at  5 29
\ellipticalarc axes ratio 2:1 -280 degrees from 12 25 center at 19 29
\ellipticalarc axes ratio 2:1 -280 degrees from 27 25 center at 34 29
\ellipticalarc axes ratio 2:1 -280 degrees from 42 25 center at 49 29
\ellipticalarc axes ratio 2:1 -180 degrees from 57 25 center at 64 29
\stoprotation
\put {$\kkg$} at -55 15
\put {$\kkg$} at -55 -15
\put {$\kkg$} at -18 0
\put {$g$} at 60 15
\put {$g$} at 60 -15
\linethickness=0pt
\putrule from 0 -52 to 0 52
\putrule from -90 0 to 90 0
\endpicture
\beginpicture
\setcoordinatesystem units <0.571\tdim,0.571\tdim>
\stpltsmbl
\plot -54 30 -54 45 /
\plot -39 30 -39 45 /
\plot -24 30 -24 45 /
\plot -9  30  -9 45 /
\plot -54 -30 -54 -45 /
\plot -39 -30 -39 -45 /
\plot -24 -30 -24 -45 /
\plot -9  -30  -9 -45 /
\plot -5  26 10  26 /
\plot -5  11 10  11 /
\plot -5  -4 10  -4 /
\plot -5 -19 10 -19 /
\plot -5 -34 10 -34 /
\ellipticalarc axes ratio 2:1 150 degrees from -62 31 center at -69 35
\ellipticalarc axes ratio 2:1 280 degrees from -47 31 center at -54 35
\ellipticalarc axes ratio 2:1 280 degrees from -32 31 center at -39 35
\ellipticalarc axes ratio 2:1 280 degrees from -17 31 center at -24 35
\ellipticalarc axes ratio 2:1 220 degrees from -0 35 center at -10 35
\ellipticalarc axes ratio 2:1 -150 degrees from -62 -31 center at -69 -35
\ellipticalarc axes ratio 2:1 -280 degrees from -47 -31 center at -54 -35
\ellipticalarc axes ratio 2:1 -280 degrees from -32 -31 center at -39 -35
\ellipticalarc axes ratio 2:1 -280 degrees from -17 -31 center at -24 -35
\ellipticalarc axes ratio 2:1 -220 degrees from 0 -35 center at -10 -35
\startrotation by 0 -1 about 0 30
\ellipticalarc axes ratio 2:1 -220 degrees from -5 29 center at  5 29
\ellipticalarc axes ratio 2:1 -280 degrees from 12 25 center at 19 29
\ellipticalarc axes ratio 2:1 -280 degrees from 27 25 center at 34 29
\ellipticalarc axes ratio 2:1 -280 degrees from 42 25 center at 49 29
\ellipticalarc axes ratio 2:1 -150 degrees from 57 25 center at 64 29
\stoprotation
\startrotation by 0.7 -0.75 about 0 30
\ellipticalarc axes ratio 2:1 -220 degrees from 0 35 center at 10 35
\ellipticalarc axes ratio 2:1 -280 degrees from 17 31 center at 24 35
\ellipticalarc axes ratio 2:1 -280 degrees from 32 31 center at 39 35
\ellipticalarc axes ratio 2:1 -280 degrees from 47 31 center at 54 35
\ellipticalarc axes ratio 2:1 -280 degrees from 62 31 center at 69 35
\ellipticalarc axes ratio 2:1 -280 degrees from 77 31 center at 84 35
\ellipticalarc axes ratio 2:1 -150 degrees from 92 31 center at 99 35
\stoprotation
\startrotation by 0.7 0.75 about 0 -30
\ellipticalarc axes ratio 2:1 220 degrees from 0 -35 center at 10 -35
\ellipticalarc axes ratio 2:1 280 degrees from 17 -31 center at 24 -35
\ellipticalarc axes ratio 2:1 280 degrees from 32 -31 center at 39 -35
\ellipticalarc axes ratio 2:1 280 degrees from 47 -31 center at 54 -35
\ellipticalarc axes ratio 2:1 280 degrees from 62 -31 center at 69 -35
\ellipticalarc axes ratio 2:1 280 degrees from 77 -31 center at 84 -35
\ellipticalarc axes ratio 2:1 150 degrees from 92 -31 center at 99 -35
\stoprotation
\put {$\kkg$} at -55 15
\put {$\kkg$} at -55 -15
\put {$\kkg$} at -18 0
\put {$g$} at 75 15
\put {$g$} at 75 -15
\linethickness=0pt
\putrule from 0 -52 to 0 52
\putrule from -90 0 to 90 0
\endpicture
\beginpicture
\setcoordinatesystem units <0.571\tdim,0.571\tdim>
\stpltsmbl
\startrotation by 0.6 -0.6 about -100 35
\plot -84 30 -84 45 /
\plot -69 30 -69 45 /
\plot -54 30 -54 45 /
\ellipticalarc axes ratio 2:1 150 degrees from -92 31 center at -99 35
\ellipticalarc axes ratio 2:1 280 degrees from -77 31 center at -84 35
\ellipticalarc axes ratio 2:1 280 degrees from -62 31 center at -69 35
\ellipticalarc axes ratio 2:1 280 degrees from -47 31 center at -54 35
\stoprotation
\startrotation by 0.6 0.6 about -100 -35
\plot -84 -30 -84 -45 /
\plot -69 -30 -69 -45 /
\plot -54 -30 -54 -45 /
\ellipticalarc axes ratio 2:1 -150 degrees from -92 -31 center at -99 -35
\ellipticalarc axes ratio 2:1 -280 degrees from -77 -31 center at -84 -35
\ellipticalarc axes ratio 2:1 -280 degrees from -62 -31 center at -69 -35
\ellipticalarc axes ratio 2:1 -280 degrees from -47 -31 center at -54 -35
\stoprotation
\ellipticalarc axes ratio 2:1 220 degrees from  0  0 center at -10 0
\ellipticalarc axes ratio 2:1 280 degrees from -17 -4 center at -24 0
\ellipticalarc axes ratio 2:1 280 degrees from -32 -4 center at -39 0
\ellipticalarc axes ratio 2:1 260 degrees from -47 -4 center at -54 0
\startrotation by 0.6 0.6 about 37 35
\ellipticalarc axes ratio 2:1 -150 degrees from  29 31 center at  36 35
\ellipticalarc axes ratio 2:1 -280 degrees from  14 31 center at  21 35
\ellipticalarc axes ratio 2:1 -280 degrees from  -1 31 center at   6 35
\ellipticalarc axes ratio 2:1 -280 degrees from -16 31 center at  -9 35
\stoprotation
\startrotation by 0.6 -0.6 about 37 -35
\ellipticalarc axes ratio 2:1 150 degrees from  29 -31 center at  36 -35
\ellipticalarc axes ratio 2:1 280 degrees from  14 -31 center at  21 -35
\ellipticalarc axes ratio 2:1 280 degrees from  -1 -31 center at   6 -35
\ellipticalarc axes ratio 2:1 280 degrees from -16 -31 center at  -9 -35
\stoprotation
\put {$\kkg$} at -70 35
\put {$\kkg$} at -70 -30
\put {$g$} at -30 20
\put {$g$} at 5 30
\put {$g$} at 5 -30
\linethickness=0pt
\putrule from 0 -52 to 0 52
\putrule from -130 0 to 70 0
\endpicture
$$
$$
\beginpicture
\setcoordinatesystem units <0.571\tdim,0.571\tdim>
\stpltsmbl
\plot 0 35 70 35 /
\plot 0 -35 70 -35 /
\barrow from 5 35 to 40 35
\barrow from 70 -35 to 25 -35
\plot -54 30 -54 45 /
\plot -39 30 -39 45 /
\plot -24 30 -24 45 /
\plot -9  30  -9 45 /
\plot -54 -30 -54 -45 /
\plot -39 -30 -39 -45 /
\plot -24 -30 -24 -45 /
\plot -9  -30  -9 -45 /
\ellipticalarc axes ratio 2:1 150 degrees from -62 31 center at -69 35
\ellipticalarc axes ratio 2:1 280 degrees from -47 31 center at -54 35
\ellipticalarc axes ratio 2:1 280 degrees from -32 31 center at -39 35
\ellipticalarc axes ratio 2:1 280 degrees from -17 31 center at -24 35
\ellipticalarc axes ratio 2:1 220 degrees from -0 35 center at -10 35
\ellipticalarc axes ratio 2:1 -150 degrees from -62 -31 center at -69 -35
\ellipticalarc axes ratio 2:1 -280 degrees from -47 -31 center at -54 -35
\ellipticalarc axes ratio 2:1 -280 degrees from -32 -31 center at -39 -35
\ellipticalarc axes ratio 2:1 -280 degrees from -17 -31 center at -24 -35
\ellipticalarc axes ratio 2:1 -220 degrees from 0 -35 center at -10 -35
\plot 0 35 0 -35 /
\plot -6  25 6  25 /
\plot -6  10 6  10 /
\plot -6 -10 6 -10 /
\plot -6 -25 6 -25 /
\setsolid
\barrow from 0 -8 to 0 8
\put {$\kkg$} at -55 15
\put {$\kkg$} at -55 -15
\put {$\kkq$} at 15 0
\put {$q$} at 60 20
\put {$\bar q$} at 60 -20
\linethickness=0pt
\putrule from 0 -52 to 0 52
\putrule from -90 0 to 90 0
\endpicture
\beginpicture
\setcoordinatesystem units <0.571\tdim,0.571\tdim>
\stpltsmbl
\plot 0 35 70 -35 /
\plot 0 -35 30 -5 /
\plot 40 5 70 35 /
\barrow from 70 -35 to 45 -10
\barrow from 50 15 to 60 25
\plot -54 30 -54 45 /
\plot -39 30 -39 45 /
\plot -24 30 -24 45 /
\plot -9  30  -9 45 /
\plot -54 -30 -54 -45 /
\plot -39 -30 -39 -45 /
\plot -24 -30 -24 -45 /
\plot -9  -30  -9 -45 /
\ellipticalarc axes ratio 2:1 150 degrees from -62 31 center at -69 35
\ellipticalarc axes ratio 2:1 280 degrees from -47 31 center at -54 35
\ellipticalarc axes ratio 2:1 280 degrees from -32 31 center at -39 35
\ellipticalarc axes ratio 2:1 280 degrees from -17 31 center at -24 35
\ellipticalarc axes ratio 2:1 220 degrees from -0 35 center at -10 35
\ellipticalarc axes ratio 2:1 -150 degrees from -62 -31 center at -69 -35
\ellipticalarc axes ratio 2:1 -280 degrees from -47 -31 center at -54 -35
\ellipticalarc axes ratio 2:1 -280 degrees from -32 -31 center at -39 -35
\ellipticalarc axes ratio 2:1 -280 degrees from -17 -31 center at -24 -35
\ellipticalarc axes ratio 2:1 -220 degrees from 0 -35 center at -10 -35
\plot  0  35 0 -35 /
\plot -6  25 6  25 /
\plot -6  10 6  10 /
\plot -6 -10 6 -10 /
\plot -6 -25 6 -25 /
\barrow from 0 8 to 0 -8
\put {$\kkg$} at -55 15
\put {$\kkg$} at -55 -15
\put {$\kkq$} at -15 0
\put {$q$} at 70 15
\put {$\bar q$} at 70 -15
\linethickness=0pt
\putrule from 0 -52 to 0 52
\putrule from -90 0 to 90 0
\endpicture
\beginpicture
\setcoordinatesystem units <0.571\tdim,0.571\tdim>
\stpltsmbl
\startrotation by 0.6 -0.6 about -100 35
\plot -84 30 -84 45 /
\plot -69 30 -69 45 /
\plot -54 30 -54 45 /
\ellipticalarc axes ratio 2:1 150 degrees from -92 31 center at -99 35
\ellipticalarc axes ratio 2:1 280 degrees from -77 31 center at -84 35
\ellipticalarc axes ratio 2:1 280 degrees from -62 31 center at -69 35
\ellipticalarc axes ratio 2:1 280 degrees from -47 31 center at -54 35
\stoprotation
\startrotation by 0.6 0.6 about -100 -35
\plot -84 -30 -84 -45 /
\plot -69 -30 -69 -45 /
\plot -54 -30 -54 -45 /
\ellipticalarc axes ratio 2:1 -150 degrees from -92 -31 center at -99 -35
\ellipticalarc axes ratio 2:1 -280 degrees from -77 -31 center at -84 -35
\ellipticalarc axes ratio 2:1 -280 degrees from -62 -31 center at -69 -35
\ellipticalarc axes ratio 2:1 -280 degrees from -47 -31 center at -54 -35
\stoprotation
\ellipticalarc axes ratio 2:1 220 degrees from  0  0 center at -10 0
\ellipticalarc axes ratio 2:1 280 degrees from -17 -4 center at -24 0
\ellipticalarc axes ratio 2:1 280 degrees from -32 -4 center at -39 0
\ellipticalarc axes ratio 2:1 260 degrees from -47 -4 center at -54 0
\plot 37 37 0 0 37 -37 /
\barrow from 0 0 to 25 25
\barrow from 35 -35 to 15 -15
\put {$\kkg$} at -70 35
\put {$\kkg$} at -70 -30
\put {$g$} at -30 20
\put {$q$} at 10 30
\put {$\bar q$} at 10 -30
\linethickness=0pt
\putrule from 0 -52 to 0 52
\putrule from -120 0 to 60 0
\endpicture$$
\caption{Diagrams for production or annihilation of a pair of KK gluons. For bound states, the diagrams with $s$ channel gluon do not happen to contribute.}
\label{fig-diagrams-KKgKKg}
$$\beginpicture
\setcoordinatesystem units <0.571\tdim,0.571\tdim>
\stpltsmbl
\plot -70 35 0 35 0 -35 -70 -35 /
\ellipticalarc axes ratio 2:1 150 degrees from -62 31 center at -69 35
\ellipticalarc axes ratio 2:1 280 degrees from -47 31 center at -54 35
\ellipticalarc axes ratio 2:1 280 degrees from -32 31 center at -39 35
\ellipticalarc axes ratio 2:1 280 degrees from -17 31 center at -24 35
\ellipticalarc axes ratio 2:1 220 degrees from -0 35 center at -10 35
\ellipticalarc axes ratio 2:1 -220 degrees from 0 35 center at 10 35
\ellipticalarc axes ratio 2:1 -280 degrees from 17 31 center at 24 35
\ellipticalarc axes ratio 2:1 -280 degrees from 32 31 center at 39 35
\ellipticalarc axes ratio 2:1 -280 degrees from 47 31 center at 54 35
\ellipticalarc axes ratio 2:1 -150 degrees from 62 31 center at 69 35
\ellipticalarc axes ratio 2:1 -150 degrees from -62 -31 center at -69 -35
\ellipticalarc axes ratio 2:1 -280 degrees from -47 -31 center at -54 -35
\ellipticalarc axes ratio 2:1 -280 degrees from -32 -31 center at -39 -35
\ellipticalarc axes ratio 2:1 -280 degrees from -17 -31 center at -24 -35
\ellipticalarc axes ratio 2:1 -220 degrees from 0 -35 center at -10 -35
\ellipticalarc axes ratio 2:1 220 degrees from 0 -35 center at 10 -35
\ellipticalarc axes ratio 2:1 280 degrees from 17 -31 center at 24 -35
\ellipticalarc axes ratio 2:1 280 degrees from 32 -31 center at 39 -35
\ellipticalarc axes ratio 2:1 280 degrees from 47 -31 center at 54 -35
\ellipticalarc axes ratio 2:1 150 degrees from 62 -31 center at 69 -35
\startrotation by 0 -1 about 0 30
\ellipticalarc axes ratio 2:1 -220 degrees from -5 29 center at  5 29
\ellipticalarc axes ratio 2:1 -280 degrees from 12 25 center at 19 29
\ellipticalarc axes ratio 2:1 -280 degrees from 27 25 center at 34 29
\ellipticalarc axes ratio 2:1 -280 degrees from 42 25 center at 49 29
\ellipticalarc axes ratio 2:1 -180 degrees from 57 25 center at 64 29
\stoprotation
\put {$\go$} at -55 15
\put {$\go$} at -55 -15
\put {$\go$} at -18 0
\put {$g$} at 60 15
\put {$g$} at 60 -15
\linethickness=0pt
\putrule from 0 -52 to 0 52
\putrule from -90 0 to 90 0
\endpicture
\beginpicture
\setcoordinatesystem units <0.571\tdim,0.571\tdim>
\stpltsmbl
\plot -70 35 0 35 0 -35 -70 -35 /
\ellipticalarc axes ratio 2:1 150 degrees from -62 31 center at -69 35
\ellipticalarc axes ratio 2:1 280 degrees from -47 31 center at -54 35
\ellipticalarc axes ratio 2:1 280 degrees from -32 31 center at -39 35
\ellipticalarc axes ratio 2:1 280 degrees from -17 31 center at -24 35
\ellipticalarc axes ratio 2:1 220 degrees from -0 35 center at -10 35
\ellipticalarc axes ratio 2:1 -150 degrees from -62 -31 center at -69 -35
\ellipticalarc axes ratio 2:1 -280 degrees from -47 -31 center at -54 -35
\ellipticalarc axes ratio 2:1 -280 degrees from -32 -31 center at -39 -35
\ellipticalarc axes ratio 2:1 -280 degrees from -17 -31 center at -24 -35
\ellipticalarc axes ratio 2:1 -220 degrees from 0 -35 center at -10 -35
\startrotation by 0 -1 about 0 30
\ellipticalarc axes ratio 2:1 -220 degrees from -5 29 center at  5 29
\ellipticalarc axes ratio 2:1 -280 degrees from 12 25 center at 19 29
\ellipticalarc axes ratio 2:1 -280 degrees from 27 25 center at 34 29
\ellipticalarc axes ratio 2:1 -280 degrees from 42 25 center at 49 29
\ellipticalarc axes ratio 2:1 -150 degrees from 57 25 center at 64 29
\stoprotation
\startrotation by 0.7 -0.75 about 0 30
\ellipticalarc axes ratio 2:1 -220 degrees from 0 35 center at 10 35
\ellipticalarc axes ratio 2:1 -280 degrees from 17 31 center at 24 35
\ellipticalarc axes ratio 2:1 -280 degrees from 32 31 center at 39 35
\ellipticalarc axes ratio 2:1 -280 degrees from 47 31 center at 54 35
\ellipticalarc axes ratio 2:1 -280 degrees from 62 31 center at 69 35
\ellipticalarc axes ratio 2:1 -280 degrees from 77 31 center at 84 35
\ellipticalarc axes ratio 2:1 -150 degrees from 92 31 center at 99 35
\stoprotation
\startrotation by 0.7 0.75 about 0 -30
\ellipticalarc axes ratio 2:1 220 degrees from 0 -35 center at 10 -35
\ellipticalarc axes ratio 2:1 280 degrees from 17 -31 center at 24 -35
\ellipticalarc axes ratio 2:1 280 degrees from 32 -31 center at 39 -35
\ellipticalarc axes ratio 2:1 280 degrees from 47 -31 center at 54 -35
\ellipticalarc axes ratio 2:1 280 degrees from 62 -31 center at 69 -35
\ellipticalarc axes ratio 2:1 280 degrees from 77 -31 center at 84 -35
\ellipticalarc axes ratio 2:1 150 degrees from 92 -31 center at 99 -35
\stoprotation
\put {$\go$} at -55 15
\put {$\go$} at -55 -15
\put {$\go$} at -18 0
\put {$g$} at 75 15
\put {$g$} at 75 -15
\linethickness=0pt
\putrule from 0 -52 to 0 52
\putrule from -90 0 to 90 0
\endpicture
\beginpicture
\setcoordinatesystem units <0.571\tdim,0.571\tdim>
\stpltsmbl
\plot -100 35 -65 0 /
\startrotation by 0.6 -0.6 about -100 35
\ellipticalarc axes ratio 2:1 150 degrees from -92 31 center at -99 35
\ellipticalarc axes ratio 2:1 280 degrees from -77 31 center at -84 35
\ellipticalarc axes ratio 2:1 280 degrees from -62 31 center at -69 35
\ellipticalarc axes ratio 2:1 280 degrees from -47 31 center at -54 35
\stoprotation
\plot -100 -35 -65 0 /
\startrotation by 0.6 0.6 about -100 -35
\ellipticalarc axes ratio 2:1 -150 degrees from -92 -31 center at -99 -35
\ellipticalarc axes ratio 2:1 -280 degrees from -77 -31 center at -84 -35
\ellipticalarc axes ratio 2:1 -280 degrees from -62 -31 center at -69 -35
\ellipticalarc axes ratio 2:1 -280 degrees from -47 -31 center at -54 -35
\stoprotation
\ellipticalarc axes ratio 2:1 220 degrees from  0  0 center at -10 0
\ellipticalarc axes ratio 2:1 280 degrees from -17 -4 center at -24 0
\ellipticalarc axes ratio 2:1 280 degrees from -32 -4 center at -39 0
\ellipticalarc axes ratio 2:1 260 degrees from -47 -4 center at -54 0
\startrotation by 0.6 0.6 about 37 35
\ellipticalarc axes ratio 2:1 -150 degrees from  29 31 center at  36 35
\ellipticalarc axes ratio 2:1 -280 degrees from  14 31 center at  21 35
\ellipticalarc axes ratio 2:1 -280 degrees from  -1 31 center at   6 35
\ellipticalarc axes ratio 2:1 -280 degrees from -16 31 center at  -9 35
\stoprotation
\startrotation by 0.6 -0.6 about 37 -35
\ellipticalarc axes ratio 2:1 150 degrees from  29 -31 center at  36 -35
\ellipticalarc axes ratio 2:1 280 degrees from  14 -31 center at  21 -35
\ellipticalarc axes ratio 2:1 280 degrees from  -1 -31 center at   6 -35
\ellipticalarc axes ratio 2:1 280 degrees from -16 -31 center at  -9 -35
\stoprotation
\put {$\go$} at -70 35
\put {$\go$} at -70 -30
\put {$g$} at -30 20
\put {$g$} at 5 30
\put {$g$} at 5 -30
\linethickness=0pt
\putrule from 0 -52 to 0 52
\putrule from -130 0 to 90 0
\endpicture$$
$$
\beginpicture
\setcoordinatesystem units <0.571\tdim,0.571\tdim>
\stpltsmbl
\plot -70 35 70 35 /
\plot -70 -35 70 -35 /
\barrow from 5 35 to 40 35
\barrow from 70 -35 to 25 -35
\ellipticalarc axes ratio 2:1 150 degrees from -62 31 center at -69 35
\ellipticalarc axes ratio 2:1 280 degrees from -47 31 center at -54 35
\ellipticalarc axes ratio 2:1 280 degrees from -32 31 center at -39 35
\ellipticalarc axes ratio 2:1 280 degrees from -17 31 center at -24 35
\ellipticalarc axes ratio 2:1 220 degrees from -0 35 center at -10 35
\ellipticalarc axes ratio 2:1 -150 degrees from -62 -31 center at -69 -35
\ellipticalarc axes ratio 2:1 -280 degrees from -47 -31 center at -54 -35
\ellipticalarc axes ratio 2:1 -280 degrees from -32 -31 center at -39 -35
\ellipticalarc axes ratio 2:1 -280 degrees from -17 -31 center at -24 -35
\ellipticalarc axes ratio 2:1 -220 degrees from 0 -35 center at -10 -35
\setdashes
\plot 0 35 0 -35 /
\setsolid
\barrow from 0 -8 to 0 8
\put {$\go$} at -55 15
\put {$\go$} at -55 -15
\put {$\sq$} at 15 0
\put {$q$} at 60 20
\put {$\bar q$} at 60 -20
\linethickness=0pt
\putrule from 0 -52 to 0 52
\putrule from -60 0 to 90 0
\endpicture
\beginpicture
\setcoordinatesystem units <0.571\tdim,0.571\tdim>
\stpltsmbl
\plot -70 35 0 35 70 -35 /
\plot -70 -35 0 -35 30 -5 /
\plot 40 5 70 35 /
\barrow from 70 -35 to 45 -10
\barrow from 50 15 to 60 25
\ellipticalarc axes ratio 2:1 150 degrees from -62 31 center at -69 35
\ellipticalarc axes ratio 2:1 280 degrees from -47 31 center at -54 35
\ellipticalarc axes ratio 2:1 280 degrees from -32 31 center at -39 35
\ellipticalarc axes ratio 2:1 280 degrees from -17 31 center at -24 35
\ellipticalarc axes ratio 2:1 220 degrees from -0 35 center at -10 35
\ellipticalarc axes ratio 2:1 -150 degrees from -62 -31 center at -69 -35
\ellipticalarc axes ratio 2:1 -280 degrees from -47 -31 center at -54 -35
\ellipticalarc axes ratio 2:1 -280 degrees from -32 -31 center at -39 -35
\ellipticalarc axes ratio 2:1 -280 degrees from -17 -31 center at -24 -35
\ellipticalarc axes ratio 2:1 -220 degrees from 0 -35 center at -10 -35
\setdashes
\plot 0 35 0 -35 /
\setsolid
\barrow from 0 8 to 0 -8
\put {$\go$} at -55 15
\put {$\go$} at -55 -15
\put {$\sq$} at -15 0
\put {$q$} at 70 15
\put {$\bar q$} at 70 -15
\linethickness=0pt
\putrule from 0 -52 to 0 52
\putrule from -90 0 to 90 0
\endpicture
\beginpicture
\setcoordinatesystem units <0.571\tdim,0.571\tdim>
\stpltsmbl
\plot -100 35 -65 0 /
\startrotation by 0.6 -0.6 about -100 35
\ellipticalarc axes ratio 2:1 150 degrees from -92 31 center at -99 35
\ellipticalarc axes ratio 2:1 280 degrees from -77 31 center at -84 35
\ellipticalarc axes ratio 2:1 280 degrees from -62 31 center at -69 35
\ellipticalarc axes ratio 2:1 280 degrees from -47 31 center at -54 35
\stoprotation
\plot -100 -35 -65 0 /
\startrotation by 0.6 0.6 about -100 -35
\ellipticalarc axes ratio 2:1 -150 degrees from -92 -31 center at -99 -35
\ellipticalarc axes ratio 2:1 -280 degrees from -77 -31 center at -84 -35
\ellipticalarc axes ratio 2:1 -280 degrees from -62 -31 center at -69 -35
\ellipticalarc axes ratio 2:1 -280 degrees from -47 -31 center at -54 -35
\stoprotation
\setsolid
\ellipticalarc axes ratio 2:1 220 degrees from  0  0 center at -10 0
\ellipticalarc axes ratio 2:1 280 degrees from -17 -4 center at -24 0
\ellipticalarc axes ratio 2:1 280 degrees from -32 -4 center at -39 0
\ellipticalarc axes ratio 2:1 260 degrees from -47 -4 center at -54 0
\plot 37 37 0 0 37 -37 /
\barrow from 0 0 to 25 25
\barrow from 35 -35 to 15 -15
\put {$\go$} at -70 35
\put {$\go$} at -70 -30
\put {$g$} at -30 20
\put {$q$} at 10 30
\put {$\bar q$} at 10 -30
\linethickness=0pt
\putrule from 0 -52 to 0 52
\putrule from -120 0 to 60 0
\endpicture
$$
\caption{Diagrams for production or annihilation of a pair of gluinos.}
\label{fig-diagrams-gogo}
\end{figure}

\begin{table}[t]
$$\begin{array}{|c|c|c|c|c|}\hline
\,\mbox{process}                \,&\, J = 0\,, J_z = 0 \,&\, J = 1\,, |J_z| = 1 \,&\, J = 2\,, |J_z| = 2 \,&\, J = 2\,, |J_z| = 1 \,\\\hline\hline
\,gg\to(\kkg\kkg)_\vv{1}        \,&\,729/128\,&\,         \,&\,243/8 \,& \\
\,gg\to(\kkg\kkg)_\vv{8_S}      \,&\,729/512\,&\,         \,&\,243/32\,& \\\hline
\,gg \to(\go\go)_\vv{1}         \,&\,243/64 \,&\,         \,&\,              \,& \,\\
\,gg \to(\go\go)_\vv{8_S}       \,&\,243/256\,&\,         \,&\,              \,& \,\\\hline
\,gg \to(\phi\phi)_\vv{1}       \,&\,243/128\,&\,         \,&\,              \,& \,\\
\,gg \to(\phi\phi)_\vv{8_S}     \,&\,243/512\,&\,         \,&\,              \,& \,\\\hline\hline
\,q\bar q\to(\kkg \kkg)_\vv{1}  \,&\,               \,&\,         \,&\,              \,&\, \ds 4\l(\frac{2m_\kkg^2}{m_\kkg^2 + m_\kkq^2}\r)^2 \,\\
\,q\bar q\to(\kkg \kkg)_\vv{8_S}\,&\,               \,&\,         \,&\,              \,&\, \ds\frac{5}{4}\l(\frac{2m_\kkg^2}{m_\kkg^2 + m_\kkq^2}\r)^2 \,\\\hline
\,q\bar q\to(\go\go)_\vv{8_A}   \,&\,               \,&\,\ds\frac{9}{4}\l(\frac{m_\sq^2 - m_\go^2}{m_\sq^2 + m_\go^2}\r)^2 \,&\, \,&\, \,\\\hline
\end{array}$$
\caption{KK gluonium $(\kkg\kkg)$ (UED), gluinonium $(\go\go)$ (MSSM), and octetonium $(\phi\phi)$ production cross section prefactors $P_{ij}$ of (\ref{parton-cs}). The different columns correspond to different values of the spin $J$ of the bound state and its polarization (the projection $J_z$ on the beam axis).}
\label{tab-KKgluonium}
\end{table}

\begin{figure}[t]
\begin{center}
\includegraphics[width=0.49\textwidth]{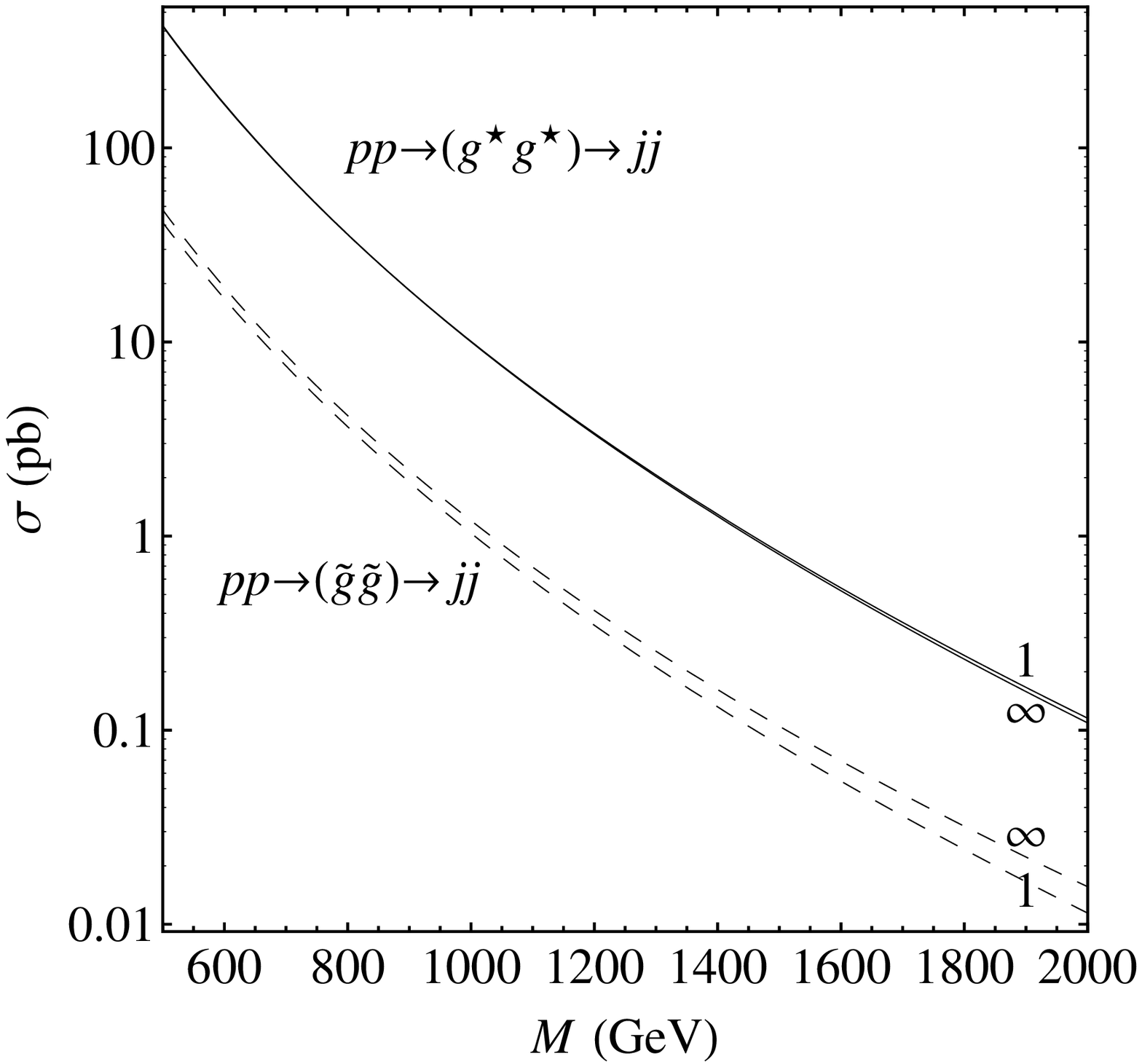}
\includegraphics[width=0.50\textwidth]{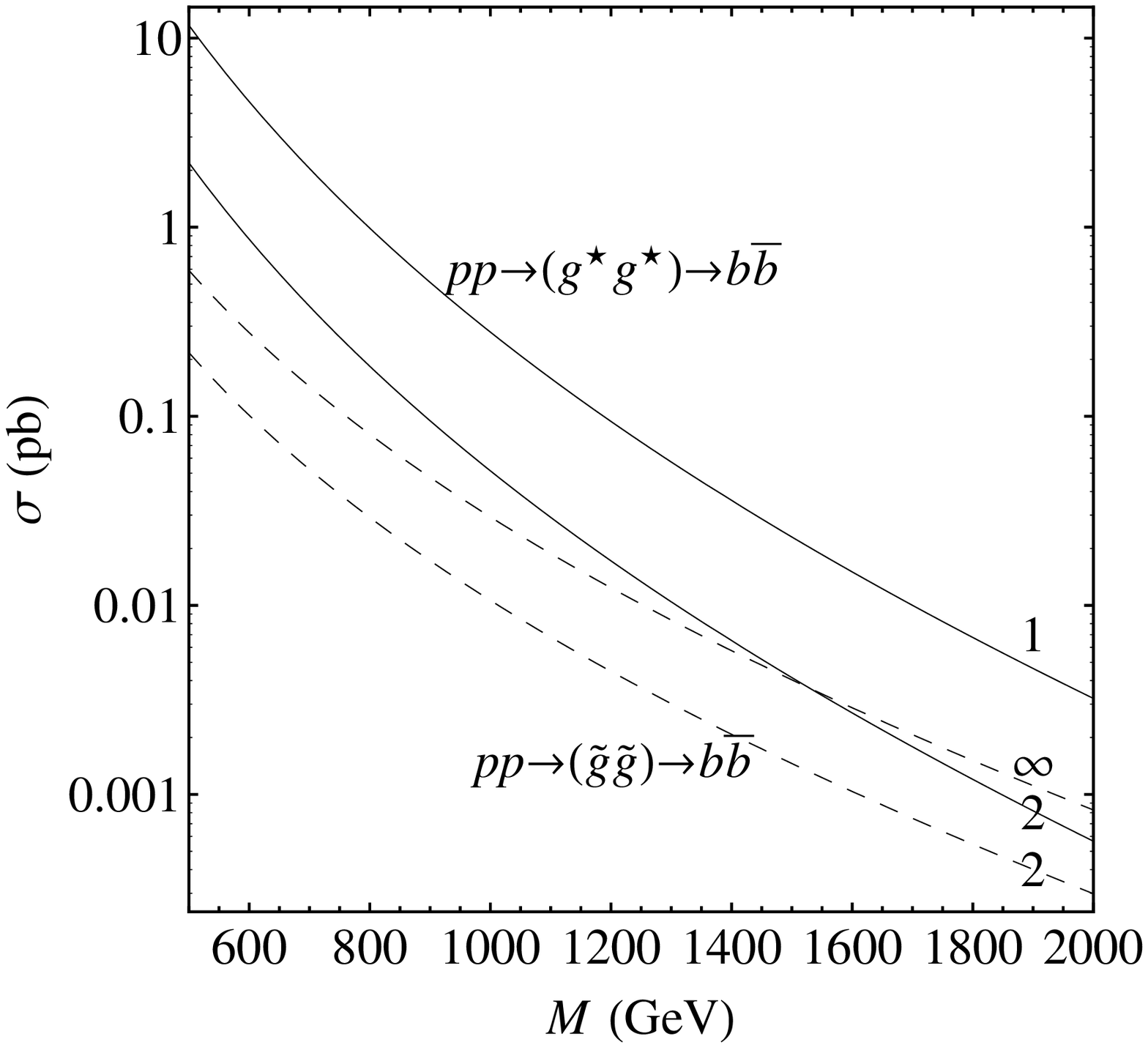}\\\vspace{3mm}
\includegraphics[width=0.50\textwidth]{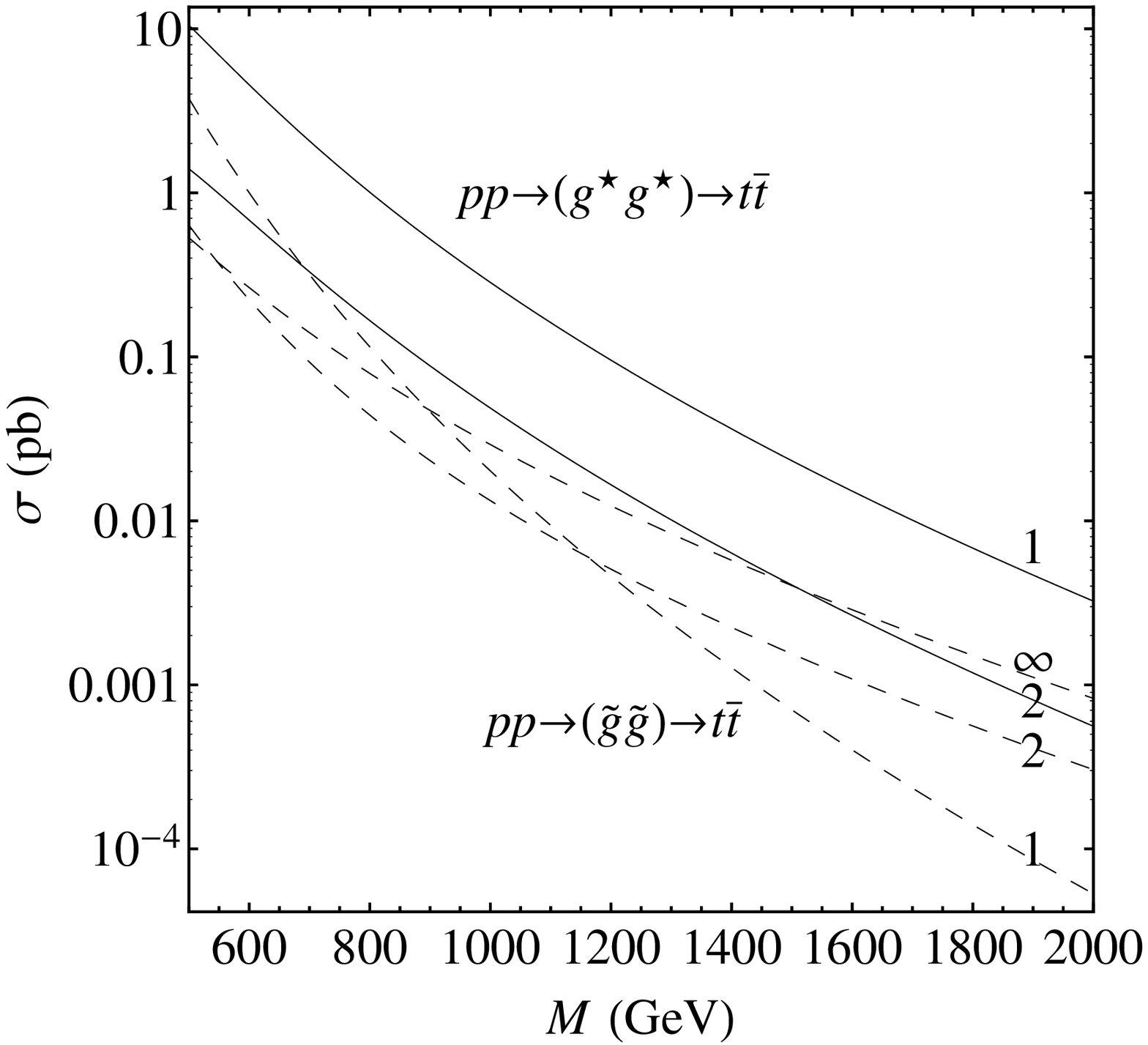}
\caption{Dijet (top left), $b\bar b$ (top right) and $t\bar t$ (bottom) signals from KK gluonium annihilation (solid lines) vs. gluinonium annihilation (dashed lines) as a function of the resonance mass ($M = 2m_\kkg$ or $2m_\go$) at the $14$ TeV LHC. The different curves correspond to different ratios of $m_\kkq/m_\kkg$ or $m_\sq/m_\go$ that are indicated next to them. The $b\bar b$ signal of $(\go\go)$ vanishes for $m_\sq = m_\go$, and the $b\bar b$ and $t\bar t$ signals of $(\kkg\kkg)$ vanish for $m_\kkq/m_\kkg \to \infty$.}
\label{fig-cs-M}
\end{center}
\end{figure}

\begin{figure}[t]
\begin{center}
\includegraphics[width=0.43\textwidth]{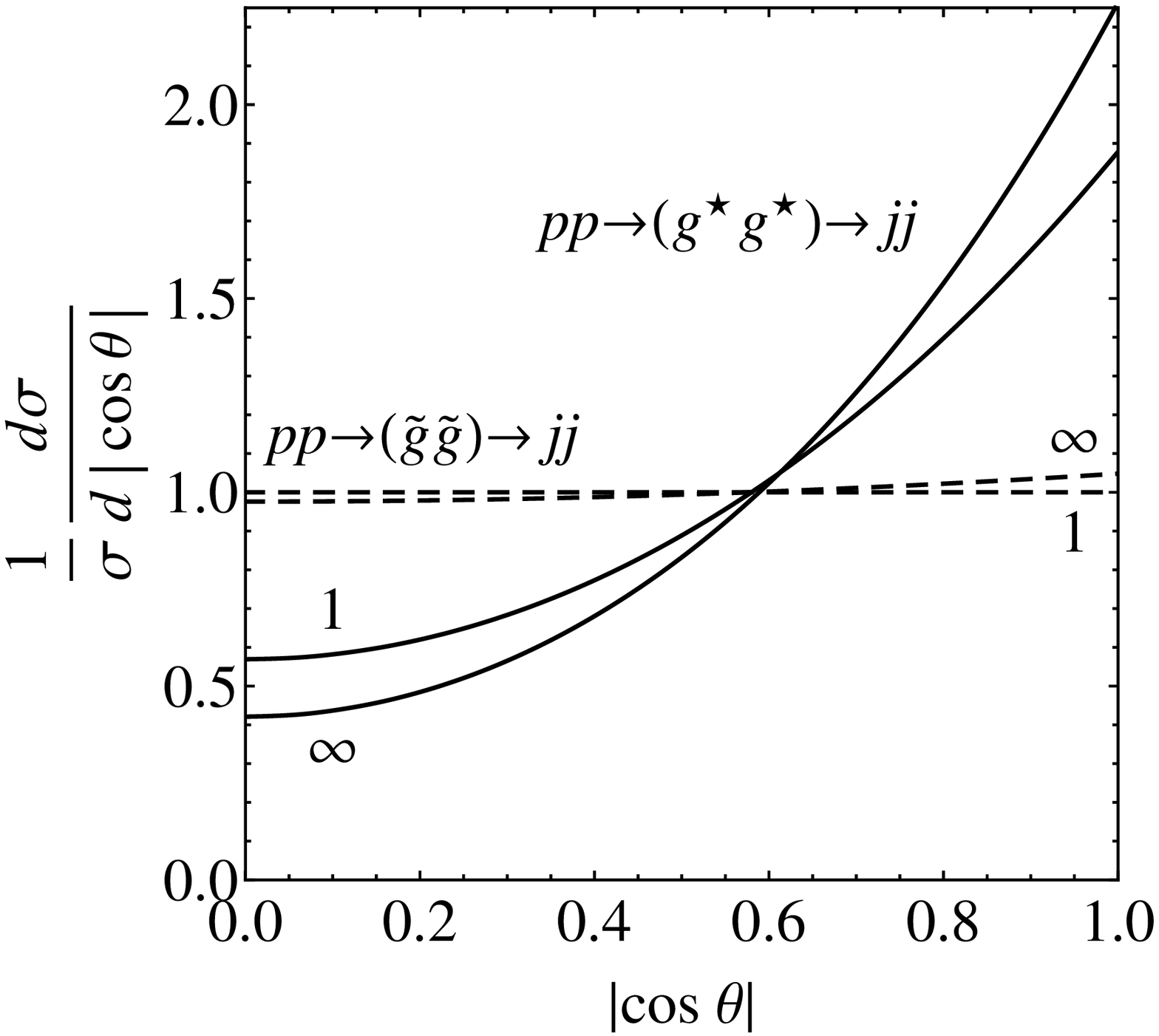}
\includegraphics[width=0.43\textwidth]{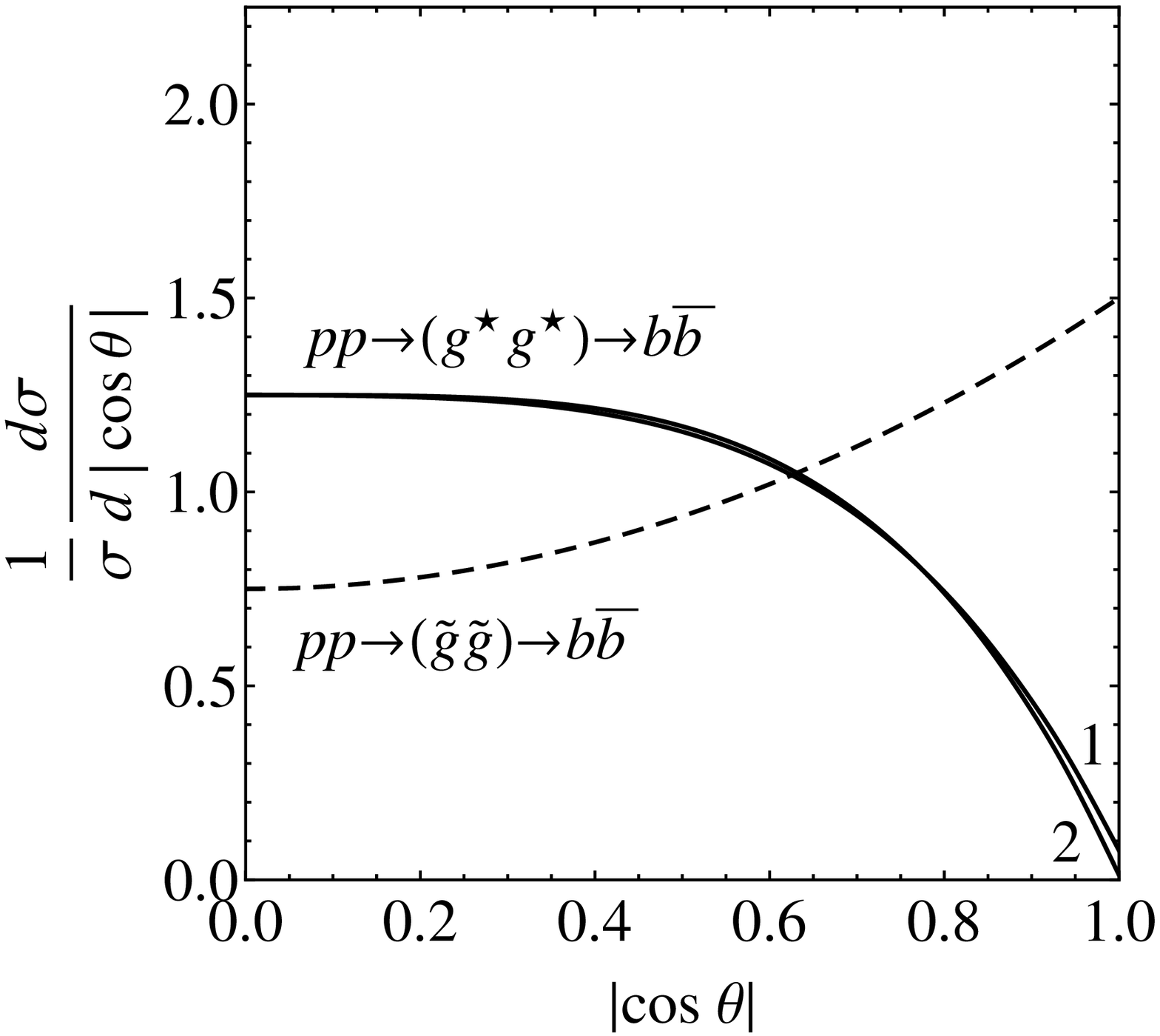}\\\vspace{3mm}
\includegraphics[width=0.43\textwidth]{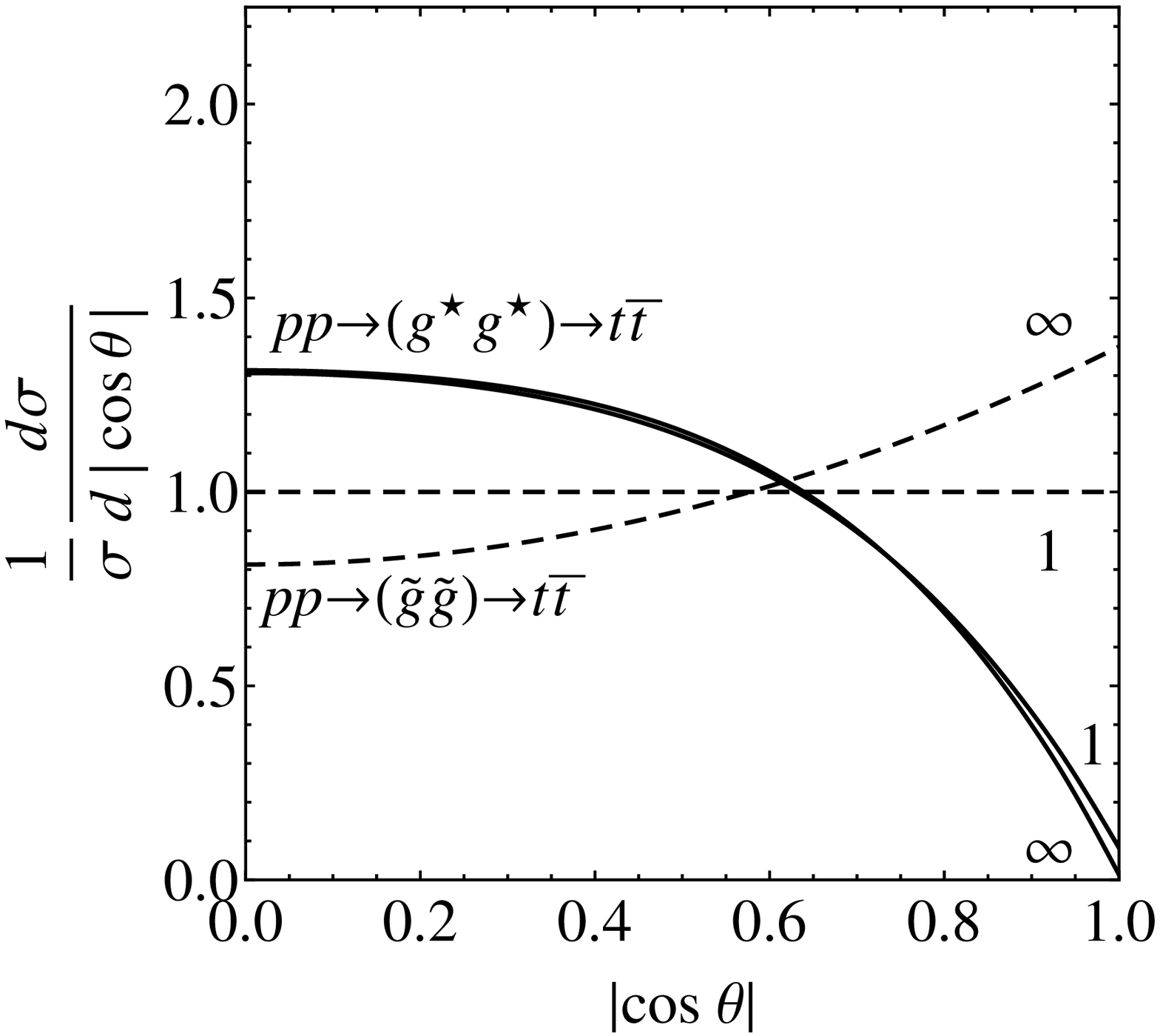}
\caption{Angular distributions in annihilation of KK gluonia and gluinonia into dijets (top left), $b\bar b$ (top right) and $t\bar t$ (bottom) at the LHC for $M = 800$~GeV (i.e., $m_\kkg = m_\go = 400$~GeV). Here $\theta$ is the angle between the beam axis and the direction of motion of the annihilation products in the center-of-mass frame. The different curves correspond to different ratios of $m_\kkq/m_\kkg$ or $m_\sq/m_\go$ that are indicated next to them (except for the $b\bar b$ signal of $(\go\go)$ where there is no such dependence).}
\label{fig-angular}
\end{center}
\end{figure}

\begin{figure}[t]
\begin{center}
\includegraphics[width=0.43\textwidth]{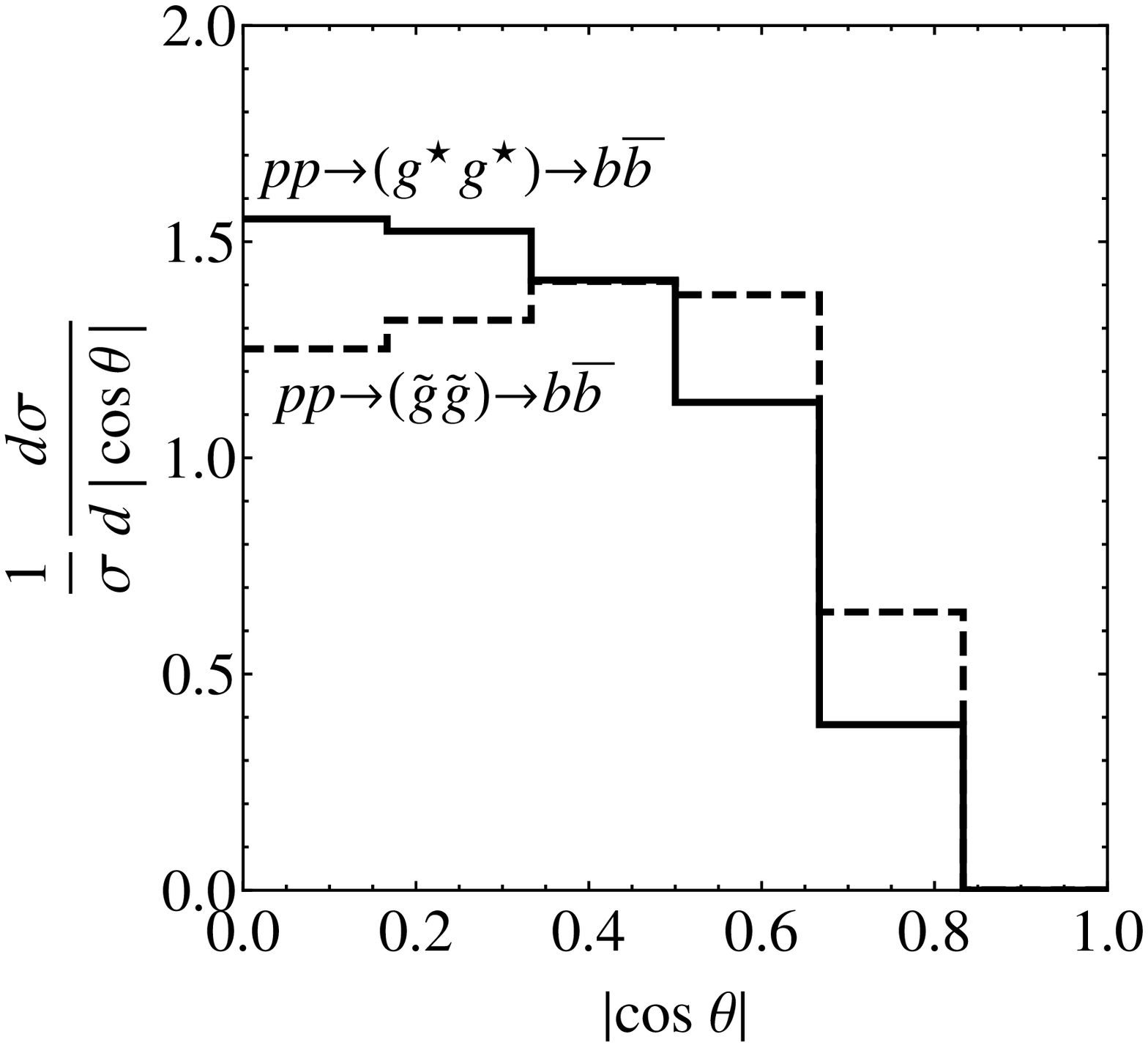}
\caption{Simulated angular distributions in annihilation of KK gluonia and gluinonia into $b\bar b$ at the LHC.}
\label{fig-angular-Pythia}
\end{center}
\end{figure}

The parton-level bound state production cross sections can be written as
\be
\hat\sigma_{{\rm bound},\,ij}(\hat s) =
P_{ij}\, \zeta(3)\,\pi^2\bar\alpha_s^3\alpha_s^2\,\delta(\hat s-M^2)
\label{parton-cs}
\ee
where $\zeta(3) = \sum_{n=1}^\infty 1/n^3 \simeq 1.2$ takes into account the contributions of the radial excitations and the prefactors $P_{ij}$ for producing the various bound states from partons $i,j$ are presented in table~\ref{tab-KKgluonium} for KK gluonium and gluinonium. These results are based on the diagrams in figures~\ref{fig-diagrams-KKgKKg} and~\ref{fig-diagrams-gogo} and factors such as (\ref{wavefunction}). More details are given in appendix~\ref{app-rates}. After convoluting (\ref{parton-cs}) with the parton distribution functions as\footnote{We are using NLO MSTW 2008 PDFs~\cite{Martin:2009iq} evaluated at the scale $M/2$, and the center-of-mass energy $\sqrt s = 14$ TeV for the LHC.}
\be
\sigma_{\rm bound} = \frac{\zeta(3)\,\pi^2\bar\alpha_s^3\alpha_s^2}{s} \sum_{i,j}P_{ij}\int_{M^2/s}^1 \frac{dx}{x}\, f_{i/p}(x)\, f_{j/p}\l(\frac{M^2}{xs}\r)
\ee
and multiplying by the appropriate branching ratios, we obtain the cross sections for dijet, $b\bar b$ and $t\bar t$ final states as shown in figure~\ref{fig-cs-M}. We also show to what extent the results depend on the mass ratios $m_\kkq/m_\kkg$ or $m_\sq/m_\go$ by varying them between $1$ and $\infty$.\footnote{In practice, the ratio must be somewhat larger than $1$ in order for the annihilation rates to dominate, and the extreme limit of $m_\kkq/m_\kkg \to \infty$ is unphysical for UED (although it may be relevant for some other theory that does not contain KK quarks).} We see that the signals will typically be an order of magnitude larger for KK gluonia than for gluinonia of the same mass (except if the KK quarks are very heavy, in which case the $b\bar b$ and $t\bar t$ signals of KK gluonia disappear).

It is interesting to ask what creates this large difference in the cross sections. Unless the KK gluons (or gluinos) are heavier than about a TeV, production via the $gg$ channel dominates due to its higher luminosity, so let us discuss it first and understand the order-of-magnitude difference seen already in the parton level expressions of table~\ref{tab-KKgluonium}. The differences should be attributed to a large extent to the more numerous spin possibilities for the KK gluonia. According to table~\ref{tab-KKgluonium}, in the $gg$ channel (in both $\vv{1}$ and $\vv{8_S}$ representations), the parton-level production cross section for $J=0$ KK gluonium is only a factor of $3/2$ larger than that for $J=0$ gluinonium. However, the KK gluonium can also be produced in $J=2$ state, whose cross section (summed over the spin projections) is larger than that of the ($J=0$) gluinonium by a factor of $8$ (or larger than that of $J=0$ KK gluonium by a factor of $16/3 \sim 5$). Overall, the $gg$ channel produces $9.5$ times more KK gluonium than gluinonium.\footnote{In this case, this happens to be exactly the ratio of the near-threshold pair production cross sections (despite the fact that for the purposes of bound states we multiply this quantity by a different $\l|\psi(\vv{0})\r|^2$ for each of the attractive color representations and exclude the repulsive ones). In general, this does not need to be the case. For example, in the $q\bar q$ channel the ratio between bound state and near-threshold pair production cross section is $3$ times bigger for the KK gluonia than for the gluinonia because the color representations in which KK gluon and gluino pairs are produced at threshold are different.} Note that these results are determined by gauge interactions alone, and therefore apply much more generally than to just UED and MSSM. In particular, the production diagrams do not involve any other new particles besides the KK gluons or gluinos.

In the $q\bar q$ channel, the comparison between KK gluonia and gluinonia is sensitive to the masses of the KK quarks (relative to the KK gluon) or squarks (relative to the gluino). If the KK quarks are very heavy the KK gluonium cross section goes to zero, while there is no such effect for the gluinonium. This is because the gluinonium can be produced via a diagram with an $s$ channel gluon that does not involve squarks, while a similar diagram for KK gluonium vanishes at threshold. On the other hand, the same $s$ channel diagram of gluinonium interferes destructively with diagrams with $t$ and $u$ channel squarks, and this makes the gluinonium cross section vanish if the squarks are degenerate with the gluino, while there is no such effect for the KK gluonium. Note however that the presence of this effect for the gluinonium depends on the existence of squarks in MSSM, and thus will not be a general feature of new physics models with color-octet spin-$1/2$ particles. Similarly, the production of KK gluonia through the $q\bar q$ channel depends on the existence of KK quarks, and thus may not be present in some other model with color-octet spin-$1$ particles. Another difference between gluinonia and KK gluonia is that even though in both cases the bound states are produced with spin component $\pm 1$ along the beam axis, it is an $\vv{8_A}$ ($J=1$) state for gluinonium and $\vv{1}$ or $\vv{8_S}$ ($J=2$) state for KK gluonium. The production of the $\vv{1}$ state (with its large color factor $(C_\vv{1}/C_\vv{8})^3 = 8$) in UED makes the cross section larger.

We also see from table~\ref{tab-KKgluonium} that KK gluonia will be produced predominantly in spin-$2$ states. If the angular distributions of the annihilation products can be measured, they can be used to distinguish the KK gluonium from the gluinonium (which is produced predominantly in spin-$0$ states, and also in the spin-$1$ state which is important for some of the annihilation channels). The angular distributions in the dijet, $b\bar b$ and $t\bar t$ channels are plotted in figure~\ref{fig-angular}. The curves measured in experiment will be further affected by detector acceptances, effects of QCD radiation, mistakes in reconstruction, cuts, and uncertainties in modeling the QCD background which needs to be subtracted from the data. However, the remarkable differences between KK gluonium and gluinonium that we see in figure~\ref{fig-angular} will hopefully remain.

To simulate the effects of some of the experimental factors on the angular distributions, we generated events with \textsc{Pythia}, and the result (for example, for the $b\bar b$ channel) is shown in figure~\ref{fig-angular-Pythia}. The details of the simulation and the imposed cuts are the same as will be described in section~\ref{sec-simulation} (except that the cut on $\cos\theta$ is not included and the $p_T$ cut is relaxed to $p_T > 0.3 M$).

\begin{figure}[t]
\begin{center}
\includegraphics[width=0.55\textwidth]{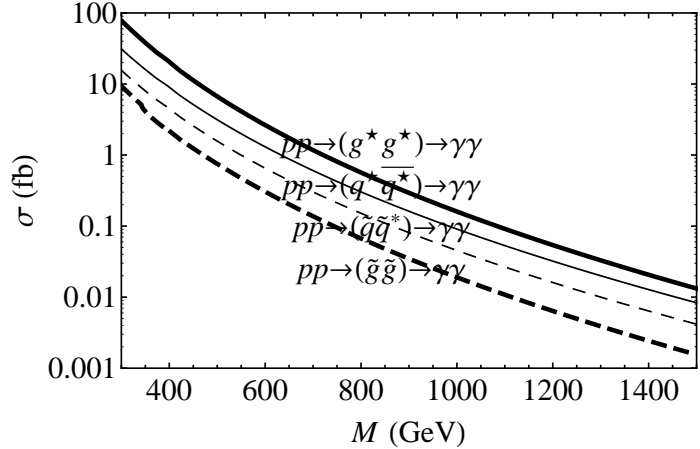}
\caption{Diphoton annihilation signal of gluinonium (thick dashed line), KK gluonium (estimate; thick solid line), squarkonium (thin dashed line) and KK quarkonium (thin solid line) at the $14$ TeV LHC as a function of the resonance mass. For gluinonium we assumed $m_\sq = m_\go$ (and similarly for KK gluonium). For squarkonium and KK quarkonium we present the contribution of a single flavor and chirality, and we assumed the charges of the constituent particles to be $|Q| = 2/3$ (for $|Q| = 1/3$, the cross section is $16$ times smaller).}
\label{fig-cs-diphoton-M}
\end{center}
\end{figure}

Another channel we can look at is the annihilation of color-singlet KK gluonia or gluinonia into a pair of photons. Even though these processes can only proceed through loop diagrams because the KK gluons and gluinos are neutral, it is still interesting to consider this signal because the background in the diphoton channel is much more favorable than in the dijet, $b\bar b$ or $t\bar t$ channels. As long as the squarks are not much heavier than the gluino, the annihilation rate of the color-singlet gluinonium into $\gamma\gamma$ is $\sim 10^{-5}$ of its annihilation rate into $gg$ (the exact expression from~\cite{Kauth:2009ud} is given in~(\ref{gogo-diphoton})). The relevant loop diagrams for KK gluonium have not been computed, but for our estimates we will assume that the diphoton branching ratio for spin-$0$ and spin-$2$ color-singlet KK gluonia is the same as for gluinonium. We expect this to be correct up to an ${\cal O}\,(1)$ factor since the relevant coupling constants and the gauge quantum numbers of the relevant particles are the same in both cases. The signal cross sections are shown in figure~\ref{fig-cs-diphoton-M}. The angular distributions in this channel may also be useful for discrimination since for gluinonia only the $J=0$ state contributes, while for KK gluonia the contribution may be coming from both $J=0$ and $J=2$.

We can also compare the KK gluonia and gluinonia (which are bound states of spin-$1$ and spin-$1/2$ color octets) with bound states of scalar color-octets. Pairs of scalar color-octets $\phi$ will be produced by gluon fusion via the same diagrams as the KK gluons in figure~\ref{fig-diagrams-KKgKKg}, although in practice only the quartic vertex is relevant to bound states -- ``octetonia'' $(\phi\phi)$ in $\vv{1}$ and $\vv{8_S}$ color representations. Their production cross sections are included in table~\ref{tab-KKgluonium}.\footnote{The color-singlet result can be found in~\cite{Kim:2008bx}, and it agrees with ours.} They are twice as small as the gluinonium cross sections. The octetonia decay rates are given in appendix~\ref{app-rates-octetonia}. Since the octetonia are scalars they will annihilate isotropically, and unless $\phi$ is charged under the electroweak group or interacts with some other particles, the only significant annihilation signal of octetonia will be $gg$ dijets. It will be easy to distinguish them from bound states of spin-$1$ color-octets (like KK gluonia) because of the factor of $\sim 20$ difference in the cross section and the very different angular distributions. It will be more difficult to distinguish them from bound states of spin-$1/2$ color-octets (like gluinonia) because the angular distributions are also almost isotropic in the spin-$1/2$ case and the difference in the cross section is only a factor of $2$. If the difference in the cross section is to be used, potential multiplicity of degenerate color-octets and higher order QCD corrections will be important. Part of the QCD corrections for gluinonium have been computed in~\cite{Hagiwara:2009hq,Kauth:2009ud}.

\section{KK quarkonia vs. squarkonia\label{sec-KKquarkonium}}
\setcounter{equation}{0}

\begin{table}[t]
$$\begin{array}{c|c|c|c}
\mbox{} & \,\mbox{color} \,&\, J^{PC} \,&\, \mbox{can couple to} \,\\\hline\hline
(\kkqR\bar \kkqR) + (\kkqL\bar \kkqL) & \vv{1} & 0^{-+} & G^{\rho\sigma}\tilde G_{\rho\sigma},\, i\bar t \gamma^5 t \\\hline
(\kkqR\bar \kkqR) - (\kkqL\bar \kkqL) & \vv{1} & 0^{+-} & - \\\hline
(\kkqL\bar \kkqR)                     & \vv{1} & 0^{-+} & i\bar q P_R q,\, i\bar t\gamma^5 t \\\hline
(\kkqR\bar \kkqR) + (\kkqL\bar \kkqL) & \vv{1} & 1^{--} &\; \bar q \gamma^\mu q,\, i\bar t\l[\gamma^\mu,\gamma^\nu\r]t,\, \bar\ell \gamma^\mu \ell,\, \bar\ell\gamma^\mu\gamma^5\ell \\\hline
(\kkqR\bar \kkqR) - (\kkqL\bar \kkqL) & \vv{1} & 1^{++} & \bar q\gamma^\mu\gamma^5q,\, \bar\ell \gamma^\mu \ell,\, \bar\ell\gamma^\mu\gamma^5\ell \\\hline
(\kkqL\bar \kkqR)                     & \vv{1} & 1^{--} & i\bar q\l[\gamma^\mu,\gamma^\nu\r]P_R q,\, \bar t \gamma^\mu t \\\hline\hline
(\sq_R\sq_R^\ast) + (\sq_L\sq_L^\ast) & \vv{1} & 0^{++} & G^{\rho\sigma}G_{\rho\sigma},\, \bar t t \\\hline
(\sq_R\sq_R^\ast) - (\sq_L\sq_L^\ast) & \vv{1} & 0^{--} & - \\\hline
(\sq_L\sq_R^\ast)                     & \vv{1} & 0^{++} & \bar q P_R q,\, \bar tt\\
\end{array}$$
\caption{KK quarkonia $(\kkq\bar\kkq)$ (UED) and squarkonia $(\sq\sq^\ast)$ (MSSM) and their possible couplings to gluons, quarks and leptons. $P_{R,L} = (1\pm\gamma^5)/2$. Antiparticle bound states, where relevant, are also present even if not listed in the table. In part of the spin-$1$ cases, the coupling is not to the bound state operator $V_\mu$ but to its derivative $\pd_\mu V_\nu$.}
\label{tab-KKqKKaq-sqasq-couplings}
\end{table}

In this section we study bound states of fundamental-antifundamental pairs of KK quarks (spin $1/2$) in UED and squarks (spin $0$) in MSSM. Later in this section we will also look at bound states of spin $1$ particles in the fundamental representation.

Table~\ref{tab-KKqKKaq-sqasq-couplings} classifies the KK quarkonia and squarkonia and their possible couplings to gluons and quarks (via the strong interactions) and leptons (via the electroweak interactions). The possible couplings to a pair of photons are like to a pair of gluons. The diagrams through which the various processes can be realized are shown in figures~\ref{fig-diagrams-KKqKKqbar} and~\ref{fig-diagrams-sqasq}.

For both KK quarkonia and squarkonia, the dominant products of the gluon fusion channel are spin-$0$ bound states in which the KK quarks or squarks have same flavors and chiralities. These are also the only bound states that can annihilate into a pair of photons, which is the most promising detection channel as we discuss in the following. The angular distributions in the $\gamma\gamma$ channel will be isotropic in both UED and MSSM but the size of the cross section can still be used for discrimination.

\begin{figure}[t]
$$
\beginpicture
\setcoordinatesystem units <0.571\tdim,0.571\tdim>
\stpltsmbl
\plot -70 37 0 37 0 -37 -70 -37 /
\barrow from -37 37 to -25 37
\barrow from -25 -37 to -37 -37
\barrow from 0 5 to 0 -5
\plot -54 30 -54 45 /
\plot -39 30 -39 45 /
\plot -24 30 -24 45 /
\plot -9  30  -9 45 /
\plot -54 -30 -54 -45 /
\plot -39 -30 -39 -45 /
\plot -24 -30 -24 -45 /
\plot -9  -30  -9 -45 /
\plot -6  25 6  25 /
\plot -6  10 6  10 /
\plot -6 -10 6 -10 /
\plot -6 -25 6 -25 /
\ellipticalarc axes ratio 2:1 -220 degrees from 0 35 center at 10 35
\ellipticalarc axes ratio 2:1 -280 degrees from 17 31 center at 24 35
\ellipticalarc axes ratio 2:1 -280 degrees from 32 31 center at 39 35
\ellipticalarc axes ratio 2:1 -280 degrees from 47 31 center at 54 35
\ellipticalarc axes ratio 2:1 -150 degrees from 62 31 center at 69 35
\ellipticalarc axes ratio 2:1 220 degrees from 0 -35 center at 10 -35
\ellipticalarc axes ratio 2:1 280 degrees from 17 -31 center at 24 -35
\ellipticalarc axes ratio 2:1 280 degrees from 32 -31 center at 39 -35
\ellipticalarc axes ratio 2:1 280 degrees from 47 -31 center at 54 -35
\ellipticalarc axes ratio 2:1 150 degrees from 62 -31 center at 69 -35
\put {$\kkq$} at -55 20
\put {$\bar\kkq$} at -55 -20
\put {$\kkq$} at 15 0
\put {$g$} at 50 20
\put {$g$} at 50 -20
\linethickness=0pt
\putrule from 0 -52 to 0 52
\putrule from -90 0 to 90 0
\endpicture
\beginpicture
\setcoordinatesystem units <0.571\tdim,0.571\tdim>
\stpltsmbl
\plot -70 37 3 37 3 -37 -70 -37 /
\barrow from -37 37 to -25 37
\barrow from -25 -37 to -37 -37
\barrow from 3 5 to 3 -5
\plot -54 30 -54 45 /
\plot -39 30 -39 45 /
\plot -24 30 -24 45 /
\plot -9  30  -9 45 /
\plot -54 -30 -54 -45 /
\plot -39 -30 -39 -45 /
\plot -24 -30 -24 -45 /
\plot -9  -30  -9 -45 /
\plot -3  25 9  25 /
\plot -3  10 9  10 /
\plot -3 -10 9 -10 /
\plot -3 -25 9 -25 /
\startrotation by 0.7 -0.75 about 0 30
\ellipticalarc axes ratio 2:1 -220 degrees from 0 35 center at 10 35
\ellipticalarc axes ratio 2:1 -280 degrees from 17 31 center at 24 35
\ellipticalarc axes ratio 2:1 -280 degrees from 32 31 center at 39 35
\ellipticalarc axes ratio 2:1 -280 degrees from 47 31 center at 54 35
\ellipticalarc axes ratio 2:1 -280 degrees from 62 31 center at 69 35
\ellipticalarc axes ratio 2:1 -280 degrees from 77 31 center at 84 35
\ellipticalarc axes ratio 2:1 -150 degrees from 92 31 center at 99 35
\stoprotation
\startrotation by 0.7 0.75 about 0 -30
\ellipticalarc axes ratio 2:1 220 degrees from 0 -35 center at 10 -35
\ellipticalarc axes ratio 2:1 280 degrees from 17 -31 center at 24 -35
\ellipticalarc axes ratio 2:1 280 degrees from 32 -31 center at 39 -35
\ellipticalarc axes ratio 2:1 280 degrees from 47 -31 center at 54 -35
\ellipticalarc axes ratio 2:1 280 degrees from 62 -31 center at 69 -35
\ellipticalarc axes ratio 2:1 280 degrees from 77 -31 center at 84 -35
\ellipticalarc axes ratio 2:1 150 degrees from 92 -31 center at 99 -35
\stoprotation
\put {$\kkq$} at -55 20
\put {$\bar\kkq$} at -55 -20
\put {$\kkq$} at -15 0
\put {$g$} at 75 15
\put {$g$} at 75 -15
\linethickness=0pt
\putrule from 0 -52 to 0 52
\putrule from -90 0 to 90 0
\endpicture
\beginpicture
\setcoordinatesystem units <0.571\tdim,0.571\tdim>
\stpltsmbl
\plot -100 35 -65 0 -100 -35 /
\barrow from -90 25 to -80 15
\barrow from -80 -15 to -90 -25
\startrotation by 0.6 -0.6 about -100 35
\plot -84 30 -84 45 /
\plot -69 30 -69 45 /
\plot -54 30 -54 45 /
\stoprotation
\startrotation by 0.6 0.6 about -100 -35
\plot -84 -30 -84 -45 /
\plot -69 -30 -69 -45 /
\plot -54 -30 -54 -45 /
\stoprotation
\ellipticalarc axes ratio 2:1 220 degrees from  0  0 center at -10 0
\ellipticalarc axes ratio 2:1 280 degrees from -17 -4 center at -24 0
\ellipticalarc axes ratio 2:1 280 degrees from -32 -4 center at -39 0
\ellipticalarc axes ratio 2:1 260 degrees from -47 -4 center at -54 0
\startrotation by 0.6 0.6 about 37 35
\ellipticalarc axes ratio 2:1 -150 degrees from  29 31 center at  36 35
\ellipticalarc axes ratio 2:1 -280 degrees from  14 31 center at  21 35
\ellipticalarc axes ratio 2:1 -280 degrees from  -1 31 center at   6 35
\ellipticalarc axes ratio 2:1 -280 degrees from -16 31 center at  -9 35
\stoprotation
\startrotation by 0.6 -0.6 about 37 -35
\ellipticalarc axes ratio 2:1 150 degrees from  29 -31 center at  36 -35
\ellipticalarc axes ratio 2:1 280 degrees from  14 -31 center at  21 -35
\ellipticalarc axes ratio 2:1 280 degrees from  -1 -31 center at   6 -35
\ellipticalarc axes ratio 2:1 280 degrees from -16 -31 center at  -9 -35
\stoprotation
\put {$\kkq$} at -70 35
\put {$\bar\kkq$} at -65 -30
\put {$g$} at -30 20
\put {$g$} at 5 30
\put {$g$} at 5 -30
\linethickness=0pt
\putrule from 0 -52 to 0 52
\putrule from -130 0 to 60 0
\endpicture
$$
$$\beginpicture
\setcoordinatesystem units <0.571\tdim,0.571\tdim>
\stpltsmbl
\plot -70 37 70 37 /
\plot -70 -39 70 -39 /
\barrow from -37 37 to -25 37
\barrow from -25 -39 to -37 -39
\plot -54 30 -54 45 /
\plot -39 30 -39 45 /
\plot -24 30 -24 45 /
\plot -9  30  -9 45 /
\plot -54 -32 -54 -47 /
\plot -39 -32 -39 -47 /
\plot -24 -32 -24 -47 /
\plot -9  -32  -9 -47 /
\plot -5  26 10  26 /
\plot -5  11 10  11 /
\plot -5  -4 10  -4 /
\plot -5 -19 10 -19 /
\plot -5 -34 10 -34 /
\barrow from 35 37 to 45 37
\barrow from 45 -39 to 35 -39
\startrotation by 0 -1 about 0 30
\ellipticalarc axes ratio 2:1 -220 degrees from -5 29 center at  5 29
\ellipticalarc axes ratio 2:1 -280 degrees from 12 25 center at 19 29
\ellipticalarc axes ratio 2:1 -280 degrees from 27 25 center at 34 29
\ellipticalarc axes ratio 2:1 -280 degrees from 42 25 center at 49 29
\ellipticalarc axes ratio 2:1 -160 degrees from 57 25 center at 64 29
\stoprotation
\put {$\kkq$} at -50 20
\put {$\bar\kkq$} at -50 -20
\put {$\kkg$} at -18 0
\put {$q$} at 60 20
\put {$\bar q$} at 60 -20
\linethickness=0pt
\putrule from 0 -52 to 0 52
\putrule from -90 0 to 100 0
\endpicture
\beginpicture
\setcoordinatesystem units <0.571\tdim,0.571\tdim>
\stpltsmbl
\plot -100 35 -65 0 -100 -35 /
\barrow from -90 25 to -80 15
\barrow from -80 -15 to -90 -25
\startrotation by 0.6 -0.6 about -100 35
\plot -84 30 -84 45 /
\plot -69 30 -69 45 /
\plot -54 30 -54 45 /
\stoprotation
\startrotation by 0.6 0.6 about -100 -35
\plot -84 -30 -84 -45 /
\plot -69 -30 -69 -45 /
\plot -54 -30 -54 -45 /
\stoprotation
\ellipticalarc axes ratio 2:1 220 degrees from  0  0 center at -10 0
\ellipticalarc axes ratio 2:1 280 degrees from -17 -4 center at -24 0
\ellipticalarc axes ratio 2:1 280 degrees from -32 -4 center at -39 0
\ellipticalarc axes ratio 2:1 260 degrees from -47 -4 center at -54 0
\plot 37 37 0 0 37 -37 /
\barrow from 0 0 to 25 25
\barrow from 35 -35 to 15 -15
\put {$\kkq$} at -70 35
\put {$\bar\kkq$} at -65 -30
\put {$g$} at -30 20
\put {$q$} at 10 30
\put {$\bar q$} at 10 -30
\linethickness=0pt
\putrule from 0 -52 to 0 52
\putrule from -130 0 to 60 0
\endpicture
$$
$$
\beginpicture
\setcoordinatesystem units <0.571\tdim,0.571\tdim>
\stpltsmbl
\plot -70 37 0 37 0 -37 -70 -37 /
\barrow from -37 37 to -25 37
\barrow from -25 -37 to -37 -37
\barrow from 0 5 to 0 -5
\plot -54 30 -54 45 /
\plot -39 30 -39 45 /
\plot -24 30 -24 45 /
\plot -9  30  -9 45 /
\plot -54 -30 -54 -45 /
\plot -39 -30 -39 -45 /
\plot -24 -30 -24 -45 /
\plot -9  -30  -9 -45 /
\plot -6  25 6  25 /
\plot -6  10 6  10 /
\plot -6 -10 6 -10 /
\plot -6 -25 6 -25 /
\ellipticalarc axes ratio 2:1 180 degrees from 12 35 center at 6 35
\ellipticalarc axes ratio 2:1 180 degrees from 12 35 center at 18 35
\ellipticalarc axes ratio 2:1 180 degrees from 36 35 center at 30 35
\ellipticalarc axes ratio 2:1 180 degrees from 36 35 center at 42 35
\ellipticalarc axes ratio 2:1 180 degrees from 60 35 center at 54 35
\ellipticalarc axes ratio 2:1 180 degrees from 12 -35 center at  6 -35
\ellipticalarc axes ratio 2:1 180 degrees from 12 -35 center at 18 -35
\ellipticalarc axes ratio 2:1 180 degrees from 36 -35 center at 30 -35
\ellipticalarc axes ratio 2:1 180 degrees from 36 -35 center at 42 -35
\ellipticalarc axes ratio 2:1 180 degrees from 60 -35 center at 54 -35
\put {$\kkq$} at -55 20
\put {$\bar\kkq$} at -55 -20
\put {$\kkq$} at 15 0
\put {$\gamma$} at 50 18
\put {$\gamma$} at 50 -18
\linethickness=0pt
\putrule from 0 -52 to 0 52
\putrule from -90 0 to 90 0
\endpicture
\beginpicture
\setcoordinatesystem units <0.571\tdim,0.571\tdim>
\stpltsmbl
\plot -70 37 0 37 0 -37 -70 -37 /
\barrow from -37 37 to -25 37
\barrow from -25 -37 to -37 -37
\barrow from 0 5 to 0 -5
\plot -54 30 -54 45 /
\plot -39 30 -39 45 /
\plot -24 30 -24 45 /
\plot -9  30  -9 45 /
\plot -54 -30 -54 -45 /
\plot -39 -30 -39 -45 /
\plot -24 -30 -24 -45 /
\plot -9  -30  -9 -45 /
\plot -6  25 6  25 /
\plot -6  10 6  10 /
\plot -6 -10 6 -10 /
\plot -6 -25 6 -25 /
\startrotation by 0.8 -0.85 about 0 37
\ellipticalarc axes ratio 2:1 180 degrees from 12 37 center at  6 37
\ellipticalarc axes ratio 2:1 180 degrees from 12 37 center at 18 37
\ellipticalarc axes ratio 2:1 180 degrees from 36 37 center at 30 37
\ellipticalarc axes ratio 2:1 180 degrees from 36 37 center at 42 37
\ellipticalarc axes ratio 2:1 180 degrees from 60 37 center at 54 37
\ellipticalarc axes ratio 2:1 180 degrees from 60 37 center at 66 37
\ellipticalarc axes ratio 2:1 180 degrees from 84 37 center at 78 37
\stoprotation
\startrotation by 0.8 0.9 about 0 -37
\ellipticalarc axes ratio 2:1 180 degrees from 12 -37 center at  6 -37
\ellipticalarc axes ratio 2:1 180 degrees from 12 -37 center at 18 -37
\ellipticalarc axes ratio 2:1 180 degrees from 36 -37 center at 30 -37
\ellipticalarc axes ratio 2:1 180 degrees from 60 -37 center at 54 -37
\ellipticalarc axes ratio 2:1 180 degrees from 60 -37 center at 66 -37
\ellipticalarc axes ratio 2:1 180 degrees from 84 -37 center at 78 -37
\stoprotation
\put {$\kkq$} at -55 20
\put {$\bar\kkq$} at -55 -20
\put {$\kkq$} at -15 0
\put {$\gamma$} at 75 20
\put {$\gamma$} at 75 -20
\linethickness=0pt
\putrule from 0 -52 to 0 52
\putrule from -90 0 to 90 0
\endpicture
\qqq
\beginpicture
\setcoordinatesystem units <0.571\tdim,0.571\tdim>
\stpltsmbl
\plot -100 35 -65 0 -100 -35 /
\barrow from -90 25 to -80 15
\barrow from -80 -15 to -90 -25
\startrotation by 0.6 -0.6 about -100 35
\plot -84 30 -84 45 /
\plot -69 30 -69 45 /
\plot -54 30 -54 45 /
\stoprotation
\startrotation by 0.6 0.6 about -100 -35
\plot -84 -30 -84 -45 /
\plot -69 -30 -69 -45 /
\plot -54 -30 -54 -45 /
\stoprotation
\ellipticalarc axes ratio 2:1 180 degrees from -48 0 center at -56 0
\ellipticalarc axes ratio 2:1 180 degrees from -48 0 center at -42 0
\ellipticalarc axes ratio 2:1 180 degrees from -24 0 center at -30 0
\ellipticalarc axes ratio 2:1 180 degrees from -24 0 center at -18 0
\ellipticalarc axes ratio 2:1 180 degrees from   0 0 center at  -6 0
\plot 37 37 0 0 37 -37 /
\barrow from 0 0 to 25 25
\barrow from 35 -35 to 15 -15
\put {$\kkq$} at -70 35
\put {$\bar\kkq$} at -65 -30
\put {$\gamma,Z$} at -30 20
\put {$\ell^-$} at 55 33
\put {$\ell^+$} at 55 -33
\linethickness=0pt
\putrule from 0 -52 to 0 52
\putrule from -130 0 to 60 0
\endpicture
$$
\caption{Diagrams for production or annihilation of a KK quark-KK antiquark pair. For bound states, the diagrams with $s$-channel gluon do not actually contribute.}
\label{fig-diagrams-KKqKKqbar}
$$
\beginpicture
\setcoordinatesystem units <0.571\tdim,0.571\tdim>
\stpltsmbl
\setdashes
\plot -37 37 0 0 -37 -37 /
\setsolid
\barrow from -25 25 to -15 15
\barrow from -15 -15 to -25 -25
\startrotation by 0.8 0.5 about 0 0
\ellipticalarc axes ratio 2:1 -220 degrees from 0   0 center at 10 0
\ellipticalarc axes ratio 2:1 -280 degrees from 17 -4 center at 24 0
\ellipticalarc axes ratio 2:1 -280 degrees from 32 -4 center at 39 0
\ellipticalarc axes ratio 2:1 -150 degrees from 47 -4 center at 54 0
\stoprotation
\startrotation by 0.8 -0.5 about 0 0
\ellipticalarc axes ratio 2:1 220 degrees from 0  0 center at 10 0
\ellipticalarc axes ratio 2:1 280 degrees from 17 4 center at 24 0
\ellipticalarc axes ratio 2:1 280 degrees from 32 4 center at 39 0
\ellipticalarc axes ratio 2:1 150 degrees from 47 4 center at 54 0
\stoprotation
\put {$\sq$} at -10 32
\put {$\sq^\ast$} at -10 -32
\put {$g$} at 20 31
\put {$g$} at 20 -35
\linethickness=0pt
\putrule from 0 -52 to 0 52
\putrule from -50 0 to 60 0
\endpicture
\beginpicture
\setcoordinatesystem units <0.571\tdim,0.571\tdim>
\stpltsmbl
\setdashes
\plot -70 37 0 37 0 -37 -70 -37 /
\setsolid
\barrow from -45 37 to -35 37
\barrow from -35 -37 to -45 -37
\barrow from 0 5 to 0 -5
\ellipticalarc axes ratio 2:1 -220 degrees from 0 35 center at 10 35
\ellipticalarc axes ratio 2:1 -280 degrees from 17 31 center at 24 35
\ellipticalarc axes ratio 2:1 -280 degrees from 32 31 center at 39 35
\ellipticalarc axes ratio 2:1 -280 degrees from 47 31 center at 54 35
\ellipticalarc axes ratio 2:1 -150 degrees from 62 31 center at 69 35
\ellipticalarc axes ratio 2:1 220 degrees from 0 -35 center at 10 -35
\ellipticalarc axes ratio 2:1 280 degrees from 17 -31 center at 24 -35
\ellipticalarc axes ratio 2:1 280 degrees from 32 -31 center at 39 -35
\ellipticalarc axes ratio 2:1 280 degrees from 47 -31 center at 54 -35
\ellipticalarc axes ratio 2:1 150 degrees from 62 -31 center at 69 -35
\put {$\sq$} at -45 20
\put {$\sq^\ast$} at -45 -20
\put {$\sq$} at 15 0
\put {$g$} at 50 20
\put {$g$} at 50 -20
\linethickness=0pt
\putrule from 0 -52 to 0 52
\putrule from -90 0 to 90 0
\endpicture
\beginpicture
\setcoordinatesystem units <0.571\tdim,0.571\tdim>
\stpltsmbl
\setdashes
\plot -70 37 0 37 0 -37 -70 -37 /
\setsolid
\barrow from -45 37 to -35 37
\barrow from -35 -37 to -45 -37
\barrow from 0 5 to 0 -5
\setsolid
\startrotation by 0.7 -0.75 about 0 30
\ellipticalarc axes ratio 2:1 -220 degrees from 0 35 center at 10 35
\ellipticalarc axes ratio 2:1 -280 degrees from 17 31 center at 24 35
\ellipticalarc axes ratio 2:1 -280 degrees from 32 31 center at 39 35
\ellipticalarc axes ratio 2:1 -280 degrees from 47 31 center at 54 35
\ellipticalarc axes ratio 2:1 -280 degrees from 62 31 center at 69 35
\ellipticalarc axes ratio 2:1 -280 degrees from 77 31 center at 84 35
\ellipticalarc axes ratio 2:1 -150 degrees from 92 31 center at 99 35
\stoprotation
\startrotation by 0.7 0.75 about 0 -30
\ellipticalarc axes ratio 2:1 220 degrees from 0 -35 center at 10 -35
\ellipticalarc axes ratio 2:1 280 degrees from 17 -31 center at 24 -35
\ellipticalarc axes ratio 2:1 280 degrees from 32 -31 center at 39 -35
\ellipticalarc axes ratio 2:1 280 degrees from 47 -31 center at 54 -35
\ellipticalarc axes ratio 2:1 280 degrees from 62 -31 center at 69 -35
\ellipticalarc axes ratio 2:1 280 degrees from 77 -31 center at 84 -35
\ellipticalarc axes ratio 2:1 150 degrees from 92 -31 center at 99 -35
\stoprotation
\put {$\sq$} at -45 20
\put {$\sq^\ast$} at -45 -20
\put {$\sq$} at -15 0
\put {$g$} at 75 15
\put {$g$} at 75 -15
\linethickness=0pt
\putrule from 0 -52 to 0 52
\putrule from -90 0 to 90 0
\endpicture
\beginpicture
\setcoordinatesystem units <0.571\tdim,0.571\tdim>
\stpltsmbl
\setdashes
\plot -100 35 -65 0 -100 -35 /
\setsolid
\barrow from -90 25 to -80 15
\barrow from -80 -15 to -90 -25
\ellipticalarc axes ratio 2:1 220 degrees from  0  0 center at -10 0
\ellipticalarc axes ratio 2:1 280 degrees from -17 -4 center at -24 0
\ellipticalarc axes ratio 2:1 280 degrees from -32 -4 center at -39 0
\ellipticalarc axes ratio 2:1 260 degrees from -47 -4 center at -54 0
\startrotation by 0.6 0.6 about 37 35
\ellipticalarc axes ratio 2:1 -150 degrees from  29 31 center at  36 35
\ellipticalarc axes ratio 2:1 -280 degrees from  14 31 center at  21 35
\ellipticalarc axes ratio 2:1 -280 degrees from  -1 31 center at   6 35
\ellipticalarc axes ratio 2:1 -280 degrees from -16 31 center at  -9 35
\stoprotation
\startrotation by 0.6 -0.6 about 37 -35
\ellipticalarc axes ratio 2:1 150 degrees from  29 -31 center at  36 -35
\ellipticalarc axes ratio 2:1 280 degrees from  14 -31 center at  21 -35
\ellipticalarc axes ratio 2:1 280 degrees from  -1 -31 center at   6 -35
\ellipticalarc axes ratio 2:1 280 degrees from -16 -31 center at  -9 -35
\stoprotation
\put {$\sq$} at -75 35
\put {$\sq^\ast$} at -70 -30
\put {$g$} at -30 20
\put {$g$} at 5 30
\put {$g$} at 5 -30
\linethickness=0pt
\putrule from 0 -52 to 0 52
\putrule from -130 0 to 60 0
\endpicture
$$
$$\beginpicture
\setcoordinatesystem units <0.571\tdim,0.571\tdim>
\stpltsmbl
\plot 70 37 0 37 0 -39 70 -39 /
\setdashes
\plot -70 37 0 37 /
\plot -70 -39 0 -39 /
\setsolid
\barrow from -45 37 to -35 37
\barrow from 35 37 to 45 37
\barrow from -35 -39 to -45 -39
\barrow from 45 -39 to 35 -39
\setsolid
\startrotation by 0 -1 about 0 30
\ellipticalarc axes ratio 2:1 -220 degrees from -5 29 center at  5 29
\ellipticalarc axes ratio 2:1 -280 degrees from 12 25 center at 19 29
\ellipticalarc axes ratio 2:1 -280 degrees from 27 25 center at 34 29
\ellipticalarc axes ratio 2:1 -280 degrees from 42 25 center at 49 29
\ellipticalarc axes ratio 2:1 -180 degrees from 57 25 center at 64 29
\stoprotation
\put {$\sq$} at -50 20
\put {$\sq^\ast$} at -50 -20
\put {$\go$} at -18 0
\put {$q$} at 60 20
\put {$\bar q$} at 60 -20
\linethickness=0pt
\putrule from 0 -52 to 0 52
\putrule from -90 0 to 100 0
\endpicture
\beginpicture
\setcoordinatesystem units <0.571\tdim,0.571\tdim>
\stpltsmbl
\setdashes
\plot -100 35 -65 0 -100 -35 /
\setsolid
\barrow from -90 25 to -80 15
\barrow from -80 -15 to -90 -25
\setsolid
\ellipticalarc axes ratio 2:1 220 degrees from  0  0 center at -10 0
\ellipticalarc axes ratio 2:1 280 degrees from -17 -4 center at -24 0
\ellipticalarc axes ratio 2:1 280 degrees from -32 -4 center at -39 0
\ellipticalarc axes ratio 2:1 260 degrees from -47 -4 center at -54 0
\plot 37 37 0 0 37 -37 /
\barrow from 0 0 to 25 25
\barrow from 35 -35 to 15 -15
\put {$\sq$} at -70 35
\put {$\sq^\ast$} at -70 -30
\put {$g$} at -30 20
\put {$q$} at 10 30
\put {$\bar q$} at 10 -30
\linethickness=0pt
\putrule from 0 -52 to 0 52
\putrule from -130 0 to 60 0
\endpicture
$$
$$
\beginpicture
\setcoordinatesystem units <0.571\tdim,0.571\tdim>
\stpltsmbl
\setdashes
\plot -37 37 0 0 -37 -37 /
\setsolid
\barrow from -25 25 to -15 15
\barrow from -15 -15 to -25 -25
\startrotation by 0.8 0.6 about 0 0
\ellipticalarc axes ratio 2:1 180 degrees from 12 0 center at  6 0
\ellipticalarc axes ratio 2:1 180 degrees from 12 0 center at 18 0
\ellipticalarc axes ratio 2:1 180 degrees from 36 0 center at 30 0
\ellipticalarc axes ratio 2:1 180 degrees from 36 0 center at 42 0
\ellipticalarc axes ratio 2:1 180 degrees from 60 0 center at 54 0
\stoprotation
\startrotation by 0.8 -0.6 about 0 0
\ellipticalarc axes ratio 2:1 180 degrees from 12 0 center at  6 0
\ellipticalarc axes ratio 2:1 180 degrees from 12 0 center at 18 0
\ellipticalarc axes ratio 2:1 180 degrees from 36 0 center at 30 0
\ellipticalarc axes ratio 2:1 180 degrees from 36 0 center at 42 0
\ellipticalarc axes ratio 2:1 180 degrees from 60 0 center at 54 0
\stoprotation
\put {$\sq$} at -10 32
\put {$\sq^\ast$} at -10 -32
\put {$\gamma$} at 20 31
\put {$\gamma$} at 20 -35
\linethickness=0pt
\putrule from 0 -52 to 0 52
\putrule from -50 0 to 70 0
\endpicture
\beginpicture
\setcoordinatesystem units <0.571\tdim,0.571\tdim>
\stpltsmbl
\setdashes
\plot -70 37 0 37 0 -37 -70 -37 /
\setsolid
\barrow from -45 37 to -35 37
\barrow from -35 -37 to -45 -37
\barrow from 0 5 to 0 -5
\ellipticalarc axes ratio 2:1 180 degrees from 12 35 center at 6 35
\ellipticalarc axes ratio 2:1 180 degrees from 12 35 center at 18 35
\ellipticalarc axes ratio 2:1 180 degrees from 36 35 center at 30 35
\ellipticalarc axes ratio 2:1 180 degrees from 36 35 center at 42 35
\ellipticalarc axes ratio 2:1 180 degrees from 60 35 center at 54 35
\ellipticalarc axes ratio 2:1 180 degrees from 12 -35 center at  6 -35
\ellipticalarc axes ratio 2:1 180 degrees from 12 -35 center at 18 -35
\ellipticalarc axes ratio 2:1 180 degrees from 36 -35 center at 30 -35
\ellipticalarc axes ratio 2:1 180 degrees from 36 -35 center at 42 -35
\ellipticalarc axes ratio 2:1 180 degrees from 60 -35 center at 54 -35
\put {$\sq$} at -45 20
\put {$\sq^\ast$} at -45 -20
\put {$\sq$} at 15 0
\put {$\gamma$} at 50 18
\put {$\gamma$} at 50 -18
\linethickness=0pt
\putrule from 0 -52 to 0 52
\putrule from -90 0 to 90 0
\endpicture
\beginpicture
\setcoordinatesystem units <0.571\tdim,0.571\tdim>
\stpltsmbl
\setdashes
\plot -70 37 0 37 0 -37 -70 -37 /
\setsolid
\barrow from -45 37 to -35 37
\barrow from -35 -37 to -45 -37
\barrow from 0 5 to 0 -5
\setsolid
\startrotation by 0.8 -0.85 about 0 37
\ellipticalarc axes ratio 2:1 180 degrees from 12 37 center at  6 37
\ellipticalarc axes ratio 2:1 180 degrees from 12 37 center at 18 37
\ellipticalarc axes ratio 2:1 180 degrees from 36 37 center at 30 37
\ellipticalarc axes ratio 2:1 180 degrees from 36 37 center at 42 37
\ellipticalarc axes ratio 2:1 180 degrees from 60 37 center at 54 37
\ellipticalarc axes ratio 2:1 180 degrees from 60 37 center at 66 37
\ellipticalarc axes ratio 2:1 180 degrees from 84 37 center at 78 37
\stoprotation
\startrotation by 0.8 0.9 about 0 -37
\ellipticalarc axes ratio 2:1 180 degrees from 12 -37 center at  6 -37
\ellipticalarc axes ratio 2:1 180 degrees from 12 -37 center at 18 -37
\ellipticalarc axes ratio 2:1 180 degrees from 36 -37 center at 30 -37
\ellipticalarc axes ratio 2:1 180 degrees from 60 -37 center at 54 -37
\ellipticalarc axes ratio 2:1 180 degrees from 60 -37 center at 66 -37
\ellipticalarc axes ratio 2:1 180 degrees from 84 -37 center at 78 -37
\stoprotation
\put {$\sq$} at -45 20
\put {$\sq^\ast$} at -45 -20
\put {$\sq$} at -15 0
\put {$\gamma$} at 75 20
\put {$\gamma$} at 75 -20
\linethickness=0pt
\putrule from 0 -52 to 0 52
\putrule from -90 0 to 90 0
\endpicture
$$
\caption{Diagrams for production or annihilation of a squark-antisquark pair. For bound states, only the first diagram in each row actually contributes.}
\label{fig-diagrams-sqasq}
\end{figure}

\begin{table}[t]
$$\begin{array}{|c|c|c|c|}\hline
\,\mbox{process} &\,J = 0\,, J_z = 0 \,&\, J = 1\,,|J_z| = 1     \,&\, J = 1\,,J_z = 0 \,\\\hline\hline
\begin{array}{c}gg\to(\kkqL\bar \kkqL)\\\q\mbox{ or }(\kkqR\bar \kkqR)\end{array} &\,\ds\frac{4}{81}\,&\,         \,&\,          \\ \hline
\begin{array}{c}gg\to(\sq_L\sq_L^\ast)\\\q\mbox{ or }(\sq_R\sq_R^\ast)\end{array}                                        &\,\ds\frac{2}{81}\,&\,         \,&\,        \,\\ \hline\hline
\begin{array}{c}q\bar q^{(\prime)}\to(\kkqL \bar \kkqL^{(\prime)})\\\qq\mbox{ or }(\kkqR \bar \kkqR^{(\prime)})\end{array} &\,       \,&\frac{128}{2187}\l(\frac{2m_\kkq^2}{m_\kkq^2 + m_\kkg^2}\r)^2\l(2 + \frac{m_\kkq^2}{m_\kkg^2}\r)^2 &\,\\
\begin{array}{c}q\bar q^{(\prime)}\to(\kkqL\bar \kkqR^{(\prime)})\\\qq\mbox{ or }(\kkqR\bar \kkqL^{(\prime)})\end{array} &\frac{64}{2187}\l(\frac{2m_\kkq^2}{m_\kkq^2 + m_\kkg^2}\r)^2 \l(4 + \frac{m_\kkq^2}{m_\kkg^2}\r)^2 &\,        \,&\frac{64}{2187}\l(\frac{2m_\kkq^2}{m_\kkq^2 + m_\kkg^2}\r)^2 \l(\frac{m_\kkq^2}{m_\kkg^2}\r)^2 \\\hline
\begin{array}{c}q\bar q^{(\prime)}\to(\sq_L\sq^{(\prime)\ast}_R)\\\qq\mbox{ or }(\sq_R\sq^{(\prime)\ast}_L)\end{array} &\ds\frac{512}{2187}\l(\frac{2m_\go m_\sq}{m_\sq^2+m_\go^2}\r)^2 &\,         \,&\,          \\\hline
\end{array}$$
\caption{KK quarkonium $(\kkq\bar\kkq)$ (UED) and squarkonium $(\sq\sq^\ast)$ (MSSM) production cross section prefactors $P_{ij}$ of (\ref{parton-cs}). All the bound states are color-singlets. We use $q$ and $q'$ to refer to two different flavors of quarks (and similarly for KK quarks and squarks), while the notation $q^{(\prime)}$ means that the flavor can be either the same or different from that of $q$. All the numbers refer to a single choice of flavors and chiralities. For simplicity, we assumed the masses of the different flavors and chiralities of the KK quarks or squarks to be the same.}
\label{tab-KKquarkonium}
\end{table}

In the $q\bar q$ fusion channel, the most interesting products are spin-$1$ bound states of KK quarks with same flavors and chiralities since they can annihilate into a pair of leptons, while there is no such process for squarkonia. A disadvantage of this production mechanism (which goes through the diagram with a $t$ channel KK gluon from figure~\ref{fig-diagrams-KKqKKqbar}) is that it only has access to KK quarks of the flavors that are present in the colliding protons, so it will not be useful in the likely case that the lightest KK quark is a KK top. This leads us to consider also subleading production mechanisms which do not suffer from this problem. In particular, while spin-$1$ KK quarkonia cannot couple to a pair of gluons, they can be produced, similarly to $J/\psi$ and $\Upsilon$~\cite{Barger:1984qg}, in the process
\be
gg \to g(q^\star_\chi \bar {q^\star_\chi})_{\vv{1},J=1}\,,\qq \chi = L \mbox{ or } R
\label{bleaching}
\ee
which is flavor-universal.

The $gg$ and $\gamma\gamma$ KK quarkonium diagrams in figure~\ref{fig-diagrams-KKqKKqbar} describe the same processes as one would have for heavy quarkonia (see~\cite{Kats:2009bv} and references therein for the discussion of toponium and~\cite{Barger:1987xg,Arik:2002nd} for quarkonia of new heavy quarks). The production mechanism~(\ref{bleaching}) is also model-independent (it is determined by interactions with gluons). On the other hand, the $q\bar q$ channel is dominated by a diagram involving the KK gluon which does not exist for quarkonia. Another important difference is that the dominant annihilation process of spin-$1$ heavy quarkonia will be into pairs of longitudinal $W$ bosons because of the quarks' large coupling to the Higgs which is responsible for their large masses~\cite{Barger:1987xg}.

\begin{figure}[t]
\begin{center}
\includegraphics[width=0.49\textwidth]{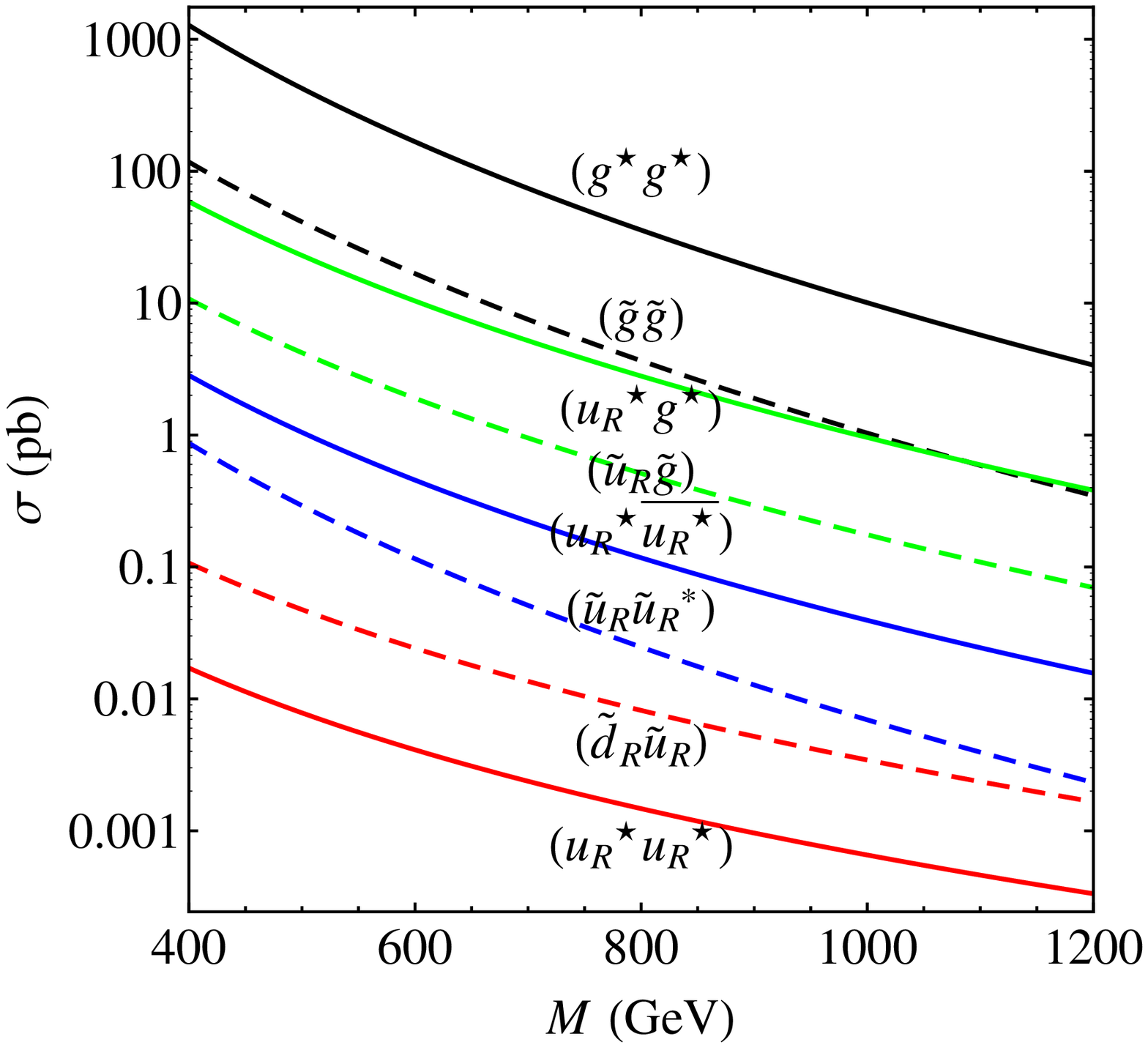}
\includegraphics[width=0.49\textwidth]{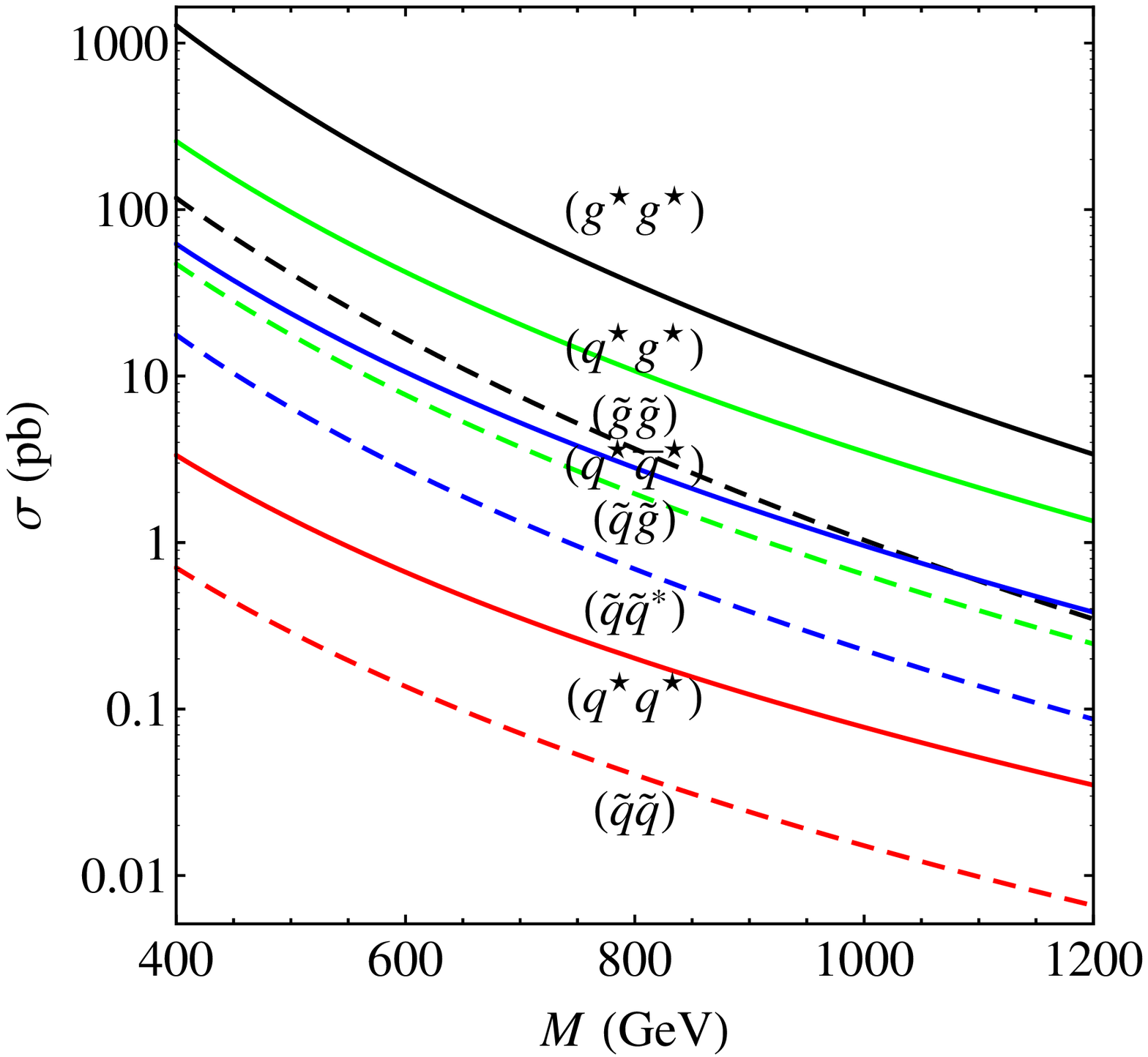}
\caption{Dijet signal of the various bound states at the $14$ TeV LHC as a function of the bound state mass ($M = 2m$), where all the particles are assumed to have the same mass $m$. The contributions of antiparticle bound states are included wherever relevant. On the left, we assume that the signal comes from particular flavors and chiralities of KK quarks or squarks and present several examples, while on the right we sum over all the possible combinations.}
\label{fig-dijet-all}
\end{center}
\begin{center}
\includegraphics[width=0.49\textwidth]{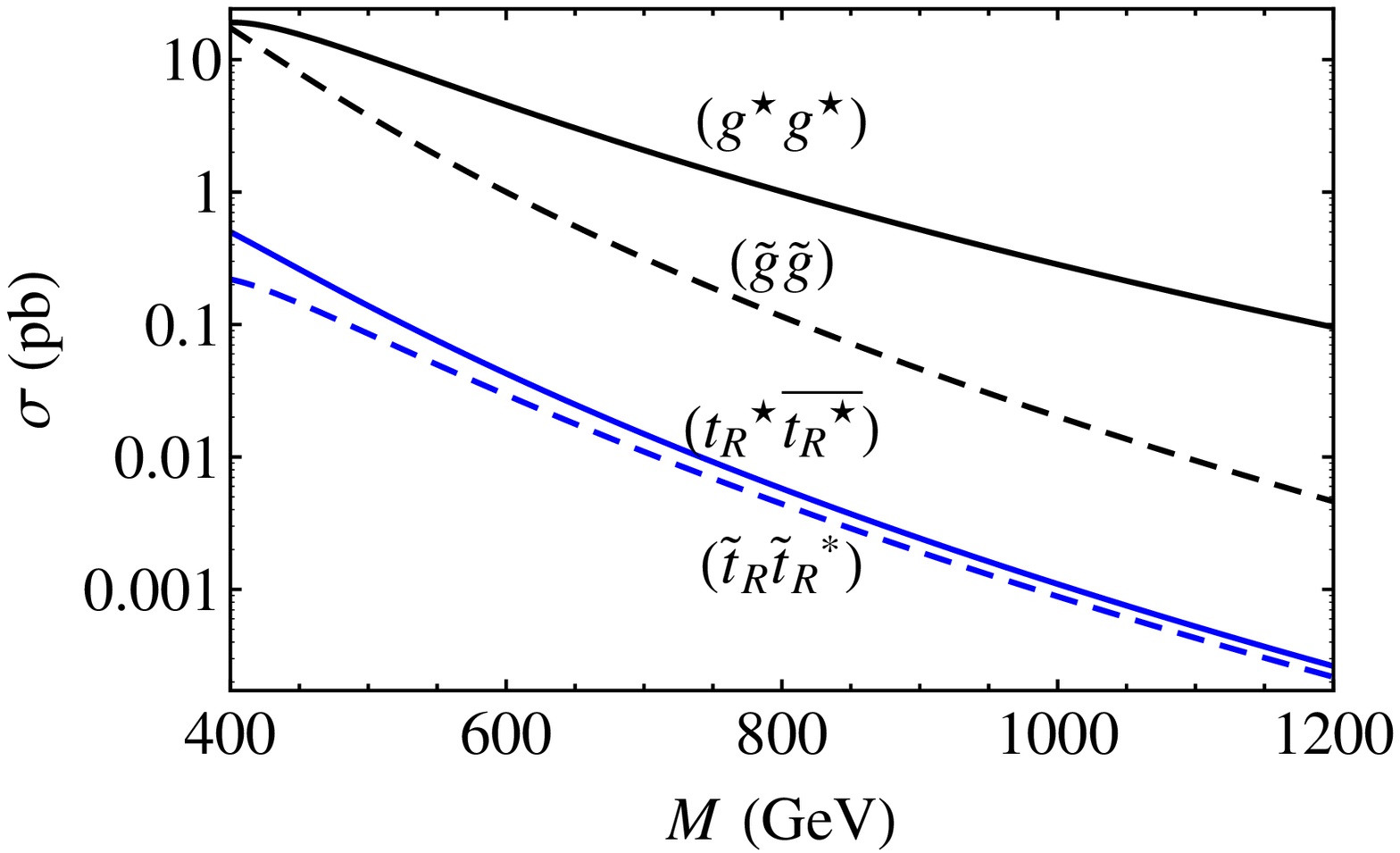}
\includegraphics[width=0.49\textwidth]{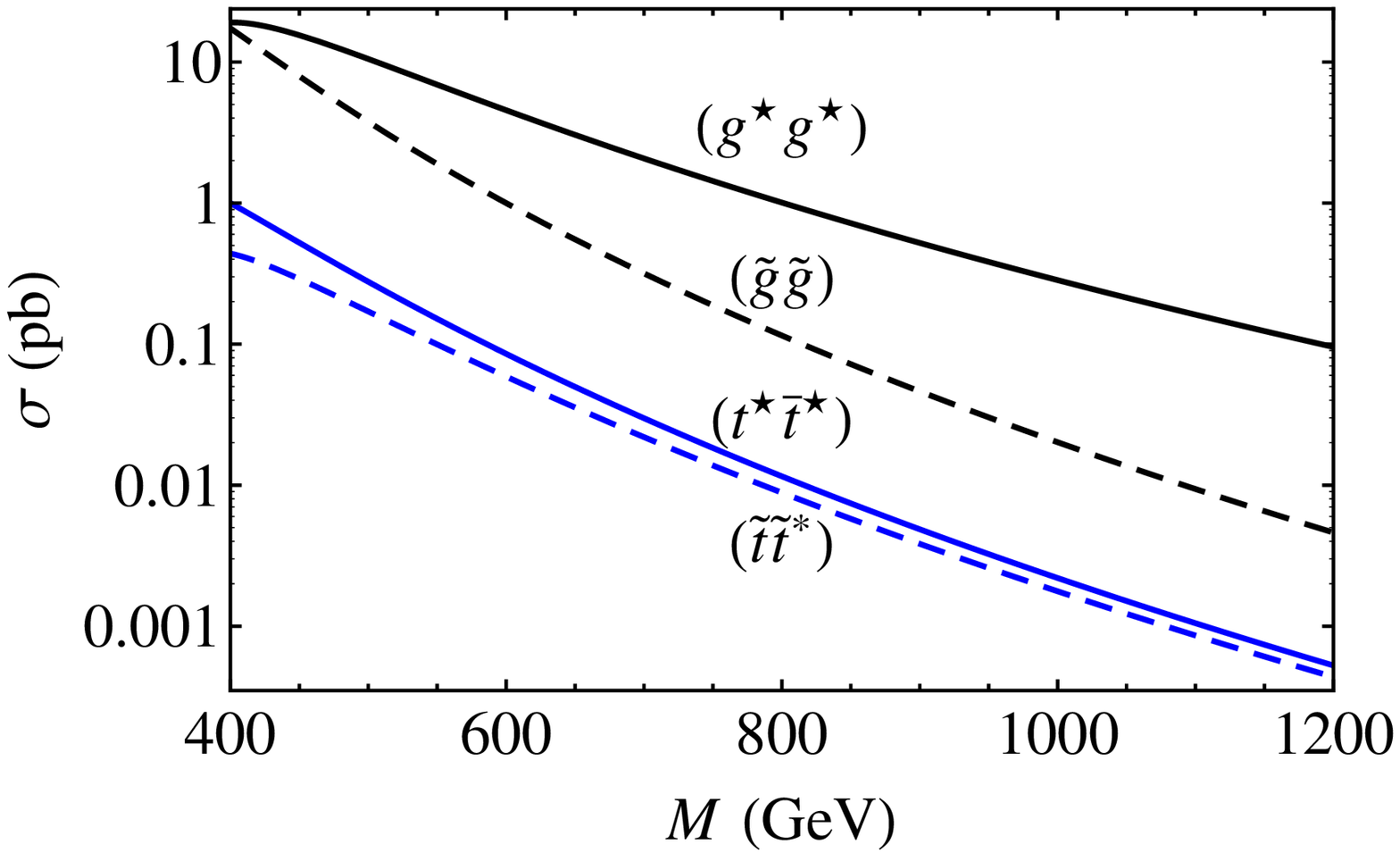}
\caption{$t\bar t$ signal of the various bound states at the $14$ TeV LHC as a function of the bound state mass ($M = 2m$), where all the particles are assumed to have the same mass $m$. For KK quarkonia and squarkonia, on the left we present examples of the contribution from KK tops and stops of a single chirality (the results for the other chirality are the same), while on the right we assume that both chiralities are close in mass and contribute.}
\label{fig-ttbar-all}
\end{center}
\end{figure}

The production cross sections for KK quarkonia and squarkonia are given by the expressions in appendices~\ref{app-rates-KKqKKq} and~\ref{app-rates-sqsq} and shown in table~\ref{tab-KKquarkonium}. Overall, the cross sections are about $2$ orders of magnitude below those of gluinonia of the same mass. This is primarily due to the smaller color factors both in $\l|\psi(\vv{0})\r|^2 \propto C^3$ and in the short-distance matrix element. As a result, the dijet and $t\bar t$ annihilation channels, whose cross sections are included in figures~\ref{fig-dijet-all} and~\ref{fig-ttbar-all}, are not promising.

However, the possibility of tree-level annihilation into $\gamma\gamma$ (which has low standard model background) is much more attractive despite the $(\alpha^2/\alpha_s^2)$-suppressed branching ratio of this mode. The cross section for the $\gamma\gamma$ signal is shown in figure~\ref{fig-cs-diphoton-M} as a function of the bound state mass. The angular distributions in this channel are isotropic for both KK quarkonium and squarkonium but the signal will be twice larger in UED (when comparing equal KK quark and squark masses) due to the twice larger production cross section. This property can be used as a discriminator, although it may be important to take higher-order QCD corrections into account. Such corrections for the squarkonium (stoponium) have been studied in~\cite{Martin:2009dj,Younkin:2009zn}. Also, we have assumed the total annihilation rate of squarkonium to be dominated by $gg$, while in some cases annihilation into pairs of $W$, $Z$ or Higgs bosons cannot be neglected~\cite{Herrero:1987df,Barger:1988sp,Drees:1993uw,Martin:2008sv}.

\begin{figure}[t]
\begin{center}
\includegraphics[width=0.48\textwidth]{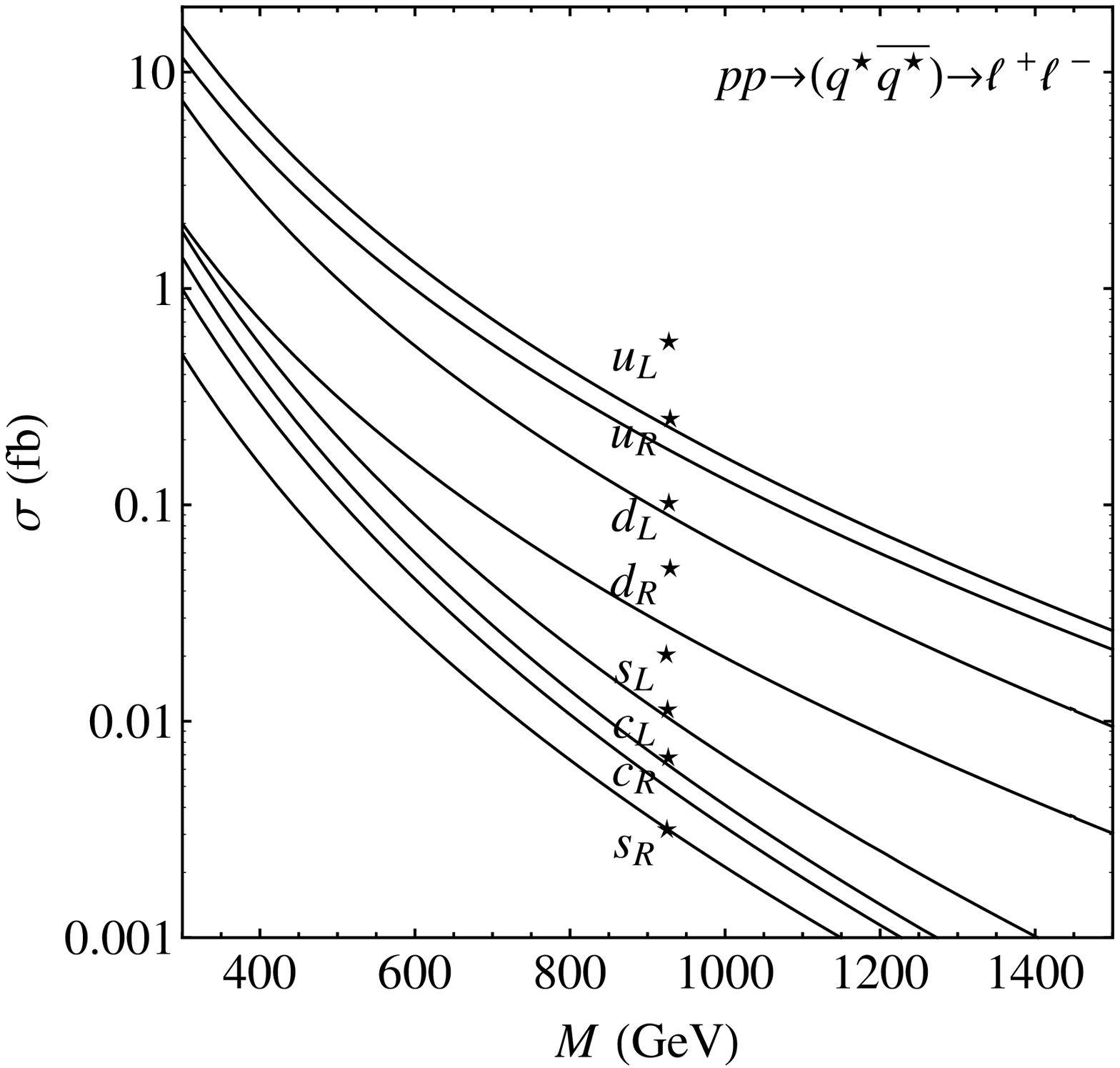}\;\;
\includegraphics[width=0.48\textwidth]{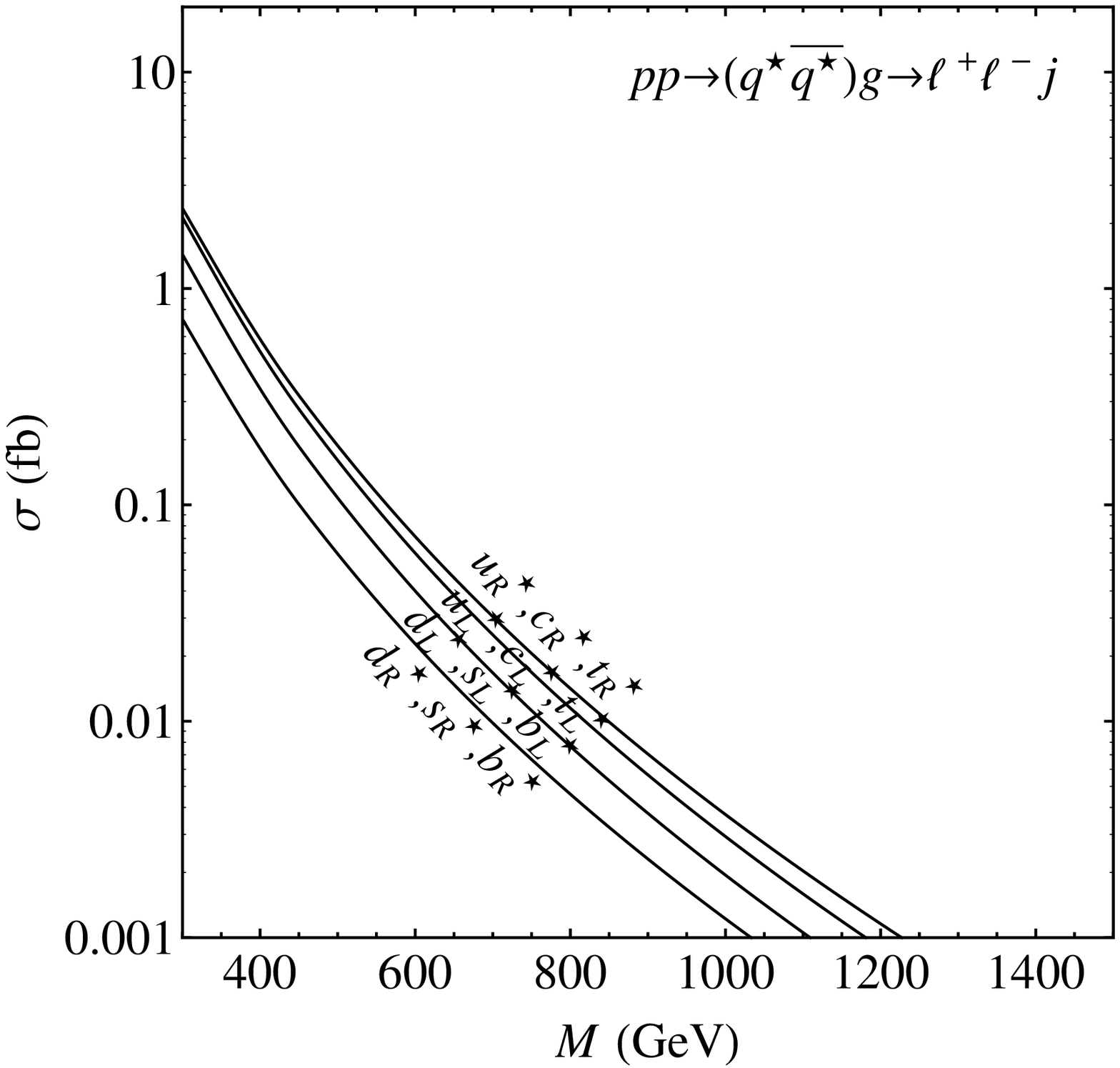}
\caption{Dilepton annihilation signal of KK quarkonium at the $14$ TeV LHC as a function of the resonance mass. The cross sections apply for any single flavor of leptons. Left: production via $q\bar q\to\kkq\bar\kkq$ (result independent of $m_\kkg$ unless $m_\kkg \gg m_\kkq$). Right: production via $gg\to\kkq\bar\kkq g$ for $m_\kkg = 2m_\kkq$ (for $m_\kkg = m_\kkq$ the branching ratios would be an order of magnitude smaller).}
\label{fig-cs-dilepton-M}
\end{center}
\end{figure}

The cross sections for the dilepton annihilation channel of spin-$1$ KK quarkonia are shown in figure~\ref{fig-cs-dilepton-M}. The right plot refers to KK quarkonia produced through the subleading process~(\ref{bleaching}) which has
\be
\hat\sigma_{\rm bound}\l(\hat s\r)
= \frac{5\,\zeta(3)\,\pi}{243\,m_\kkq^2}\,\alpha_s^3\bar\alpha_s^3\, I\l(\frac{\hat s}{M^2}\r)
\ee
where as in~\cite{Barger:1984qg}
\be
I(\gamma) = \theta(\gamma-1)\l[\frac{2}{\gamma^2}\l(\frac{\gamma+1}{\gamma-1} - \frac{2\gamma\ln \gamma}{(\gamma-1)^2}\r) + \frac{2(\gamma-1)}{\gamma(\gamma+1)^2} + \frac{4\ln \gamma}{(\gamma+1)^3}\r]
\ee
On the other hand, squarkonia do not give rise to a dilepton signal.

\begin{table}[t]
$$\begin{array}{|c|c|c|c|c|}\hline
\,\mbox{process} &\,J = 0,\, J_z = 0 \,&\, J = 2,\, |J_z| = 2 \,\\\hline\hline
\; gg\to (WW^\ast) \;&\,2/27\,&\,32/81\,\\ \hline
\end{array}$$
\caption{Tripletonium production cross section prefactors $P_{ij}$ of (\ref{parton-cs}).}
\label{tab-tripletonium}
\end{table}

While squarks and KK quarks are examples of spin-$0$ and spin-$1/2$ particles in the $\vv{3}$ representation, it may be interesting to consider also bound states of spin-$1$ particles in the same representation. Such vector color-triplets $W^\mu$ couple to gluons via
\be
\cL = -\frac12W_{\mu\nu}^\ast W^{\mu\nu} - ig_s W_\mu^\ast T^a W_\nu G^{\mu\nu a} + m_W^2 W_\mu^\ast W^\mu
\ee
where $W_{\mu\nu} = D_\mu W_\nu - D_\nu W_\mu$ with $D_\mu = \pd_\mu - ig_s A_\mu^a T^a$ and $G^a_{\mu\nu}$ is the gluon field strength. The second term here is chosen such that tree-level unitarity in the production of vector pairs from $gg$ is preserved at high energies, like in the situation when $W^\mu$ is a gauge boson of an extended gauge group~\cite{Borisov:1986ev}. The $W^\mu$ particles may also couple to photons, which can be described by including the $-ieQ A_\mu$ term in $D_\mu$ and adding an electromagnetic term analogous to the second term above. The diagrams coupling these vector particles to gluons or photons are the same as for squark-antisquark pairs, figure~\ref{fig-diagrams-sqasq}. The $s$-channel diagram does not actually contribute for the bound states. We do not consider production from $q\bar q$ since it would depend on the (model-dependent) couplings of the quarks to the vector bosons. We present the cross sections for the resulting bound states, ``tripletonia'', in table~\ref{tab-tripletonium}. The production cross section from gluons is an order of magnitude bigger than that of squarkonia and KK quarkonia. Furthermore, most of the tripletonia are produced with $J=2$ rather than $J=0$, which leads to different angular distributions of the annihilation products. More details are given in appendix~\ref{app-rates-tripletonia}.

\section{Di-KK quarks vs. di-squarks and KK quark-KK gluon vs. squark-gluino bound states\label{sec-diKKquarks-KKqKKg}}
\setcounter{equation}{0}

\begin{figure}[t]
$$\beginpicture
\setcoordinatesystem units <0.571\tdim,0.571\tdim>
\stpltsmbl
\plot -70 37 70 37 /
\plot -70 -39 70 -39 /
\barrow from 35  37 to 45  37
\barrow from 35 -39 to 45 -39
\barrow from -37  37 to -25  37
\barrow from -37 -39 to -25 -39
\plot -54 30 -54 45 /
\plot -39 30 -39 45 /
\plot -24 30 -24 45 /
\plot -9  30  -9 45 /
\plot -54 -32 -54 -47 /
\plot -39 -32 -39 -47 /
\plot -24 -32 -24 -47 /
\plot -9  -32  -9 -47 /
\plot -5  26 10  26 /
\plot -5  11 10  11 /
\plot -5  -4 10  -4 /
\plot -5 -19 10 -19 /
\plot -5 -34 10 -34 /
\startrotation by 0 -1 about 0 30
\ellipticalarc axes ratio 2:1 -220 degrees from -5 29 center at  5 29
\ellipticalarc axes ratio 2:1 -280 degrees from 12 25 center at 19 29
\ellipticalarc axes ratio 2:1 -280 degrees from 27 25 center at 34 29
\ellipticalarc axes ratio 2:1 -280 degrees from 42 25 center at 49 29
\ellipticalarc axes ratio 2:1 -180 degrees from 57 25 center at 64 29
\stoprotation
\put {$\kkq$} at -50 20
\put {$\kkq$} at -50 -20
\put {$\kkg$} at -18 0
\put {$q$} at 60 20
\put {$q$} at 60 -20
\linethickness=0pt
\putrule from 0 -52 to 0 52
\putrule from -90 0 to 100 0
\endpicture
\beginpicture
\setcoordinatesystem units <0.571\tdim,0.571\tdim>
\stpltsmbl
\plot -70  37 0  37 70 -39 /
\plot -70 -39 0 -39 30  -5 /
\plot 40 5 70 37 /
\barrow from  0 37 to 55 -23
\barrow from 40  5 to 55  21
\barrow from -37  37 to -25  37
\barrow from -37 -39 to -25 -39
\plot -54 30 -54 45 /
\plot -39 30 -39 45 /
\plot -24 30 -24 45 /
\plot -9  30  -9 45 /
\plot -54 -32 -54 -47 /
\plot -39 -32 -39 -47 /
\plot -24 -32 -24 -47 /
\plot -9  -32  -9 -47 /
\plot -5  26 10  26 /
\plot -5  11 10  11 /
\plot -5  -4 10  -4 /
\plot -5 -19 10 -19 /
\plot -5 -34 10 -34 /
\startrotation by 0 -1 about 0 30
\ellipticalarc axes ratio 2:1 -220 degrees from -5 29 center at  5 29
\ellipticalarc axes ratio 2:1 -280 degrees from 12 25 center at 19 29
\ellipticalarc axes ratio 2:1 -280 degrees from 27 25 center at 34 29
\ellipticalarc axes ratio 2:1 -280 degrees from 42 25 center at 49 29
\ellipticalarc axes ratio 2:1 -180 degrees from 57 25 center at 64 29
\stoprotation
\put {$\kkq$} at -50 20
\put {$\kkq$} at -50 -20
\put {$\kkg$} at -18 0
\put {$q$} at 70 20
\put {$q$} at 70 -20
\linethickness=0pt
\putrule from 0 -52 to 0 52
\putrule from -90 0 to 100 0
\endpicture
$$
\caption{Diagrams for production or annihilation of a pair of KK quarks. The second diagram is relevant only if the flavors are equal.}
\label{fig-diagrams-KKqKKq}
$$\beginpicture
\setcoordinatesystem units <0.571\tdim,0.571\tdim>
\stpltsmbl
\plot 70 37 0 37 0 -39 70 -39 /
\setdashes
\plot -70 37 0 37 /
\plot -70 -39 0 -39 /
\setsolid
\barrow from -45 37 to -35 37
\barrow from  35 37 to  45 37
\barrow from -45 -39 to -35 -39
\barrow from  35 -39 to  45 -39
\setsolid
\startrotation by 0 -1 about 0 30
\ellipticalarc axes ratio 2:1 -220 degrees from -5 29 center at  5 29
\ellipticalarc axes ratio 2:1 -280 degrees from 12 25 center at 19 29
\ellipticalarc axes ratio 2:1 -280 degrees from 27 25 center at 34 29
\ellipticalarc axes ratio 2:1 -280 degrees from 42 25 center at 49 29
\ellipticalarc axes ratio 2:1 -180 degrees from 57 25 center at 64 29
\stoprotation
\put {$\sq$} at -60 20
\put {$\sq$} at -60 -20
\put {$\go$} at -18 0
\put {$q$} at 60 20
\put {$q$} at 60 -20
\linethickness=0pt
\putrule from 0 -52 to 0 52
\putrule from -90 0 to 100 0
\endpicture
\beginpicture
\setcoordinatesystem units <0.571\tdim,0.571\tdim>
\stpltsmbl
\plot 70 37 40 5 /
\plot 30 -5 0 -39 0 37 70 -39 /
\setdashes
\plot -70 37 0 37 /
\plot -70 -39 0 -39 /
\setsolid
\barrow from  0 37 to 55 -23
\barrow from 40  5 to 55  21
\barrow from -37  37 to -25  37
\barrow from -37 -39 to -25 -39
\setsolid
\startrotation by 0 -1 about 0 30
\ellipticalarc axes ratio 2:1 -220 degrees from -5 29 center at  5 29
\ellipticalarc axes ratio 2:1 -280 degrees from 12 25 center at 19 29
\ellipticalarc axes ratio 2:1 -280 degrees from 27 25 center at 34 29
\ellipticalarc axes ratio 2:1 -280 degrees from 42 25 center at 49 29
\ellipticalarc axes ratio 2:1 -180 degrees from 57 25 center at 64 29
\stoprotation
\put {$\sq$} at -60 20
\put {$\sq$} at -60 -20
\put {$\go$} at -18 0
\put {$q$} at 70 20
\put {$q$} at 70 -20
\linethickness=0pt
\putrule from 0 -52 to 0 52
\putrule from -90 0 to 100 0
\endpicture
$$
\caption{Diagrams for production or annihilation of a pair of squarks. The second diagram is relevant only if the flavors are equal.}
\label{fig-diagrams-sqsq}
\end{figure}

\begin{table}[t]
$$\begin{array}{|c|c|c|c|}\hline
\,\mbox{process} &\,J = 0,\, J_z = 0 \,&\, J = 1,\, |J_z| = 1 \,&\, J = 1,\, J_z = 0 \,\\\hline\hline
\begin{array}{c}q q\to(\kkqL \kkqL)\\\q\mbox{ or }(\kkqR \kkqR)\end{array} &\,       \,&  &\, \frac{4}{729}\l(\frac{2m_\kkq^2}{m_\kkq^2 + m_\kkg^2}\r)^2\l(\frac{m_\kkq^2}{m_\kkg^2}\r)^2 \\
\begin{array}{c}q q'\to(\kkqL \kkqL')\\\q\mbox{ or }(\kkqR \kkqR')\end{array} &\, \frac{2}{729}\l(\frac{2m_\kkq^2}{m_\kkq^2 + m_\kkg^2}\r)^2\l(4 + \frac{m_\kkq^2}{m_\kkg^2}\r)^2 \,&  &\, \frac{2}{729}\l(\frac{2m_\kkq^2}{m_\kkq^2 + m_\kkg^2}\r)^2\l(\frac{m_\kkq^2}{m_\kkg^2}\r)^2 \\
q q\to(\kkqL \kkqR) & &\, \frac{8}{729}\l(\frac{2m_\kkq^2}{m_\kkq^2 + m_\kkg^2}\r)^2\l(2 + \frac{m_\kkq^2}{m_\kkg^2}\r)^2 \,& \\
\begin{array}{c}q q'\to(\kkqL \kkqR')\\\q\mbox{ or }(\kkqR \kkqL')\end{array} & &\, \frac{4}{729}\l(\frac{2m_\kkq^2}{m_\kkq^2 + m_\kkg^2}\r)^2\l(2 + \frac{m_\kkq^2}{m_\kkg^2}\r)^2 \,& \\\hline
\begin{array}{c}q q'\to(\sq_L\sq'_L)\\\q\mbox{ or }(\sq_R\sq'_R)\end{array} & \ds\frac{16}{729}\l(\frac{2m_\go m_\sq}{m_\go^2 + m_\sq^2}\r)^2 & &\,          \\\hline
\end{array}$$
\caption{Di-KK quark $(\kkq\kkq)$ (UED) and di-squark $(\sq\sq)$ (MSSM) production cross section prefactors $P_{ij}$ of (\ref{parton-cs}). All the bound states are in the $\vv{\bar 3}$ representation. All the numbers refer to a single flavor of that mass, and a single choice of the chiralities, and our notation assumes that $q' \neq q$. There also exist similar processes in which all the particles are replaced by their antiparticles. For simplicity, we assumed the masses of the different flavors and chiralities of the KK quarks or squarks to be the same.}
\label{tab-diKKquark}
\end{table}

\begin{figure}[t]
$$\beginpicture
\setcoordinatesystem units <0.571\tdim,0.571\tdim>
\stpltsmbl
\plot -70 0 5 0 42 37 /
\barrow from -70 0 to -25 0
\barrow from 5 0 to 28 23
\startrotation by 0.6 -0.6 about -100 35
\plot -100 35 -45 35 /
\barrow from -100 35 to -70 35
\plot -84 28 -84 42 /
\plot -69 28 -69 42 /
\plot -54 28 -54 42 /
\stoprotation
\startrotation by 0.6 0.6 about -100 -35
\plot -84 -30 -84 -45 /
\plot -69 -30 -69 -45 /
\plot -54 -30 -54 -45 /
\ellipticalarc axes ratio 2:1 -150 degrees from -92 -31 center at -99 -35
\ellipticalarc axes ratio 2:1 -280 degrees from -77 -31 center at -84 -35
\ellipticalarc axes ratio 2:1 -280 degrees from -62 -31 center at -69 -35
\ellipticalarc axes ratio 2:1 -280 degrees from -47 -31 center at -54 -35
\stoprotation
\startrotation by 0.6 -0.6 about 37 -35
\ellipticalarc axes ratio 2:1 150 degrees from  29 -31 center at  36 -35
\ellipticalarc axes ratio 2:1 280 degrees from  14 -31 center at  21 -35
\ellipticalarc axes ratio 2:1 280 degrees from  -1 -31 center at   6 -35
\ellipticalarc axes ratio 2:1 280 degrees from -16 -31 center at  -9 -35
\stoprotation
\put {$\kkq$} at -70 35
\put {$\kkg$} at -65 -30
\put {$q$} at -35 15
\put {$q$} at 10 30
\put {$g$} at 5 -30
\linethickness=0pt
\putrule from 0 -52 to 0 52
\putrule from -130 0 to 70 0
\endpicture
\beginpicture
\setcoordinatesystem units <0.571\tdim,0.571\tdim>
\stpltsmbl
\plot -70 37 70 37 /
\barrow from -70 37 to -25 37
\barrow from -70 37 to 45 37
\plot -54 30 -54 45 /
\plot -39 30 -39 45 /
\plot -24 30 -24 45 /
\plot -9  30  -9 45 /
\plot -54 -30 -54 -45 /
\plot -39 -30 -39 -45 /
\plot -24 -30 -24 -45 /
\plot -9  -30  -9 -45 /
\plot -5  26 10  26 /
\plot -5  11 10  11 /
\plot -5  -4 10  -4 /
\plot -5 -19 10 -19 /
\plot -5 -34 10 -34 /
\ellipticalarc axes ratio 2:1 -150 degrees from -62 -31 center at -69 -35
\ellipticalarc axes ratio 2:1 -280 degrees from -47 -31 center at -54 -35
\ellipticalarc axes ratio 2:1 -280 degrees from -32 -31 center at -39 -35
\ellipticalarc axes ratio 2:1 -280 degrees from -17 -31 center at -24 -35
\ellipticalarc axes ratio 2:1 -220 degrees from 0 -35 center at -10 -35
\ellipticalarc axes ratio 2:1 220 degrees from 0 -35 center at 10 -35
\ellipticalarc axes ratio 2:1 280 degrees from 17 -31 center at 24 -35
\ellipticalarc axes ratio 2:1 280 degrees from 32 -31 center at 39 -35
\ellipticalarc axes ratio 2:1 280 degrees from 47 -31 center at 54 -35
\ellipticalarc axes ratio 2:1 150 degrees from 62 -31 center at 69 -35
\startrotation by 0 -1 about 0 30
\ellipticalarc axes ratio 2:1 -220 degrees from -5 29 center at  5 29
\ellipticalarc axes ratio 2:1 -280 degrees from 12 25 center at 19 29
\ellipticalarc axes ratio 2:1 -280 degrees from 27 25 center at 34 29
\ellipticalarc axes ratio 2:1 -280 degrees from 42 25 center at 49 29
\ellipticalarc axes ratio 2:1 -180 degrees from 57 25 center at 64 29
\stoprotation
\put {$\kkq$} at -55 15
\put {$\kkg$} at -55 -15
\put {$\kkg$} at -18 0
\put {$q$} at 60 20
\put {$g$} at 60 -15
\linethickness=0pt
\putrule from 0 -52 to 0 52
\putrule from -90 0 to 90 0
\endpicture
\beginpicture
\setcoordinatesystem units <0.571\tdim,0.571\tdim>
\stpltsmbl
\plot -70 37 2 37 2 -39 25 -10 /
\plot 45 10 70 37 /
\barrow from -70 37 to -25 37
\barrow from 2 37 to 2 -2
\barrow from 45 10 to 60 26
\plot -54 30 -54 45 /
\plot -39 30 -39 45 /
\plot -24 30 -24 45 /
\plot -9  30  -9 45 /
\plot -54 -30 -54 -45 /
\plot -39 -30 -39 -45 /
\plot -24 -30 -24 -45 /
\plot -9  -30  -9 -45 /
\plot -5  26 9  26 /
\plot -5  11 9  11 /
\plot -5  -4 9  -4 /
\plot -5 -19 9 -19 /
\plot -5 -34 9 -34 /
\ellipticalarc axes ratio 2:1 -150 degrees from -62 -31 center at -69 -35
\ellipticalarc axes ratio 2:1 -280 degrees from -47 -31 center at -54 -35
\ellipticalarc axes ratio 2:1 -280 degrees from -32 -31 center at -39 -35
\ellipticalarc axes ratio 2:1 -280 degrees from -17 -31 center at -24 -35
\ellipticalarc axes ratio 2:1 -220 degrees from 0 -35 center at -10 -35
\startrotation by 0.7 -0.75 about 0 30
\ellipticalarc axes ratio 2:1 -220 degrees from 0 35 center at 10 35
\ellipticalarc axes ratio 2:1 -280 degrees from 17 31 center at 24 35
\ellipticalarc axes ratio 2:1 -280 degrees from 32 31 center at 39 35
\ellipticalarc axes ratio 2:1 -280 degrees from 47 31 center at 54 35
\ellipticalarc axes ratio 2:1 -280 degrees from 62 31 center at 69 35
\ellipticalarc axes ratio 2:1 -280 degrees from 77 31 center at 84 35
\ellipticalarc axes ratio 2:1 -150 degrees from 92 31 center at 99 35
\stoprotation
\put {$\kkq$} at -55 15
\put {$\kkg$} at -55 -15
\put {$\kkq$} at -18 0
\put {$q$} at 75 20
\put {$g$} at 75 -15
\linethickness=0pt
\putrule from 0 -52 to 0 52
\putrule from -90 0 to 90 0
\endpicture
$$
\caption{Diagrams for production or annihilation of a KK quark-KK gluon pair.}
\label{fig-diagrams-KKqKKg}
$$\beginpicture
\setcoordinatesystem units <0.571\tdim,0.571\tdim>
\stpltsmbl
\setdashes
\plot -100 35 -65 0 /
\setsolid
\plot -100 -35 -65 0 5 0 40 35 /
\barrow from -70 0 to -25 0
\barrow from 5 0 to 28 23
\barrow from -85 20 to -80 15
\startrotation by 0.6 0.6 about -100 -35
\ellipticalarc axes ratio 2:1 -150 degrees from -92 -31 center at -99 -35
\ellipticalarc axes ratio 2:1 -280 degrees from -77 -31 center at -84 -35
\ellipticalarc axes ratio 2:1 -280 degrees from -62 -31 center at -69 -35
\ellipticalarc axes ratio 2:1 -280 degrees from -47 -31 center at -54 -35
\stoprotation
\startrotation by 0.6 -0.6 about 37 -35
\ellipticalarc axes ratio 2:1 150 degrees from  29 -31 center at  36 -35
\ellipticalarc axes ratio 2:1 280 degrees from  14 -31 center at  21 -35
\ellipticalarc axes ratio 2:1 280 degrees from  -1 -31 center at   6 -35
\ellipticalarc axes ratio 2:1 280 degrees from -16 -31 center at  -9 -35
\stoprotation
\put {$\sq$} at -75 35
\put {$\go$} at -70 -30
\put {$q$} at -35 15
\put {$q$} at 10 30
\put {$g$} at 5 -30
\linethickness=0pt
\putrule from 0 -52 to 0 52
\putrule from -130 0 to 70 0
\endpicture
\beginpicture
\setcoordinatesystem units <0.571\tdim,0.571\tdim>
\stpltsmbl
\plot 70 35 0 35 0 -35 -70 -35 /
\setdashes
\plot -70 35 0 35 /
\setsolid
\barrow from -45 35 to -35 35
\barrow from 35 35 to 45 35
\ellipticalarc axes ratio 2:1 -150 degrees from -62 -31 center at -69 -35
\ellipticalarc axes ratio 2:1 -280 degrees from -47 -31 center at -54 -35
\ellipticalarc axes ratio 2:1 -280 degrees from -32 -31 center at -39 -35
\ellipticalarc axes ratio 2:1 -280 degrees from -17 -31 center at -24 -35
\ellipticalarc axes ratio 2:1 -220 degrees from 0 -35 center at -10 -35
\ellipticalarc axes ratio 2:1 220 degrees from 0 -35 center at 10 -35
\ellipticalarc axes ratio 2:1 280 degrees from 17 -31 center at 24 -35
\ellipticalarc axes ratio 2:1 280 degrees from 32 -31 center at 39 -35
\ellipticalarc axes ratio 2:1 280 degrees from 47 -31 center at 54 -35
\ellipticalarc axes ratio 2:1 150 degrees from 62 -31 center at 69 -35
\startrotation by 0 -1 about 0 30
\ellipticalarc axes ratio 2:1 -220 degrees from -5 29 center at  5 29
\ellipticalarc axes ratio 2:1 -280 degrees from 12 25 center at 19 29
\ellipticalarc axes ratio 2:1 -280 degrees from 27 25 center at 34 29
\ellipticalarc axes ratio 2:1 -280 degrees from 42 25 center at 49 29
\ellipticalarc axes ratio 2:1 -180 degrees from 57 25 center at 64 29
\stoprotation
\put {$\sq$} at -55 15
\put {$\go$} at -55 -15
\put {$\go$} at -18 0
\put {$q$} at 60 15
\put {$g$} at 60 -15
\linethickness=0pt
\putrule from 0 -52 to 0 52
\putrule from -90 0 to 90 0
\endpicture
\beginpicture
\setcoordinatesystem units <0.571\tdim,0.571\tdim>
\stpltsmbl
\plot -70 -35 0 -35 25 -10 /
\plot 45 10 70 35 /
\setdashes
\plot -70 35 0 35 0 -35 /
\setsolid
\barrow from -45 35 to -35 35
\barrow from 0 7 to 0 -10
\barrow from 45 10 to 55 20
\ellipticalarc axes ratio 2:1 -150 degrees from -62 -31 center at -69 -35
\ellipticalarc axes ratio 2:1 -280 degrees from -47 -31 center at -54 -35
\ellipticalarc axes ratio 2:1 -280 degrees from -32 -31 center at -39 -35
\ellipticalarc axes ratio 2:1 -280 degrees from -17 -31 center at -24 -35
\ellipticalarc axes ratio 2:1 -220 degrees from 0 -35 center at -10 -35
\startrotation by 0.7 -0.75 about 0 30
\ellipticalarc axes ratio 2:1 -220 degrees from 0 35 center at 10 35
\ellipticalarc axes ratio 2:1 -280 degrees from 17 31 center at 24 35
\ellipticalarc axes ratio 2:1 -280 degrees from 32 31 center at 39 35
\ellipticalarc axes ratio 2:1 -280 degrees from 47 31 center at 54 35
\ellipticalarc axes ratio 2:1 -280 degrees from 62 31 center at 69 35
\ellipticalarc axes ratio 2:1 -280 degrees from 77 31 center at 84 35
\ellipticalarc axes ratio 2:1 -150 degrees from 92 31 center at 99 35
\stoprotation
\put {$\sq$} at -55 15
\put {$\go$} at -55 -15
\put {$\sq$} at -18 0
\put {$q$} at 75 15
\put {$g$} at 75 -15
\linethickness=0pt
\putrule from 0 -52 to 0 52
\putrule from -90 0 to 90 0
\endpicture
$$
\caption{Diagrams for production or annihilation of a squark-gluino pair. For bound states, the last diagram does not actually contribute.}
\label{fig-diagrams-sqgo}
\end{figure}

\begin{table}[t]
$$\begin{array}{|c|c|c|}\hline
\,\mbox{process} & J = \frac{3}{2},\, |J_z| = \frac{3}{2} & J = \frac{1}{2},\, |J_z| = \frac{1}{2} \\\hline\hline
\,qg\to(\kkq\kkg)_\vv{3} \,&\, \ds\frac{3\,m_\kkq\l(m_\kkg + 9m_\kkq\r)^2}{32\l(m_\kkg+m_\kkq\r)^3} \,&\, \ds\frac{9\,m_\kkq^3\l(m_\kkg + 9m_\kkq\r)^2}{64\l(m_\kkg+m_\kkq\r)^5}\, \\[12pt]
\,qg\to(\kkq\kkg)_\vv{\bar 6} \,&\, \ds\frac{m_\kkq}{16\l(m_\kkg+m_\kkq\r)} \,&\, \ds \frac{3\,m_\kkq^3}{32\l(m_\kkg+m_\kkq\r)^3} \\[15pt]\hline
\,qg\to(\sq\go)_\vv{3}   \,&\,  \,&\, \ds\frac{3\,m_\go m_\sq^2\l(m_\go+9m_\sq\r)^2}{32\l(m_\go+m_\sq\r)^5}  \,\\[12pt]
\,qg\to(\sq\go)_\vv{\bar 6}  \,&\,  \,&\, \ds\frac{m_\go m_\sq^2}{16\l(m_\go+m_\sq\r)^3} \,\\\hline
\end{array}$$
\caption{KK quark-KK gluon $(\kkq\kkg)$ (UED) and squark-gluino $(\sq\go)$ (MSSM) bound states production cross section prefactors $P_{ij}$ of (\ref{parton-cs}). All the numbers refer to a single flavor and chirality. There also exist similar processes in which all the particles are replaced by their antiparticles.}
\label{tab-KKquarkKKgluon}
\end{table}

The possible diagrams for the remaining bound states are shown in figures~\ref{fig-diagrams-KKqKKq}, \ref{fig-diagrams-sqsq}, \ref{fig-diagrams-KKqKKg} and~\ref{fig-diagrams-sqgo}. The resulting cross section prefactors are given in tables~\ref{tab-diKKquark} and~\ref{tab-KKquarkKKgluon}. For all of these bound states, the annihilation will be almost entirely into dijets, without any cleaner channels (even with a small branching ratio) to consider. Furthermore, the dijet signal (see figure~\ref{fig-dijet-all}) is typically even smaller than that of the gluinonium (whose signal we will analyze in more detail in the next section), and therefore cannot be seen on top of the QCD background in most scenarios. The only exception is if the mass spectrum of the model is very degenerate to the extent that signals from many different bound states merge into one. This situation is exemplified in the right plot of figure~\ref{fig-dijet-all} which sums over the flavors and chiralities of the KK quarks or squarks, which corresponds to the overly optimistic scenario in which all the flavors and chiralities are sufficiently long-lived and close in mass within $\sim 5\%$ so that the dijet signals of all their bound states merge into a single peak in the invariant mass distribution. Similarly, we may further sum the curves corresponding to different types of bound states. Even then, the dijet signal will be extremely challenging.

\section{Detection prospects\label{sec-simulation}}
\setcounter{equation}{0}

We will now analyze to what extent the bound state signals discussed in the previous sections will be detectable at the LHC.\footnote{LHC signals of gluinonium have been also studied in~\cite{Chikovani:1996bk,Cheung:2004ad,BouhovaThacker:2004nh,BouhovaThacker:2006pj,Kats:2009bv} and squarkonium in~\cite{Drees:1993yr,Drees:1993uw,Martin:2008sv,Younkin:2009zn}.} To that end, it is informative to consider the existing experimental constraints on the masses of the various UED and MSSM particles. Direct limits from the Tevatron Run I constrain the size of the extra dimension in UED as $1/R \gtrsim 300$~GeV~\cite{Macesanu:2002db,Lin:2005ix} (the KK modes masses are $m_1 \sim 1/R$). A much stronger limit can probably be obtained from the analysis of data available today. The most stringent indirect limit, $1/R \gtrsim 600$~GeV, arises from the inclusive radiative $\bar{B}\to X_s\gamma$ decay~\cite{Haisch:2007vb}. With the assumption that the Higgs is not much heavier than $115$~GeV, the same limit is obtained from electroweak observables based on LEP data~\cite{Gogoladze:2006br}. For the MSSM, the gluino mass is constrained by collider searches to $m_\go \gtrsim 400$~GeV, and squarks are constrained to $m_\sq \gtrsim 600$~GeV, except for the sbottom (which can be lighter than $300$~GeV) and the stop (which can even be under $200$~GeV)~\cite{Feng:2009te,Khachatryan:2011tk,Collaboration:2011hh,daCosta:2011qk,Chatrchyan:2011wc,Chatrchyan:2011bz,Aad:2011yf}. It should be remembered though that MSSM and UED have multiple free parameters (the soft SUSY-breaking parameters in MSSM and the boundary terms in UED), and experimental bounds usually depend on certain arbitrary assumptions about them and thus apply in only part of the parameter space (for an example, see~\cite{Alwall:2008ve}). We therefore find it useful to consider also particles that are lighter than the bounds quoted above. Note also that many of our results do not depend on the full particle content of MSSM or UED and can be relevant to bound states in other models that are less constrained.

To simulate the bound state signals and the dominant standard model backgrounds we used \textsc{Pythia} (version 8.120)~\cite{Sjostrand:2006za,Sjostrand:2007gs} with NLO MSTW 2008 PDFs~\cite{Martin:2009iq} and \textsc{SISCone} jet algorithm (version 2.0.1)~\cite{Salam:2007xv} with cone radius $R = 1$ (except for the $t\bar t$ analysis where we used $R=0.5$), overlap parameter $f = 0.75$, no $p_T$ threshold on stable cones, and an infinite number of passes. We selected several particles defined in \textsc{Pythia} and modified their coupling constants and branching ratios such that they would behave according to the bound state effective vertices from tables~\ref{tab-KKgKKg-gogo-couplings} and~\ref{tab-KKqKKaq-sqasq-couplings}. For the purpose of simulation, we pretended all the bound states to be color singlets. We used the BSM Higgs for simulating spin-$0$ bound states, the $Z'$ for spin-$1$ bound states (and $\Upsilon$ for simulating~(\ref{bleaching})), and the KK graviton for spin-$2$ bound states. We have simulated the backgrounds without any $K$-factors since our signals do not include higher-order QCD corrections either. Such corrections to the pair production processes, the bound state wavefunctions and the annihilation processes can be large and sometimes even change the cross section by a factor of $\sim 2$. Part of these corrections have already been computed for some of the MSSM bound states~\cite{Hagiwara:2009hq,Kauth:2009ud,Martin:2009dj,Younkin:2009zn} but none for UED. Our results will need to be re-examined once these corrections are known.

\begin{figure}[t]
\begin{center}
\includegraphics[width=0.50\textwidth]{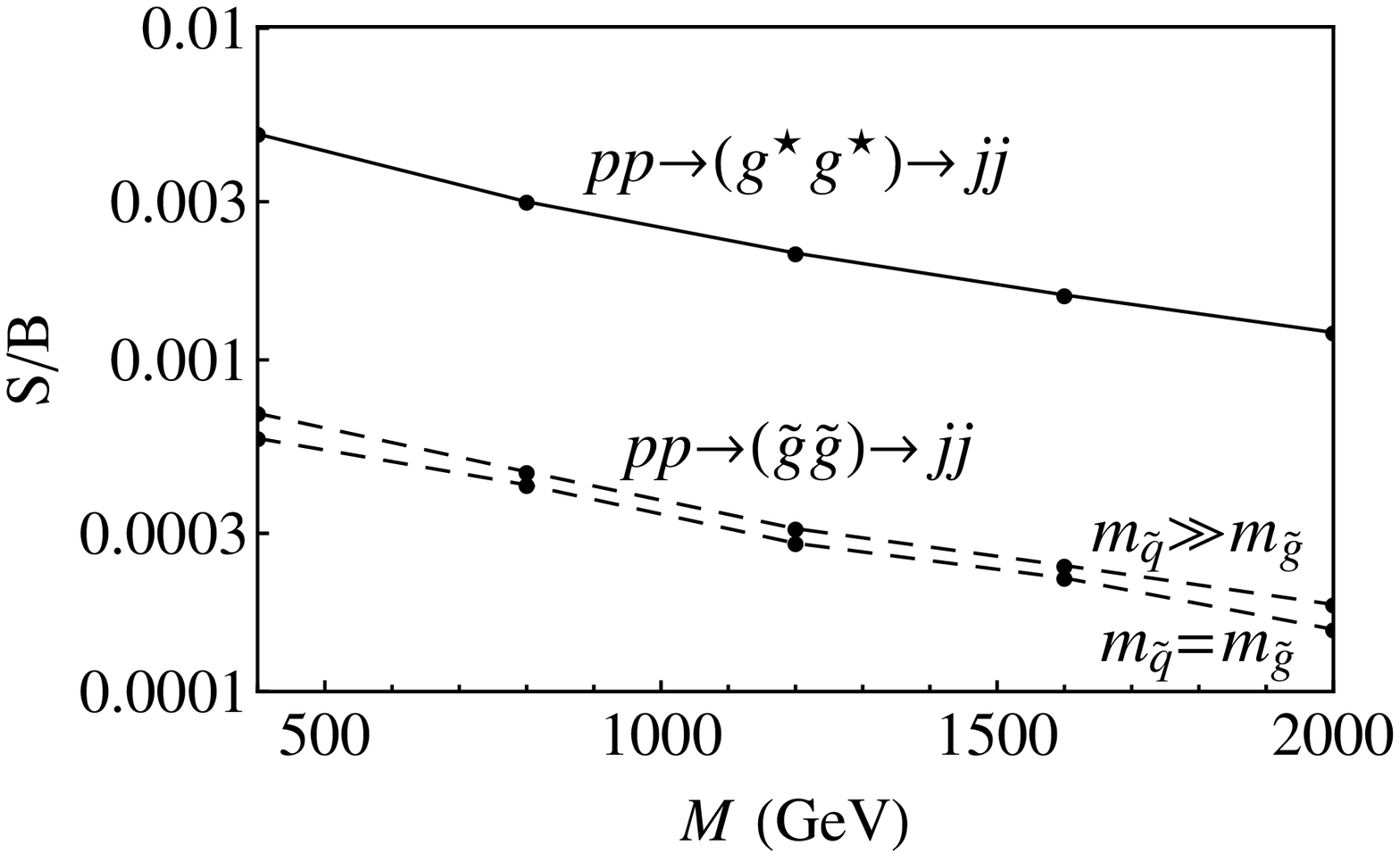}
\includegraphics[width=0.49\textwidth]{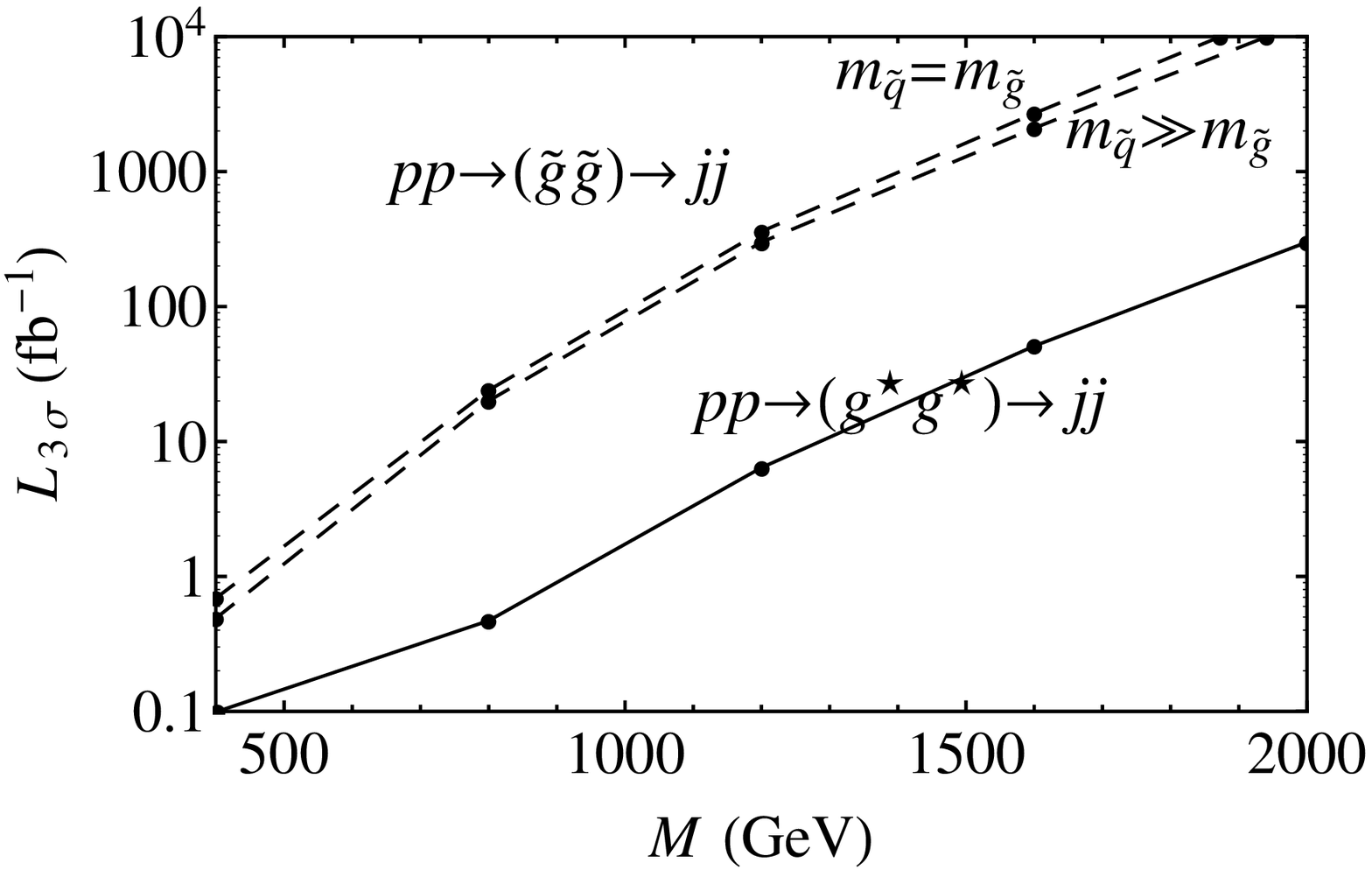}
\caption{Dijet channel: signal-to-background ratio (left) and the luminosity required for $3\sigma$ significance (right) for gluinonium (dashed lines) and KK gluonium (solid line) at the $14$ TeV LHC as a function of the resonance mass. For KK gluonium we assume $m_\kkq = m_\kkg$ while for gluinonium we present two cases.}
\label{fig-SB-dijet-M}
\end{center}
\end{figure}

\begin{figure}[t]
\begin{center}
\includegraphics[width=0.49\textwidth]{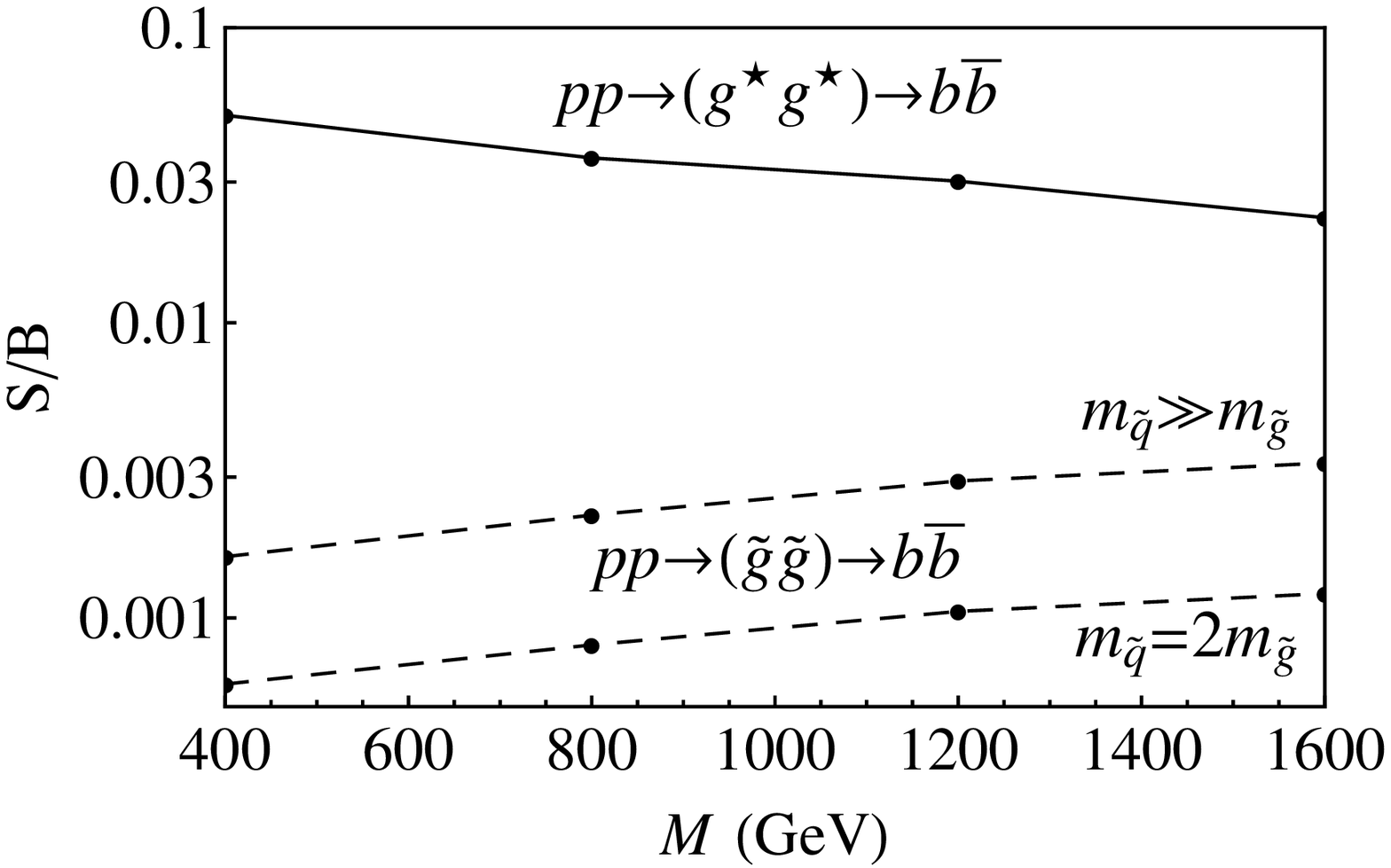}
\includegraphics[width=0.49\textwidth]{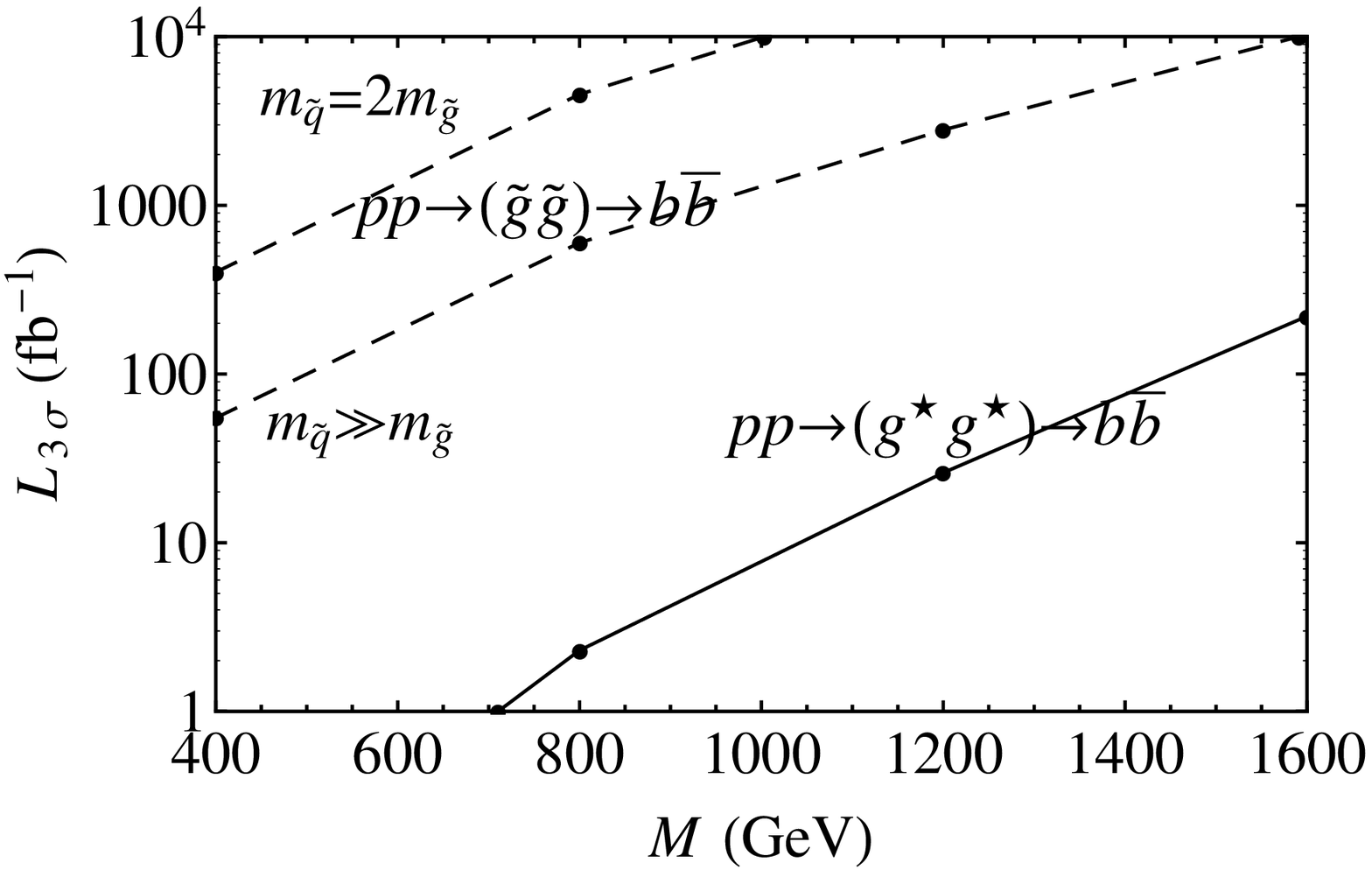}
\caption{$b\bar b$ channel: signal-to-background ratio (left) and the luminosity required for $3\sigma$ significance (right) for gluinonium (dashed lines) and KK gluonium (solid line) at the $14$ TeV LHC as a function of the resonance mass. For KK gluonium we assume $m_\kkq = m_\kkg$ while for gluinonium we present two cases.}
\label{fig-SB-bbbar-M}
\end{center}
\end{figure}

\begin{figure}[t]
\begin{center}
\includegraphics[width=0.49\textwidth]{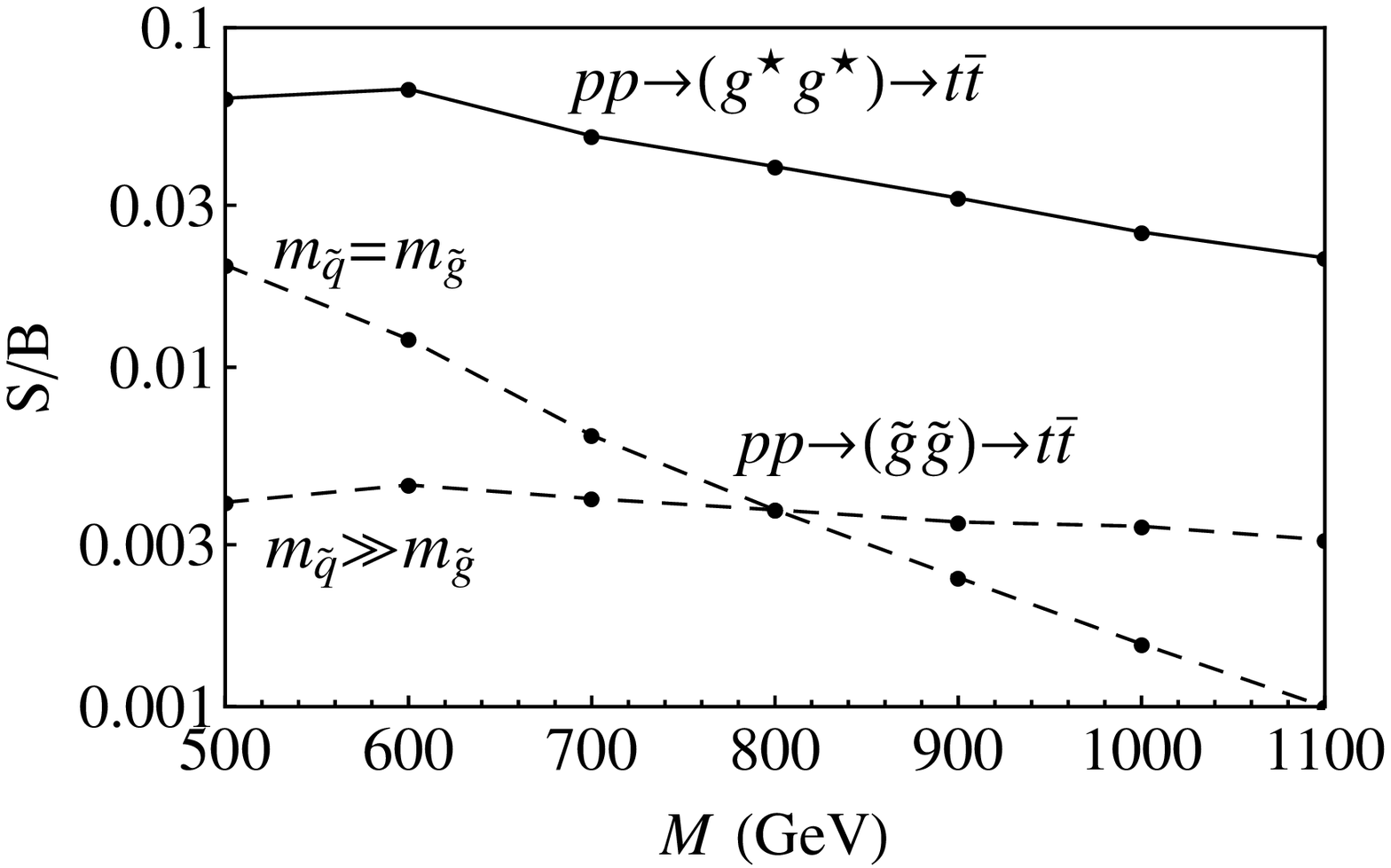}
\includegraphics[width=0.49\textwidth]{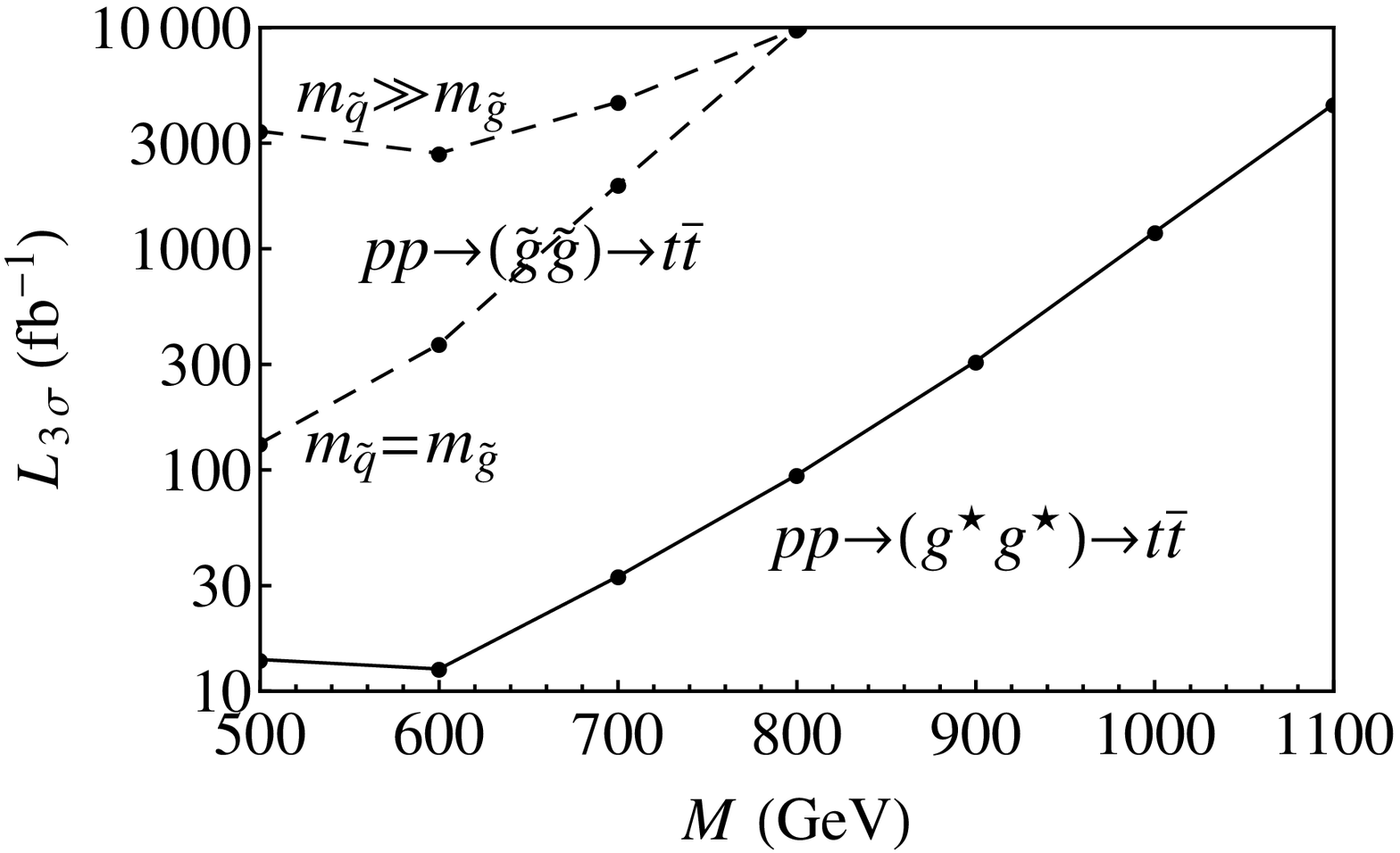}
\caption{Semileptonic $t\bar t$ channel: signal-to-background ratio (left) and the luminosity required for $3\sigma$ significance (right) for gluinonium (dashed lines) and KK gluonium (solid line) at the $14$ TeV LHC as a function of the resonance mass. For KK gluonium we assume $m_\kkq = m_\kkg$ while for gluinonium we present two cases.}
\label{fig-SB-ttbar-M}
\end{center}
\end{figure}

In the analysis of the dijet channel, we require the two hardest jets within $|\eta| < 2.5$ to have $p_T > 2M/5$, the scattering angle in the partonic collision frame to satisfy $\l|\cos\theta\r| < 0.5$, and the dijet invariant mass to be within $\pm 15\%$ from the mass of the resonance. In the $b\bar b$ channel, we require two tagged jets and assume $60\%$ tagging efficiency for $b$-jets, $1\%$ mistag rate for gluon and light quark jets and $15\%$ mistag rate for $c$-jets. Based on~\cite{ATL-PHYS-PUB-2009-018,CMS-PAS-BTV-09-001}, we believe this level of $b$ tagging performance will be realistic at least for resonances with $M \lesssim 1$~TeV. This leaves the standard model $b\bar b$ production as the dominant background. Besides the tagging efficiency factor, we use the same cuts as in the dijet channel. In the $t\bar t$ channel, we consider the standard model top production processes to be the background and use the procedure described in appendix~C of~\cite{Kats:2009bv} to reconstruct the $4$-momenta of the two tops from their semileptonic decay products (which is the situation where one $W$ from $t\to Wb$ decays leptonically and the other hadronically). We then count events in which the $t\bar t$ invariant mass is within $\pm 15\%$ from the mass of the resonance, the tops have $p_T > M/3$, and $\l|\cos\theta\r| < 0.5$.

The results for the dijet, $b\bar b$ and $t\bar t$ channels, respectively, are presented figures~\ref{fig-SB-dijet-M}, \ref{fig-SB-bbbar-M} and~\ref{fig-SB-ttbar-M}. The dijet channel is unlikely to be realistic because of the very small signal-to-background ratio $S/B$ (which implies high sensitivity to systematic errors in modeling the background). In the $b\bar b$ and $t\bar t$ channels the situation regarding $S/B$ is more promising, especially for KK gluonium. As for the fluctuations in the background, the luminosity required for the statistical significance of the signal\footnote{We compute the significance as $S/\sqrt B$, and present the value of integrated luminosity for which $S/\sqrt B = 3$.} is achievable for $m_\go \lesssim 450$~GeV (if $m_\sq \gg m_\go$) or $m_\kkg \lesssim 800$~GeV in the $b\bar b$ channel, and for $m_\go \lesssim 300$~GeV (if $m_\sq \sim m_\go$) or $m_\kkg \lesssim 500$~GeV in the $t\bar t$ channel.

\begin{figure}[t]
\begin{center}
\includegraphics[width=0.49\textwidth]{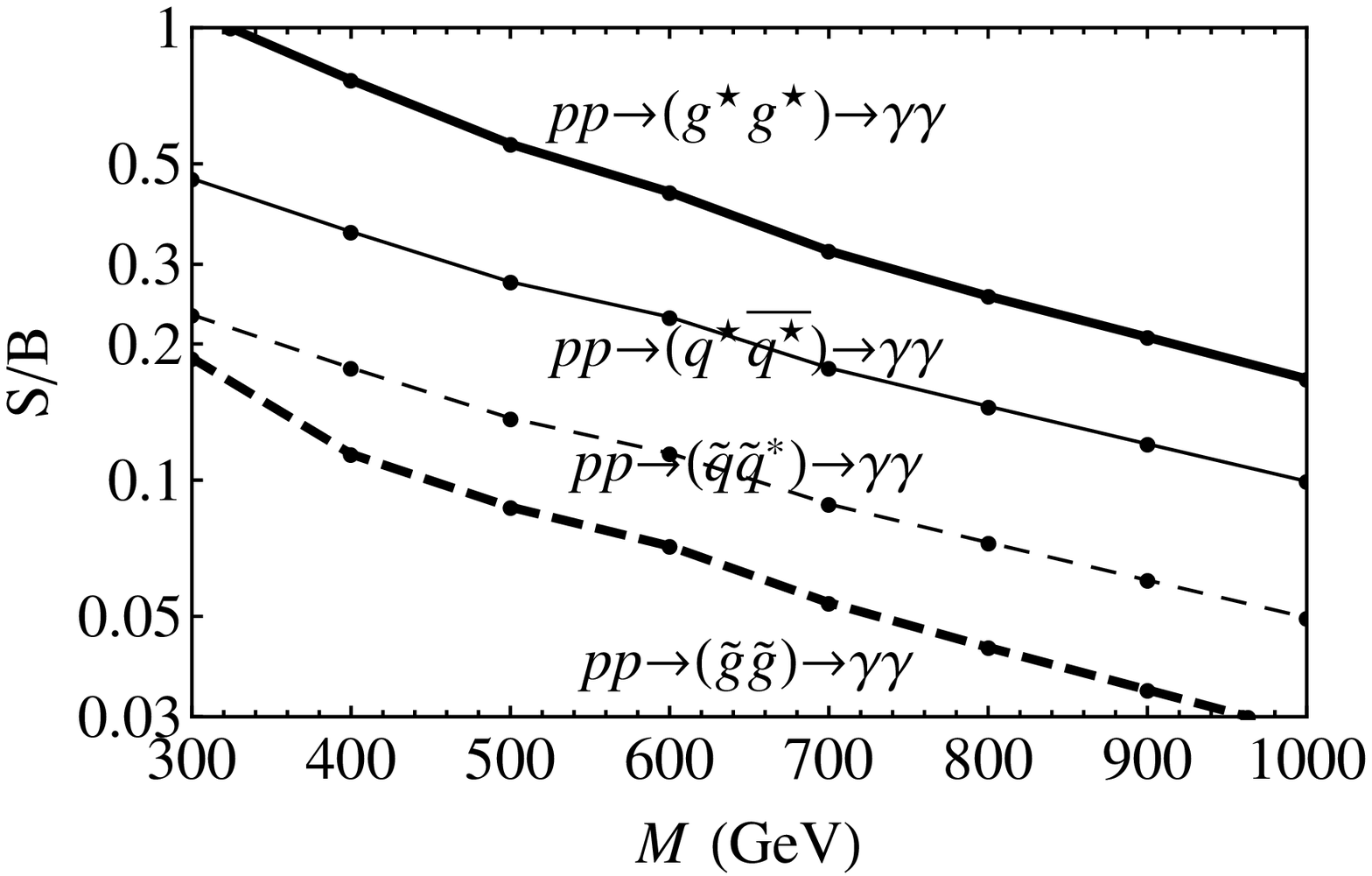}
\includegraphics[width=0.49\textwidth]{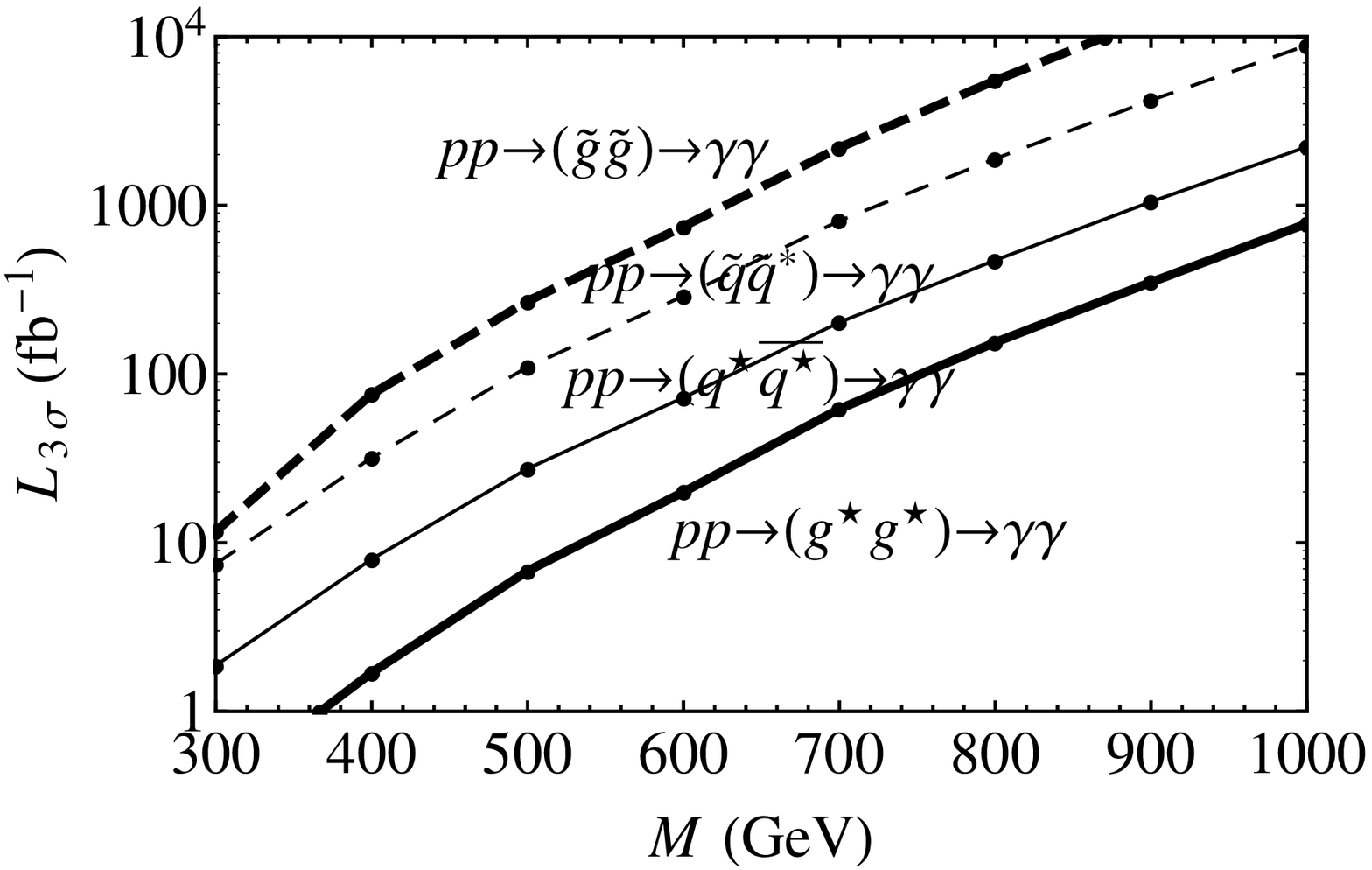}
\caption{Diphoton channel: signal-to-background ratio (left) and the luminosity required for $3\sigma$ significance (right) for gluinonium (thick dashed line), KK gluonium (estimate; thick solid line), squarkonium (thin dashed line) and KK quarkonium (thin solid line) at the $14$ TeV LHC as a function of the resonance mass. For gluinonium we assumed $m_\sq = m_\go$ (and similarly for KK gluonium).}
\label{fig-SB-diphoton-M}
\end{center}
\end{figure}

In the diphoton channel we consider $q\bar q\to\gamma\gamma$ and $gg\to\gamma\gamma$ to be the dominant backgrounds.\footnote{There will also be a contribution from $\gamma$+jet and dijet events in which jets are misidentified as photons. However, for the heavy resonances considered here, this background can be made subdominant by the tight photon identification criteria with only a mild reduction of the signal (see~\cite{CMS-PAS-EXO-10-019} for a recent study by CMS).} We select events in which the two hardest photons within $|\eta| < 1.5$ have $p_T > 50$~GeV and look in the invariant mass window of $\pm 2\%$ around the mass of the resonance. The results are presented in figure~\ref{fig-SB-diphoton-M}. Since gluinonium is unlikely to be lighter than $600$~GeV (corresponding to a $300$~GeV gluino), the luminosity required to see its $\gamma\gamma$ signal is too high. On the other hand, for KK gluonium the signal is promising up to KK gluon masses of $\sim 500$~GeV. For squarkonium, the signal is viable for squark masses $\lesssim 350$~GeV. While typical squarks are expected to be at least as heavy as $\sim 600$~GeV, a stop can be much lighter. Luckily, the stop is also the squark that has the largest chance for being sufficiently long-lived so that stoponium will decay primarily by annihilation. Several models in which this happens were discussed in~\cite{Martin:2008sv}. The stoponium $\gamma\gamma$ signal can thus be within the reach of the LHC. Similarly, a KK quark with mass $\lesssim 400$~GeV can have a viable $\gamma\gamma$ signal from KK quarkonium. Here we assumed that the KK quarks or squarks that form the bound state have charge $Q = 2/3$. For KK quarks or squarks with charge $Q = -1/3$ the cross sections will be $16$ times smaller.

\begin{figure}[t]
\begin{center}
\includegraphics[width=0.49\textwidth]{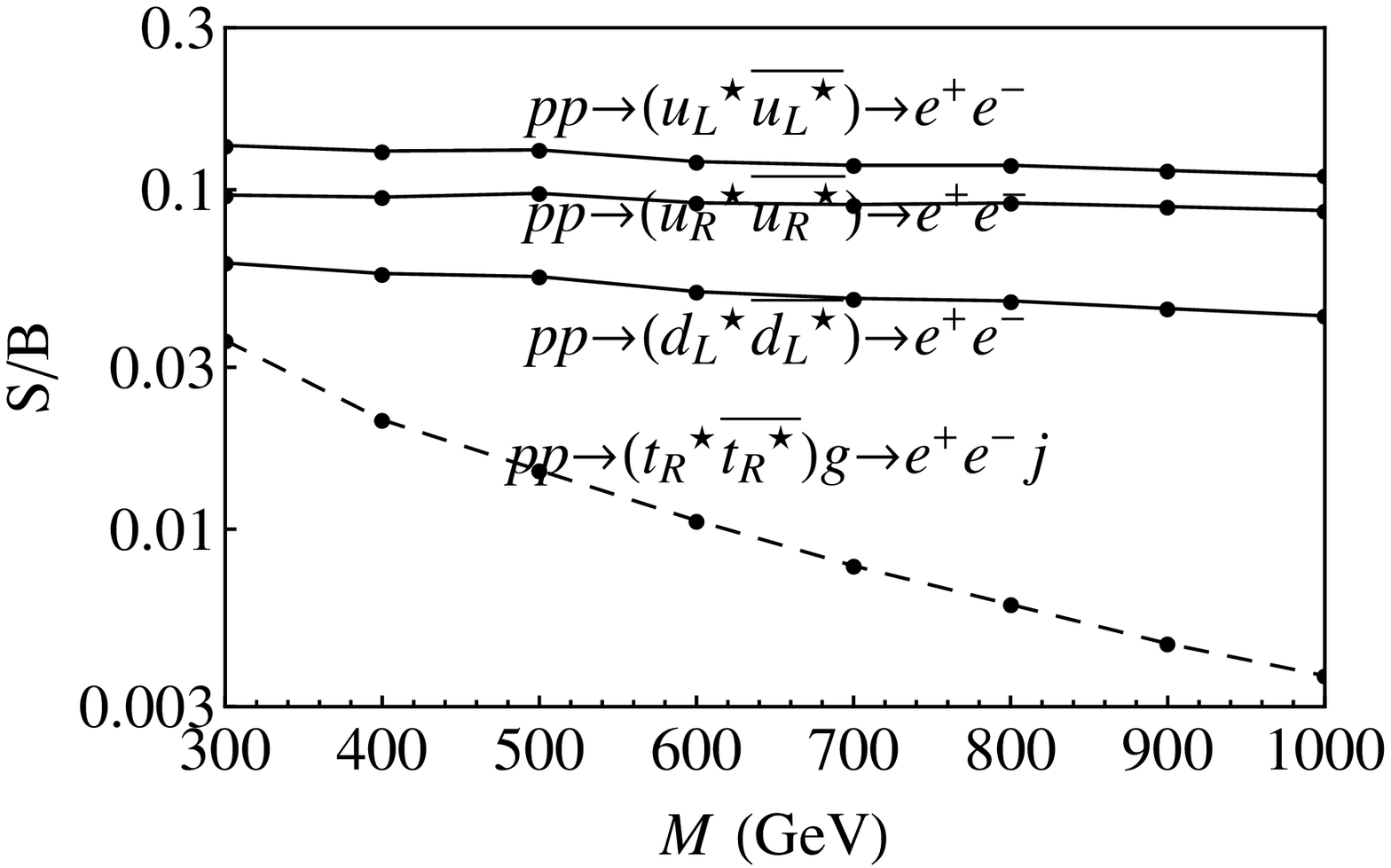}
\includegraphics[width=0.49\textwidth]{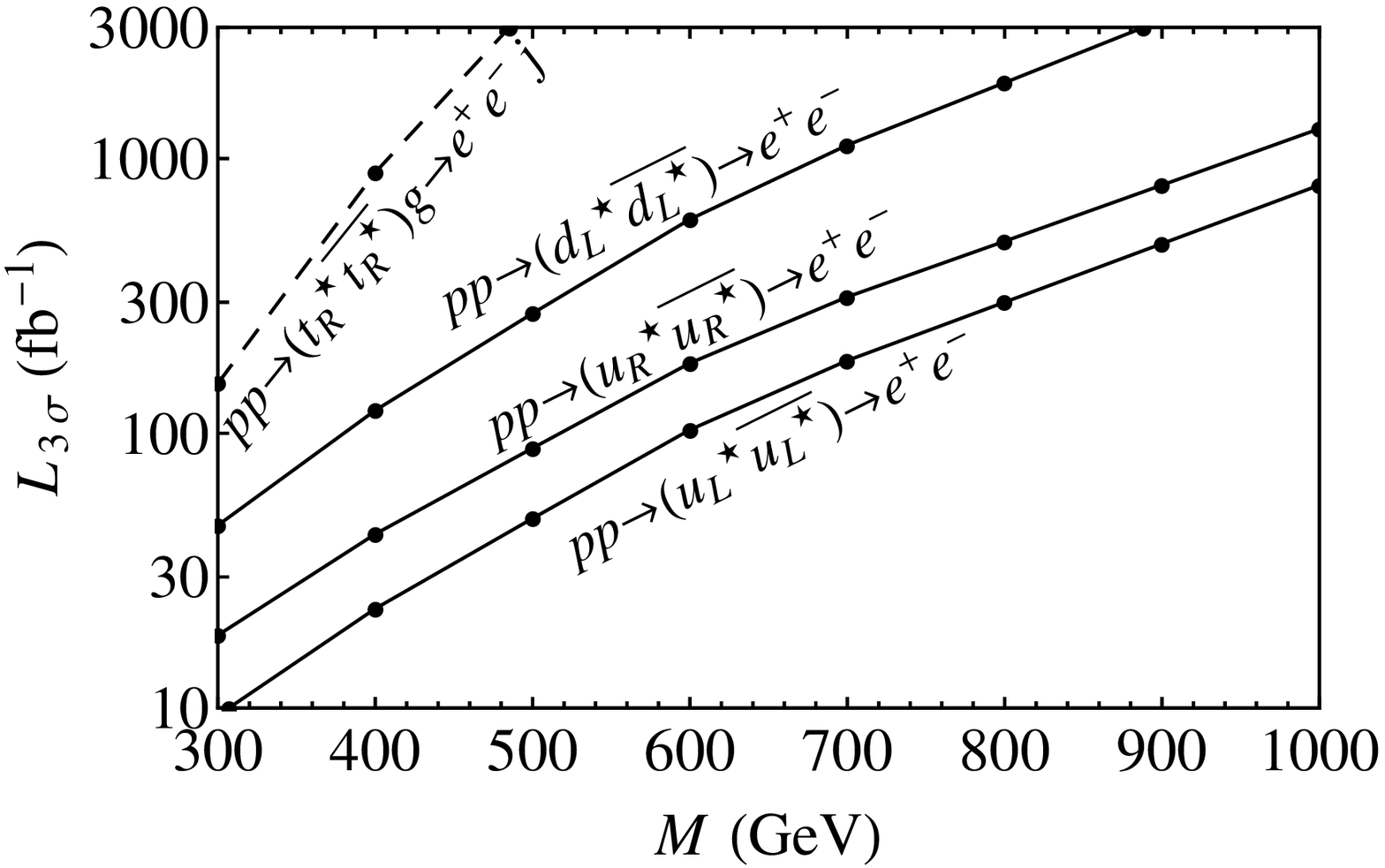}
\caption{Dielectron channel: signal-to-background ratio (left) and the luminosity required for $3\sigma$ significance (right) for several examples of spin-$1$ KK quarkonia at the $14$ TeV LHC as a function of the resonance mass. The solid curves correspond to the left plot of figure~\ref{fig-cs-dilepton-M} and the dashed curves to the right plot.}
\label{fig-SB-dilepton-M}
\end{center}
\end{figure}

In the dilepton channel, we focus on $e^+e^-$ processes since they will have the best mass resolution. We include Drell-Yan processes as the background. We select events which include two electrons with $|\eta| < 2.5$ and $p_T > 60$~GeV and look in the invariant mass window of $\pm 2\%$ around the mass of the resonance. The resulting reach is shown in figure~\ref{fig-SB-dilepton-M} for several cases of KK quarkonium. The signal is detectable for some KK quarkonia with KK quarks corresponding to the light quark flavors if $m_\kkq \lesssim 500$~GeV.  On the other hand, for a perhaps more likely case that $t^\star_R$ is light, the $e^+e^-+\mbox{jet}$ signal of $(t^\star_R\bar{t^\star_R})$ is much less promising, especially since for this case, as in the right plot of figure~\ref{fig-cs-dilepton-M}, we are already considering the more favorable situation of $m_\kkg = 2m_\kkq$ (rather than $m_\kkg \approx m_\kkq$).

\section{Conclusions\label{sec-conclusions}}
\setcounter{equation}{0}

In this paper we have addressed all the possible $S$-wave QCD bound states of pairs of particles from the UED scenario and compared them with analogous bound states of MSSM particles. KK gluonium and gluinonium were also compared with bound states of adjoint scalars, and KK quarkonium and squarkonium with bound states of color-triplet spin-$1$ particles.

We have found that if the KK gluon is sufficiently stable (which is likely to be the case if it is lighter than the KK quarks), its bound states signals at the LHC will have cross sections that are significantly larger than those of the gluinonium of the MSSM, mostly due to the larger number of spin states for KK gluonia. Besides the order-of-magnitude difference in the size of the cross sections, the angular distributions of the annihilation products can be used as a discriminator since KK gluonia will predominantly form with spin $2$ (and some spin $0$), while gluinonia will be mostly spin $0$ (and some spin $1$).

The diphoton signal of spin-$0$ KK quarkonium is twice as large as that of the squarkonium, and is potentially detectable if the constituent KK quarks are sufficiently stable and light. This difference in the cross section can help distinguish between KK quarks and squarks. Another potential discriminator is the dilepton annihilation signal of the vector KK quarkonium which might be measurable.

Despite the fact that bound state cross sections are very small, we found that in some cases the detection of the signal is realistic. For example, a $600$~GeV KK gluon can give rise to a $1.2$~TeV resonance in the $b\bar b$ channel, with $S/B \sim 3\%$ and $3\sigma$ significance reached at $\sim 30\mbox{ fb}^{-1}$. With even more luminosity, the angular distribution can hopefully be extracted and indicate the spin-$2$ nature of the bound state, confirming that the KK gluon has spin $1$. Another example is a $400$~GeV KK top whose bound state will appear as an $800$~GeV resonance in the $\gamma\gamma$ channel with $S/B \sim 10\%$ and $3\sigma$ significance at $\sim 600\mbox{ fb}^{-1}$. Furthermore, many of our results remain valid for other models that contain pair-produced particles with color representations and spins that our study has covered, but which can otherwise be slightly or completely different from the standard MSSM and UED scenarios and have weaker experimental bounds on the masses, which opens up additional possibilities.

Overall, our study has demonstrated how processes of bound state formation and annihilation, which are easy to compute for a given model and easy to reconstruct at the collider, provide additional channels for studying new particles at the LHC.

\section*{Acknowledgments}
We would like to thank Matthew Schwartz for giving us the idea for this project and for many useful discussions. We are also grateful to Matthew Strassler for several important suggestions. Our research is supported by NSF grant PHY-0804450 and DOE grant DE-FG02-96ER40959. Some of the computations in this paper were performed on the Odyssey cluster supported by the FAS Research Computing Group at Harvard University. DK is supported by the General Sir John Monash Award.

\appendix

\section{Cross sections, annihilation rates, angular distributions\label{app-rates}}
\setcounter{equation}{0}

In this appendix we give the expressions for the parton-level near-threshold pair-production cross section $\hat\sigma_0(\hat s)$ which enters the calculation of the bound state production cross section (\ref{sigma-bound}). For UED processes, we used Feynman rules from~\cite{Dicus:2000hm,Macesanu:2002db}. We separate the results by the color representation of the pair, its spin $J$, and the spin component $J_z$ along the direction of motion of the first parton (where by, e.g., $|J_z| = 1$, we will refer to the sum of the contributions with $J_z = 1$ and $J_z = -1$ in cases when the two contributions are equal). We consider all the attractive color states and all the possible spins that can be produced from incoming quarks and/or gluons. We use $q$ and $q'$ to refer to two different flavors of quarks (and similarly for KK quarks and squarks), while the notation $q^{(\prime)}$ means that the flavor can be either the same or different from $q$. We use the subscript $\chi$ (as in $\sq_\chi$) to denote the chirality of the particle ($\chi = L\mbox { or } R$) while $\slash\chi$ refers to the other chirality. Our expressions for the cross sections refer to a single flavor and chirality choice for each of the two produced particles (for squarks and KK quarks). Processes that are not listed have vanishing cross sections at threshold.

We also list the annihilation rates of the bound states into various final states and the corresponding angular distributions (where $\theta$ is the angle between the momenta of the annihilation products and the spin quantization axis, in the center-of-mass frame).

When a $t$ quark appears as an annihilation product we take its mass $m_t$ into account, but otherwise the masses of the quarks are set to zero. For simplicity, we assume that all the KK quarks have a common mass $m_\kkq$ and all the squarks have mass $m_\sq$.

\subsection{KK gluonia\label{app-rates-KKgKKg}}

The cross sections for KK gluon pair production processes in the near-threshold limit, for the attractive color representations, are
\bea
\hat\sigma_0\l(gg \to \kkg \kkg\,;\,\vv{1},\,J=0\r) &=& \frac{27\pi\alpha_s^2\beta}{256m_\kkg^2}\\
\hat\sigma_0\l(gg \to \kkg \kkg\,;\,\vv{8_S},\,J=0\r) &=& \frac{27\pi\alpha_s^2\beta}{128m_\kkg^2}\\
\hat\sigma_0\l(gg \to \kkg \kkg\,;\,\vv{1},\,J=2,|J_z|=2\r) &=& \frac{9\pi\alpha_s^2\beta}{16m_\kkg^2}\\
\hat\sigma_0\l(gg \to \kkg \kkg\,;\,\vv{8_S},\,J=2,|J_z|=2\r) &=& \frac{9\pi\alpha_s^2\beta}{8m_\kkg^2} \\
\hat\sigma_0\l(q\bar q \to \kkg \kkg\,;\,\vv{1},\,J=2,|J_z|=1\r) &=& \l(\frac{2m_\kkg^2}{m_\kkg^2 + m_\kkq^2}\r)^2 \frac{2\pi\alpha_s^2\beta}{27m_\kkg^2} \\
\hat\sigma_0\l(q\bar q \to \kkg \kkg\,;\,\vv{8_S},\,J=2,|J_z|=1\r) &=& \l(\frac{2m_\kkg^2}{m_\kkg^2 + m_\kkq^2}\r)^2 \frac{5\pi\alpha_s^2\beta}{27m_\kkg^2}
\eea

The annihilation rates of the KK gluonia into the various final states are
\bea
\Gamma\l((\kkg\kkg)_{\vv{1},J=0}\to gg\r) &=& \frac{729}{8}\,\bar\alpha_s^3\alpha_s^2 m_\kkg \\
\Gamma\l((\kkg\kkg)_{\vv{8_S},J=0}\to gg\r) &=& \frac{729}{256}\,\bar\alpha_s^3\alpha_s^2 m_\kkg \\
\Gamma\l((\kkg\kkg)_{\vv{1},J=0}\to t\bar t\r) &=& \frac{3}{2}\frac{m_t^2}{m_\kkg^2}\l(1-\frac{m_t^2}{m_\kkg^2}\r)^{3/2} \l(\frac{2m_\kkg^2}{m_\kkg^2 + m_\kkt^2 - m_t^2}\r)^2 \bar\alpha_s^3\alpha_s^2 m_\kkg \\
\Gamma\l((\kkg\kkg)_{\vv{8_S},J=0}\to t\bar t\r) &=& \frac{15}{256}\frac{m_t^2}{m_\kkg^2}\l(1-\frac{m_t^2}{m_\kkg^2}\r)^{3/2} \l(\frac{2m_\kkg^2}{m_\kkg^2 + m_\kkt^2 - m_t^2}\r)^2 \bar\alpha_s^3\alpha_s^2 m_\kkg \\
\Gamma\l((\kkg\kkg)_{\vv{1},J=2}\to gg\r) &=& \frac{486}{5}\,\bar\alpha_s^3\alpha_s^2 m_\kkg \\
\Gamma\l((\kkg\kkg)_{\vv{8_S},J=2}\to gg\r) &=& \frac{243}{80}\,\bar\alpha_s^3\alpha_s^2 m_\kkg \\
\Gamma\l((\kkg\kkg)_{\vv{1},J=2}\to q\bar q\r) &=& \frac{18}{5} \l(\frac{2m_\kkg^2}{m_\kkg^2 + m_\kkq^2}\r)^2 \bar\alpha_s^3\alpha_s^2 m_\kkg \\
\Gamma\l((\kkg\kkg)_{\vv{8_S},J=2}\to q\bar q\r) &=& \frac{9}{64} \l(\frac{2m_\kkg^2}{m_\kkg^2 + m_\kkq^2}\r)^2 \bar\alpha_s^3\alpha_s^2 m_\kkg \eea
\be
\Gamma\l((\kkg\kkg)_{\vv{1},J=2}\to t\bar t\r) = \frac{18}{5} \l(1-\frac{m_t^2}{m_\kkg^2}\r)^{3/2} \l(1 + \frac{2m_t^2}{3m_\kkg^2}\r) \l(\frac{2m_\kkg^2}{m_\kkg^2 + m_\kkt^2 - m_t^2}\r)^2 \bar\alpha_s^3\alpha_s^2 m_\kkg
\ee
\be
\Gamma\l((\kkg\kkg)_{\vv{8_S},J=2}\to t\bar t\r) = \frac{9}{64} \l(1-\frac{m_t^2}{m_\kkg^2}\r)^{3/2} \l(1 + \frac{2m_t^2}{3m_\kkg^2}\r) \l(\frac{2m_\kkg^2}{m_\kkg^2 + m_\kkt^2 - m_t^2}\r)^2 \bar\alpha_s^3\alpha_s^2 m_\kkg
\ee
The annihilation rates into $q\bar q$ should be further summed over the different quark flavors.

The angular distributions in annihilation of KK gluonia (in both $\vv{1}$ and $\vv{8_S}$ representations) into $gg$ are given by
\bea
J=2,J_z=\pm 2:&\qq& \frac{1}{\Gamma}\frac{d\Gamma}{\sin\theta\,d\theta} = \frac{5}{256}\l(35 + 28 \cos 2\theta + \cos 4\theta\r) \\
J=2,J_z=\pm 1:&\qq& \frac{1}{\Gamma}\frac{d\Gamma}{\sin\theta\,d\theta} = \frac{5}{16} \l(3 + \cos 2\theta\r) \sin^2\theta \\
J=2,J_z=0:&\qq& \frac{1}{\Gamma}\frac{d\Gamma}{\sin\theta\,d\theta} = \frac{15}{16} \sin^4\theta
\eea
In annihilation into $q\bar q$, for massless quarks we have
\bea
J=2,J_z=\pm 2:&\qq& \frac{1}{\Gamma}\frac{d\Gamma}{\sin\theta\,d\theta} = \frac{5}{16} \l(3 + \cos 2\theta\r) \sin^2\theta \\
J=2,J_z=\pm 1:&\qq& \frac{1}{\Gamma}\frac{d\Gamma}{\sin\theta\,d\theta} = \frac{5}{16} \l(2 + \cos 2\theta + \cos 4\theta\r) \\
J=2,J_z=0:&\qq& \frac{1}{\Gamma}\frac{d\Gamma}{\sin\theta\,d\theta} = \frac{15}{16} \sin^2 2\theta
\eea
and for $t\bar t$
\bea
J=2,J_z=\pm 2:&& \frac{1}{\Gamma}\frac{d\Gamma}{\sin\theta\,d\theta} = \frac{5}{16}\,\frac{1}{1+2m_t^2/3m_\kkg^2}\nn\\
&&\qqqqq\times\l(3 + \frac{m_t^2}{m_\kkg^2} + \l(1 - \frac{m_t^2}{m_\kkg^2}\r)\cos 2\theta\r) \sin^2\theta \\
J=2,J_z=\pm 1:&& \frac{1}{\Gamma}\frac{d\Gamma}{\sin\theta\,d\theta} = \frac{5}{16}\,\frac{1}{1+2m_t^2/3m_\kkg^2}\nn\\
&&\qqqqq\times\l(2 + \frac{m_t^2}{m_\kkg^2} + \cos 2\theta + \l(1 - \frac{m_t^2}{m_\kkg^2}\r)\cos 4\theta\r) \\
J=2,J_z=0:&& \frac{1}{\Gamma}\frac{d\Gamma}{\sin\theta\,d\theta} = \frac{15}{16}\,\frac{1}{1+2m_t^2/3m_\kkg^2}\\
&&\qqqqq\times\l(\sin^2 2\theta + \frac{11m_t^2}{18m_\kkg^2} + \frac{2m_t^2}{3m_\kkg^2}\cos 2\theta + \frac{m_t^2}{2m_\kkg^2}\cos 4\theta\r) \nn
\eea

\subsection{Gluinonia\label{app-rates-gogo}}

The near-threshold production cross sections of attractive gluino pairs are
\bea
\hat\sigma_0\l(gg \to \go\go\,;\,\vv{1},\,J=0\r) &=& \frac{9\pi\alpha_s^2\beta}{128m_\go^2} \\
\hat\sigma_0\l(gg \to \go\go\,;\,\vv{8_S},\,J=0\r) &=& \frac{9\pi\alpha_s^2\beta}{64m_\go^2} \\
\hat\sigma_0\l(q\bar q \to \go\go\,;\,\vv{8_A},\,J=1,\,|J_z|=1\r) &=& \l(\frac{m_\sq^2-m_\go^2}{m_\sq^2+m_\go^2}\r)^2\frac{\pi\alpha_s^2\beta}{3m_\go^2}
\eea

The annihilation rates of the gluinonia are
\bea
\Gamma\l((\go\go)_{\vv{1},J=0}\to gg\r) &=& \frac{243}{4} \bar\alpha_s^3\alpha_s^2 m_\go \\
\Gamma\l((\go\go)_{\vv{8_S},J=0}\to gg\r) &=& \frac{243}{128} \bar\alpha_s^3\alpha_s^2 m_\go \\
\Gamma\l((\go\go)_{\vv{1},J=0}\to t\bar t\r)
&=& 9\,\frac{m_t^2}{m_\go^2}\sqrt{1 - \frac{m_t^2}{m_\go^2}}\,\l(\frac{2m_\go^2}{m_\go^2 + m_\st^2 - m_t^2}\r)^2 \bar\alpha_s^3\alpha_s^2 m_\go \\
\Gamma\l((\go\go)_{\vv{8_S},J=0}\to t\bar t\r)
&=& \frac{45}{128}\,\frac{m_t^2}{m_\go^2} \sqrt{1 - \frac{m_t^2}{m_\go^2}}\,\l(\frac{2m_\go^2}{m_\go^2 + m_\st^2 - m_t^2}\r)^2 \bar\alpha_s^3\alpha_s^2 m_\go \\
\Gamma\l((\go\go)_{\vv{8_A},J=1}\to q\bar q\r) &=& \frac{27}{64} \left(\frac{m_\sq^2 - m_\go^2}{m_\sq^2 + m_\go^2}\right)^2
\bar\alpha_s^3\alpha_s^2 m_\go \\
\Gamma\l((\go\go)_{\vv{8_A},J=1}\to t\bar t\r) &=& \frac{27}{64}\l(1 + \frac{m_t^2}{2m_\go^2}\r)\sqrt{1 - \frac{m_t^2}{m_\go^2}}\, \left(\frac{m_\st^2 - m_\go^2 - m_t^2}{m_\st^2 + m_\go^2 - m_t^2}\right)^2
\bar\alpha_s^3\alpha_s^2 m_\go
\eea

The branching ratio into $\gamma\gamma$ via loop diagrams is~\cite{Kauth:2009ud}
\be
\frac{\Gamma\l((\go\go)_{\vv{1},J=0}\to \gamma\gamma\r)}{\Gamma\l((\go\go)_{\vv{1},J=0}\to gg\r)} = \frac{50\alpha^2}{81\pi^2}\l|\mbox{Li}_2\l(-\frac{m_\go^2}{m_\sq^2}\r) - \mbox{Li}_2\l(\frac{m_\go^2}{m_\sq^2}\r)\r|^2 = \frac{25\pi^2}{648}\alpha^2
\label{gogo-diphoton}
\ee
where in the last expression we substituted $m_\sq = m_\go$.

The $J=1$ $\vv{8_A}$ gluinonium has the following angular distributions in the annihilation into $q\bar q$ for massless quarks:
\bea
J=1,J_z=\pm 1:&\qq& \frac{1}{\Gamma}\frac{d\Gamma}{\sin\theta\,d\theta} = \frac{3}{8} \l(1 + \cos^2\theta\r) \\
J=1,J_z=0:&\qq& \frac{1}{\Gamma}\frac{d\Gamma}{\sin\theta\,d\theta} = \frac{3}{4} \sin^2\theta
\eea
while for $t\bar t$:
\bea
J=1,J_z=\pm 1:&\qq& \frac{1}{\Gamma}\frac{d\Gamma}{\sin\theta\,d\theta}
= \frac{3}{8}\,\frac{1+m_t^2/m_\go^2}{1+m_t^2/2m_\go^2} \l(1 + \frac{1-m_t^2/m_\go^2}{1+m_t^2/m_\go^2}\,\cos^2\theta\r) \\
J=1,J_z=0:&\qq& \frac{1}{\Gamma}\frac{d\Gamma}{\sin\theta\,d\theta}
= \frac{3}{4}\,\frac{1}{1+m_t^2/2m_\go^2} \l(\sin^2\theta + \frac{m_t^2}{m_\go^2}\cos^2\theta\r)
\eea

\subsection{Octetonia\label{app-rates-octetonia}}

The near-threshold production cross sections of attractive pairs of scalar color-octets are
\be
\hat\sigma_0\l(gg \to \phi\phi\,;\,\vv{1}\r) = \frac{9\pi\alpha_s^2\beta}{256m_\phi^2}
\ee
\be
\hat\sigma_0\l(gg \to \phi\phi\,;\,\vv{8_S}\r) = \frac{9\pi\alpha_s^2\beta}{128m_\phi^2}
\ee

The annihilation rates of the octetonia are
\be
\Gamma\l((\phi\phi)_{\vv{1}}\to gg\r) = \frac{243}{8} \bar\alpha_s^3\alpha_s^2 m_\phi
\ee
\be
\Gamma\l((\phi\phi)_{\vv{8_S}}\to gg\r) = \frac{243}{256} \bar\alpha_s^3\alpha_s^2 m_\phi
\ee

\subsection{KK quarkonia\label{app-rates-KKqKKq}}

The near-threshold production cross sections of attractive KK quark-KK antiquark pairs are
\bea
\hat\sigma_0\l(gg \to \kkqX \bar \kkqX\,;\,\vv{1},\,J=0\r)
&=& \frac{\pi\alpha_s^2\beta}{96m_\kkq^2} \\
\hat\sigma_0\l(q\bar q^{(\prime)} \to \kkqX \bar \kkqXX^{(\prime)}\,;\,\vv{1},\, J=0\r)
&=& \l(\frac{2m_\kkq^2}{m_\kkq^2 + m_\kkg^2}\r)^2 \l(4 + \frac{m_\kkq^2}{m_\kkg^2}\r)^2 \frac{\pi\alpha_s^2\beta}{162m_\kkq^2}\\
\hat\sigma_0\l(q\bar q^{(\prime)} \to \kkqX \bar \kkqXX^{(\prime)}\,;\,\vv{1},\, J=1, J_z=0\r)
&=& \l(\frac{2m_\kkq^2}{m_\kkq^2 + m_\kkg^2}\r)^2\l(\frac{m_\kkq^2}{m_\kkg^2}\r)^2 \frac{\pi\alpha_s^2\beta}{162m_\kkq^2}\\
\hat\sigma_0\l(q\bar q^{(\prime)} \to \kkqL \bar \kkqL^{(\prime)}\,;\,\vv{1},\,J=1,J_z=-1\r)
&=& \hat\sigma_0\l(q\bar q^{(\prime)} \to \kkqR \bar \kkqR^{(\prime)}\,;\,\vv{1},\,J=1,J_z=1\r)\nn\\
&=& \l(\frac{2m_\kkq^2}{m_\kkq^2 + m_\kkg^2}\r)^2\l(2 + \frac{m_\kkq^2}{m_\kkg^2}\r)^2 \frac{\pi\alpha_s^2\beta}{81m_\kkq^2}
\eea

The annihilation rates of the KK quarkonia are
\bea
\Gamma\l((\kkqX\bar \kkqX)_{\vv{1},\,J=0}\to gg\r)
&=& \frac{64}{81}\,\bar\alpha_s^3\alpha_s^2 m_\kkq\\
\Gamma\l((\kktX\bar \kktX)_{\vv{1},\,J=0}\to t\bar t\r)
&=& \frac{16}{243}\frac{m_t^2}{m_\kkq^2}\sqrt{1-\frac{m_t^2}{m_\kkq^2}}\nn\\
\qqqqqq\qqqq&\;&\times \l(\frac{2m_\kkq^2}{m_\kkg^2 + m_\kkq^2 - m_t^2}\r)^2 \l(2 + \frac{m_\kkq^2}{m_\kkg^2} - \frac{m_t^2}{m_\kkg^2}\r)^2 \bar\alpha_s^3\alpha_s^2 m_\kkq \q\\
\Gamma\l((\kkqX\bar \kkqX^{(\prime)})_{\vv{1},\,J=1}\to q\bar q^{(\prime)}\r)
&=& \frac{64}{729}\l(\frac{2m_\kkq^2}{m_\kkg^2 + m_\kkq^2}\r)^2 \l(2 + \frac{m_\kkq^2}{m_\kkg^2}\r)^2\bar\alpha_s^3\alpha_s^2 m_\kkq\\
\Gamma\l((\kkqX\bar \kkqXX^{(\prime)})_{\vv{1},\,J=0}\to q\bar q^{(\prime)}\r)
&=& \frac{32}{243}\l(\frac{2m_\kkq^2}{m_\kkg^2 + m_\kkq^2}\r)^2 \l(4 + \frac{m_\kkq^2}{m_\kkg^2}\r)^2 \bar\alpha_s^3\alpha_s^2 m_\kkq \\
\Gamma\l((\kkqX\bar \kkqXX^{(\prime)})_{\vv{1},\,J=1}\to q\bar q^{(\prime)}\r)
&=& \frac{32}{729}
\l(\frac{2m_\kkq^2}{m_\kkg^2 + m_\kkq^2}\r)^2\l(\frac{m_\kkq^2}{m_\kkg^2}\r)^2 \bar\alpha_s^3\alpha_s^2 m_\kkq\\
\Gamma\l((\kkqX\bar \kkqX)_{\vv{1},\,J=0}\to \gamma\gamma\r)
&=& \frac{32}{9}\,\bar\alpha_s^3\,Q^4\alpha^2 m_\kkq
\eea
The electroweak rates into fermions $f$ (leptons or quarks) are
\be
\Gamma\l((q^\star_\chi\bar{q^\star_\chi})_{\vv{1},\,J=1}\to f_\eta\bar f_\eta\r)
= \frac{16}{27}n_c\, \bar\alpha_s^3\l(\frac{\l(Q_\eta - T^3_\eta\r)\l(Q_\chi - T^3_\chi\r)}{\cos^2\theta_W} + \frac{T^3_\eta T^3_\chi}{\sin^2\theta_W}\r)^2
\alpha^2 m_\kkq
\ee
where $\chi$ and $\eta$ describe the chirality of the KK quark and the fermion, respectively, $n_c = 1$ for leptons and $3$ for quarks, $Q$ is the electric charge and $T^3$ is the weak isospin, and we assumed $m_Z^2 \ll (2m_\kkq)^2$. More specifically, for annihilation into charged leptons
\be
\Gamma\l((q^\star_\chi\bar{q^\star_\chi})_{\vv{1},\,J=1}\to \ell_\eta^+\ell_\eta^-\r)
= \frac{16}{27}\,\bar\alpha_s^3 c_{\chi\eta}\frac{\alpha^2}{\cos^4\theta_W} m_\kkq
\ee
with
\be
c_{RR} = Q^2\,,\q
c_{RL} = \frac{Q^2}{4}\,,\q
c_{LR} = \l(Q - T^3\r)^2\,,\q
c_{LL} = \frac{1}{4}\l(Q - T^3 + T^3\cot^2\theta_W\r)^2
\ee
where one should substitute $Q = 2/3$, $T^3 = 1/2$ for up-type KK quarks and $Q = -1/3$, $T^3 = -1/2$ for down-type KK quarks.

The angular distributions for $(\kkqL\bar \kkqL^{(\prime)})_{\vv{1},\,J=1}\to q\bar q^{(\prime)}$ are
\bea
J=1,J_z=1:\qq&& \frac{1}{\Gamma}\frac{d\Gamma}{\sin\theta\,d\theta} = \frac{3}{2}\sin^4\frac{\theta}{2} \\
J=1,J_z=-1:\qq&& \frac{1}{\Gamma}\frac{d\Gamma}{\sin\theta\,d\theta} = \frac{3}{2}\cos^4\frac{\theta}{2} \\
J=1,J_z=0:\qq&& \frac{1}{\Gamma}\frac{d\Gamma}{\sin\theta\,d\theta}
= \frac{3}{4}\sin^2\theta
\eea
while for $(\kkqR\bar \kkqR^{(\prime)})_{\vv{1},\,J=1}\to q\bar q^{(\prime)}$ the same expressions hold for opposite signs of $J_z$.
For both $(\kkqL\bar \kkqR^{(\prime)})_{\vv{1},\,J=1}\to q\bar q^{(\prime)}$ and $(\kkqR\bar \kkqL^{(\prime)})_{\vv{1},\,J=1}\to q\bar q^{(\prime)}$,
\bea
J=1,J_z=\pm 1:&\qq& \frac{1}{\Gamma}\frac{d\Gamma}{\sin\theta\,d\theta} =
\frac{3}{4}\sin^2\theta \\
J=1,J_z=0:&\qq& \frac{1}{\Gamma}\frac{d\Gamma}{\sin\theta\,d\theta}
= \frac{3}{2}\cos^2\theta
\eea
For $(\kkqX\bar\kkqX)_{\vv{1},\,J=1}\to \ell^+_R\ell^-_R$:
\bea
J=1,J_z=1:&\qq& \frac{1}{\Gamma}\frac{d\Gamma}{\sin\theta\,d\theta} =
\frac{3}{2}\cos^4\frac{\theta}{2} \\
J=1,J_z=-1:&\qq& \frac{1}{\Gamma}\frac{d\Gamma}{\sin\theta\,d\theta} =
\frac{3}{2}\sin^4\frac{\theta}{2} \\
J=1,J_z=0:&\qq& \frac{1}{\Gamma}\frac{d\Gamma}{\sin\theta\,d\theta}
= \frac{3}{4}\sin^2\theta
\eea
while for $(\kkqX\bar\kkqX)_{\vv{1},\,J=1}\to \ell^+_L\ell^-_L$ the same expressions hold for opposite signs of $J_z$.

\subsection{Squarkonia\label{app-rates-sqsq}}

The near-threshold production cross sections of attractive squark-antisquark pairs are
\bea
\hat\sigma_0\l(gg \to \sq_\chi\sq^\ast_\chi;\,\,\vv{1},\,J=0\r) &=& \frac{\pi\alpha_s^2\beta}{192m_\sq^2}\\
\hat\sigma_0\l(q\bar q^{(\prime)} \to \sq_\chi\sq^{(\prime)\ast}_{\slash\chi};\,\vv{1},\,J=0\r) &=& \l(\frac{2m_\go m_\sq}{m_\sq^2+m_\go^2}\r)^2 \frac{4\pi\alpha_s^2\beta}{81m_\sq^2}
\eea

The squarkonium annihilation rates are
\bea
\Gamma\l((\sq_\chi\sq^\ast_\chi)_{\vv{1},\,J=0}\to gg\r) &=& \frac{32}{81}\, \bar\alpha_s^3\alpha_s^2 m_\sq \\
\Gamma\l((\st_\chi\st_\chi^\ast)_{\vv{1},\,J=0}\to t\bar t\r) &=& \frac{128}{243}\frac{m_t^2}{m_\sq^2} \l(1-\frac{m_t^2}{m_\sq^2}\r) \sqrt{1-\frac{m_t^2}{m_\sq^2}} \l(\frac{2m_\sq^2}{m_\go^2 + m_\sq^2 - m_t^2}\r)^2 \bar\alpha_s^3\alpha_s^2 m_\sq \qqq\\
\Gamma\l((\sq_\chi\sq_{\slash\chi}^{(\prime)\ast})_{\vv{1},\,J=0}\to q\bar q^{(\prime)}\r) &=& \frac{256}{243}\l(\frac{2m_\go m_\sq}{m_\go^2 + m_\sq^2}\r)^2 \bar\alpha_s^3\alpha_s^2 m_\sq \\
\Gamma\l((\sq_\chi\sq^\ast_\chi)_{\vv{1},\,J=0}\to \gamma\gamma\r) &=& \frac{16}{9}\, \bar\alpha_s^3\,Q^4\alpha^2 m_\sq
\eea
where $Q$ is the electric charge of the squark.

\subsection{Tripletonia\label{app-rates-tripletonia}}

The near-threshold production cross sections of attractive pairs of vector color-triplets are
\bea
\hat\sigma_0\l(gg \to WW^\ast\,;\,\vv{1},\,J=2,|J_z|=2\r) &=& \frac{\pi\alpha_s^2\beta}{12m_W^2}\\
\hat\sigma_0\l(gg \to WW^\ast\,;\,\vv{1},\,J=0\r) &=& \frac{\pi\alpha_s^2\beta}{64m_W^2}
\eea

The annihilation rates of the tripletonia into $gg$ are
\be
\Gamma\l((WW^\ast)_{\vv{1},J=2}\to gg\r) = \frac{512}{405}\,\bar\alpha_s^3\alpha_s^2 m_W
\ee
\be
\Gamma\l((WW^\ast)_{\vv{1},J=0}\to gg\r) = \frac{32}{27}\,\bar\alpha_s^3\alpha_s^2 m_W
\ee
If the vector bosons have electric charge $Q$ they can annihilate into $\gamma\gamma$ with rates which can be obtained by the replacement
\be
\alpha_s^2 \to \frac{9}{2}Q^4\alpha^2
\ee
The angular distributions of the annihilation products (both $gg$ and $\gamma\gamma$) are
\bea
J=2,J_z=\pm 2:&\qq& \frac{1}{\Gamma}\frac{d\Gamma}{\sin\theta\,d\theta} = \frac{5}{256}\l(35 + 28\cos2\theta + \cos4\theta\r) \\
J=2,J_z=\pm 1:&\qq& \frac{1}{\Gamma}\frac{d\Gamma}{\sin\theta\,d\theta} = \frac{5}{16} \l(3 + \cos 2\theta\r) \sin^2\theta \\
J=2,J_z=0:&\qq& \frac{1}{\Gamma}\frac{d\Gamma}{\sin\theta\,d\theta} = \frac{15}{16}\, \sin^4\theta
\eea

\subsection{Di-KK quarks\label{app-rates-diKKq}}

The near-threshold production cross sections of attractive KK quark pairs are
\bea
\hat\sigma_0\l(qq \to \kkqX \kkqX\,;\,\vv{\bar 3},\, J=1,J_z=0\r)
&=& \l(\frac{2m_\kkq^2}{m_\kkq^2 + m_\kkg^2}\r)^2\l(\frac{m_\kkq^2}{m_\kkg^2}\r)^2\frac{\pi\alpha_s^2\beta}{108m_\kkq^2}\\
\hat\sigma_0\l(qq' \to \kkqX \kkqX'\,;\,\vv{\bar 3},\, J=1,J_z=0\r)
&=& \l(\frac{2m_\kkq^2}{m_\kkq^2 + m_\kkg^2}\r)^2\l(\frac{m_\kkq^2}{m_\kkg^2}\r)^2\frac{\pi\alpha_s^2\beta}{216m_\kkq^2}\\
\hat\sigma_0\l(qq' \to \kkqX \kkqX'\,;\,\vv{\bar 3},\, J=0\r)
&=& \l(\frac{2m_\kkq^2}{m_\kkq^2 + m_\kkg^2}\r)^2\l(4 + \frac{m_\kkq^2}{m_\kkg^2}\r)^2\frac{\pi\alpha_s^2\beta}{216m_\kkq^2}\\
\hat\sigma_0\l(qq \to \kkqL \kkqR\,;\,\vv{\bar 3},\,J=1,|J_z|=1\r)
&=& \l(\frac{2m_\kkq^2}{m_\kkq^2 + m_\kkg^2}\r)^2\l(2 + \frac{m_\kkq^2}{m_\kkg^2}\r)^2\frac{\pi\alpha_s^2\beta}{54m_\kkq^2}\\
\hat\sigma_0\l(qq' \to \kkqL \kkqR'\,;\,\vv{\bar 3},\,J=1,J_z=-1\r)
&=& \hat\sigma_0\l(qq' \to \kkqR \kkqL'\,;\,\vv{\bar 3},\,J=1,J_z=1\r)\nn\\
&=& \l(\frac{2m_\kkq^2}{m_\kkq^2 + m_\kkg^2}\r)^2\l(2 + \frac{m_\kkq^2}{m_\kkg^2}\r)^2\frac{\pi\alpha_s^2\beta}{108m_\kkq^2}
\eea
There are also processes in which the quarks and KK quarks are replaced by their antiparticles, which have the same cross sections for opposite signs of $J_z$.

The bound states annihilate into the same quark flavors from which they were produced. The rates are
\bea
\Gamma\l((\kkqX \kkqX^{(\prime)})_{\vv{\bar 3},\,J=1}\to qq^{(\prime)}\r)
&=& \frac{1}{729}\l(\frac{2m_\kkq^2}{m_\kkq^2 + m_\kkg^2}\r)^2\l(\frac{m_\kkq^2}{m_\kkg^2}\r)^2\bar\alpha_s^3\alpha_s^2 m_\kkq\\
\Gamma\l((\kkqX \kkqX')_{\vv{\bar 3},\,J=0}\to q q'\r)
&=& \frac{1}{243}\l(\frac{2m_\kkq^2}{m_\kkq^2 + m_\kkg^2}\r)^2\l(4 + \frac{m_\kkq^2}{m_\kkg^2}\r)^2 \bar\alpha_s^3\alpha_s^2 m_\kkq\\
\Gamma\l((\kkqX \kkqXX^{(\prime)})_{\vv{\bar 3},\,J=1}\to qq^{(\prime)}\r)
&=& \frac{2}{729}\l(\frac{2m_\kkq^2}{m_\kkq^2 + m_\kkg^2}\r)^2\l(2 + \frac{m_\kkq^2}{m_\kkg^2}\r)^2\bar\alpha_s^3\alpha_s^2 m_\kkq
\eea

The angular distributions for $(\kkqX \kkqX^{(\prime)})_{\vv{\bar 3},\,J=1}\to qq^{(\prime)}$ are
\bea
J=1,J_z=\pm 1:\qq&& \frac{1}{\Gamma}\frac{d\Gamma}{\sin\theta\,d\theta} = \frac{3}{4}\sin^2\theta \\
J=1,J_z=0:\qq&& \frac{1}{\Gamma}\frac{d\Gamma}{\sin\theta\,d\theta}
= \frac{3}{2}\cos^2\theta
\eea
For $(\kkqX \kkqXX)_{\vv{\bar 3},\,J=1}\to qq$,
\bea
J=1,J_z=\pm 1:&\qq& \frac{1}{\Gamma}\frac{d\Gamma}{\sin\theta\,d\theta} =
\frac{3}{8}\l(1 + \cos^2\theta\r) \\
J=1,J_z=0:&\qq& \frac{1}{\Gamma}\frac{d\Gamma}{\sin\theta\,d\theta}
= \frac{3}{4}\sin^2\theta
\eea
For $(\kkqL \kkqR')_{\vv{\bar 3},\,J=1}\to qq^{\prime}$ (and for $(\kkqR \kkqL')_{\vv{\bar 3},\,J=1}\to qq^{\prime}$ for opposite signs of $J_z$)
\bea
J=1,J_z=1:&\qq& \frac{1}{\Gamma}\frac{d\Gamma}{\sin\theta\,d\theta} =
\frac{3}{2}\sin^4\frac{\theta}{2} \\
J=1,J_z=-1:&\qq& \frac{1}{\Gamma}\frac{d\Gamma}{\sin\theta\,d\theta} =
\frac{3}{2}\cos^4\frac{\theta}{2} \\
J=1,J_z=0:&\qq& \frac{1}{\Gamma}\frac{d\Gamma}{\sin\theta\,d\theta}
= \frac{3}{4}\sin^2\theta
\eea
For the corresponding anti di-KK quarks, the angular distributions are the same for opposite signs of $J_z$.

\subsection{Di-squarks\label{app-rates-disq}}

The near-threshold production cross section for attractive squark pairs is
\be
\hat\sigma_0\l(q q' \to \sq_\chi\sq'_\chi;\,\vv{\bar 3},\,J=0\r) = \l(\frac{2m_\go m_\sq}{m_\go^2 + m_\sq^2}\r)^2 \frac{\pi\alpha_s^2\beta}{27m_\sq^2}
\ee
There are also processes in which the quarks and squarks are replaced by their antiparticles, which have the same cross section.

The annihilation rate of di-squarks is
\be
\Gamma\l((\sq_\chi\sq_\chi')_{\vv{\bar 3},\,J=0}\to q q'\r) = \frac{8}{243} \l(\frac{2m_\go m_\sq}{m_\go^2 + m_\sq^2}\r)^2 \bar\alpha_s^3\alpha_s^2 m_\sq
\ee

\subsection{KK quark-KK gluon bound states\label{app-rates-KKqKKg}}

The near-threshold production cross sections for attractive KK quark-KK gluon pairs are
\bea
\hat\sigma_0\l(qg \to \kkqR g^\star\,;\,\vv{3},\,J = \frac{3}{2}, J_z = \frac{3}{2}\r) &=& \frac{\l(m_\kkg + 9m_\kkq\r)^2}{288\,m_\kkg^2 m_\kkq\l(m_\kkg+m_\kkq\r)}\,\pi\alpha_s^2\beta\\
\hat\sigma_0\l(qg \to \kkqR g^\star\,;\,\vv{3},\,J = \frac{1}{2}, J_z = -\frac{1}{2}\r) &=& \frac{m_\kkq\l(m_\kkg + 9m_\kkq\r)^2}{192\,m_\kkg^2\l(m_\kkg+m_\kkq\r)^3}\,\pi\alpha_s^2\beta\\
\hat\sigma_0\l(qg \to \kkqR g^\star\,;\,\vv{\bar 6},\,J = \frac{3}{2}, J_z = \frac{3}{2}\r) &=& \frac{\l(m_\kkg+m_\kkq\r)}{16\,m_\kkg^2 m_\kkq}\,\pi\alpha_s^2\beta\\
\hat\sigma_0\l(qg \to \kkqR g^\star\,;\,\vv{\bar 6},\,J = \frac{1}{2}, J_z = -\frac{1}{2}\r) &=& \frac{3\,m_\kkq}{32\,m_\kkg^2\l(m_\kkg+m_\kkq\r)}\,\pi\alpha_s^2\beta
\eea
Processes with $\kkqL$ instead of $\kkqR$ have the same cross sections for opposite signs of $J_z$. There are also processes in which the quark and KK quark are replaced by their antiparticles, which have the same cross sections for opposite signs of $J_z$.

The annihilation rates are
\bea
\Gamma\l((\kkq\kkg)_{\vv{3},\,J=3/2}\to q g\r) &=& \frac{3}{64}\, \bar\alpha_s^3\alpha_s^2\, \frac{m_\kkq\l(m_\kkg + 9m_\kkq\r)^2}{\l(m_\kkg + m_\kkq\r)^2} \\
\Gamma\l((\kkq\kkg)_{\vv{3},\,J=1/2}\to q g\r) &=& \frac{9}{64}\, \bar\alpha_s^3\alpha_s^2\, \frac{m_\kkq^3\l(m_\kkg + 9m_\kkq\r)^2}{\l(m_\kkg + m_\kkq\r)^4} \\
\Gamma\l((\kkq\kkg)_{\vv{\bar 6},\,J=3/2}\to q g\r) &=& \frac{1}{64}\, \bar\alpha_s^3\alpha_s^2\, m_\kkq \\
\Gamma\l((\kkq\kkg)_{\vv{\bar 6},\,J=1/2}\to q g\r) &=& \frac{3}{64}\, \bar\alpha_s^3\alpha_s^2\, \frac{m_\kkq^3}{\l(m_\kkg + m_\kkq\r)^2}
\eea
The angular distributions for $(\kkqR g^\star)\to qg$ (in both $\vv{3}$ and $\vv{\bar 6}$ representations) are
\bea
J=\frac{3}{2},\,J_z=+\frac{3}{2}:&\qq& \frac{1}{\Gamma}\frac{d\Gamma}{\sin\theta\,d\theta} =
2\cos^6\frac\theta2 \\
J=\frac{3}{2},\,J_z=+\frac{1}{2}:&\qq& \frac{1}{\Gamma}\frac{d\Gamma}{\sin\theta\,d\theta} =
6\cos^4\frac\theta2\sin^2\frac\theta2 \\
J=\frac{3}{2},\,J_z=-\frac{1}{2}:&\qq& \frac{1}{\Gamma}\frac{d\Gamma}{\sin\theta\,d\theta} =
6\cos^2\frac\theta2\sin^4\frac\theta2 \\
J=\frac{3}{2},\,J_z=-\frac{3}{2}:&\qq& \frac{1}{\Gamma}\frac{d\Gamma}{\sin\theta\,d\theta} =
2\sin^6\frac\theta2
\eea
\bea
J=\frac{1}{2},\,J_z=+\frac{1}{2}:&\qq& \frac{1}{\Gamma}\frac{d\Gamma}{\sin\theta\,d\theta} =
\sin^2\frac\theta2 \\
J=\frac{1}{2},\,J_z=-\frac{1}{2}:&\qq& \frac{1}{\Gamma}\frac{d\Gamma}{\sin\theta\,d\theta} =
\cos^2\frac\theta2
\eea
Processes with $\kkqL$ instead of $\kkqR$ have the same angular distributions for opposite signs of $J_z$. For the corresponding anti KK quark-KK gluon bound states, the angular distributions are again the same for opposite signs of $J_z$.

\subsection{Squark-gluino bound states\label{app-rates-sqgo}}

The near-threshold production cross sections for attractive squark-gluino pairs are
\bea
\hat\sigma_0\l(qg \to \sq_R\go\,;\,\vv{3},\,J = \frac{1}{2}, J_z = -\frac{1}{2}\r) &=& \frac{(m_\go + 9m_\sq)^2}{288\,m_\go\l(m_\go + m_\sq\r)^3}\,\pi\alpha_s^2\beta\\
\hat\sigma_0\l(qg \to \sq_R\go\,;\,\vv{\bar 6},\,J = \frac{1}{2}, J_z = -\frac{1}{2}\r) &=& \frac{1}{16\,m_\go\l(m_\go + m_\sq\r)}\,\pi\alpha_s^2\beta
\eea
Processes with $\sq_L$ instead of $\sq_R$ have the same cross sections for opposite signs of $J_z$. There are also processes in which the quark and squark are replaced by their antiparticles, which have the same cross sections for opposite signs of $J_z$.

The annihilation rates are
\bea
\Gamma\l((\sq\go)_{\vv{3},\,J=1/2}\to q g\r) &=& \frac{3}{32}\, \bar\alpha_s^3\alpha_s^2\, \frac{m_\go m_\sq^2\l(m_\go + 9m_\sq\r)^2}{\l(m_\go + m_\sq\r)^4} \\
\Gamma\l((\sq\go)_{\vv{\bar 6},\,J=1/2}\to q g\r) &=& \frac{1}{32}\, \bar\alpha_s^3\alpha_s^2\, \frac{m_\go m_\sq^2}{\l(m_\go + m_\sq\r)^2}
\eea
The angular distributions for $(\sq_R\go)\to qg$ (in both $\vv{3}$ and $\vv{\bar 6}$ representations) are
\bea
J_z=\frac{1}{2}:&\qq& \frac{1}{\Gamma}\frac{d\Gamma}{\sin\theta\,d\theta} =
\sin^2\frac{\theta}{2} \\
J_z=-\frac{1}{2}:&\qq& \frac{1}{\Gamma}\frac{d\Gamma}{\sin\theta\,d\theta} =
\cos^2\frac{\theta}{2}
\eea
The angular distributions for $(\sq_L\go)\to qg$ are the same for opposite signs of $J_z$. For the corresponding anti squark-gluino bound states, the angular distributions are the same, again for opposite signs of $J_z$.

\bibliography{ued}

\providecommand{\href}[2]{#2}\begingroup\raggedright\begin{thebibliography}{10}

\bibitem{Smillie:2005ar}
J.~M. Smillie and B.~R. Webber, ``{Distinguishing spins in supersymmetric and
  universal extra dimension models at the Large Hadron Collider},''
  \href{http://dx.doi.org/10.1088/1126-6708/2005/10/069}{{\em JHEP} {\bf 10}
  (2005)  069}, \href{http://arxiv.org/abs/hep-ph/0507170}{{\tt
  arXiv:hep-ph/0507170}}.

\bibitem{Datta:2005zs}
A.~Datta, K.~Kong, and K.~T. Matchev, ``{Discrimination of supersymmetry and
  universal extra dimensions at hadron colliders},''
  \href{http://dx.doi.org/10.1103/PhysRevD.72.096006}{{\em Phys. Rev.} {\bf
  D72} (2005)  096006}, \href{http://arxiv.org/abs/hep-ph/0509246}{{\tt
  arXiv:hep-ph/0509246}}. Erratum \textit{ibid.} \textbf{D72} (2005) 119901.

\bibitem{Barr:2005dz}
A.~J. Barr, ``{Measuring slepton spin at the LHC},''
  \href{http://dx.doi.org/10.1088/1126-6708/2006/02/042}{{\em JHEP} {\bf 02}
  (2006)  042}, \href{http://arxiv.org/abs/hep-ph/0511115}{{\tt
  arXiv:hep-ph/0511115}}.

\bibitem{Alves:2006df}
A.~Alves, O.~Eboli, and T.~Plehn, ``{It's a gluino},''
  \href{http://dx.doi.org/10.1103/PhysRevD.74.095010}{{\em Phys. Rev.} {\bf
  D74} (2006)  095010}, \href{http://arxiv.org/abs/hep-ph/0605067}{{\tt
  arXiv:hep-ph/0605067}}.

\bibitem{Athanasiou:2006ef}
C.~Athanasiou, C.~G. Lester, J.~M. Smillie, and B.~R. Webber, ``{Distinguishing
  spins in decay chains at the Large Hadron Collider},''
  \href{http://dx.doi.org/10.1088/1126-6708/2006/08/055}{{\em JHEP} {\bf 08}
  (2006)  055}, \href{http://arxiv.org/abs/hep-ph/0605286}{{\tt
  arXiv:hep-ph/0605286}}.

\bibitem{Athanasiou:2006hv}
C.~Athanasiou, C.~G. Lester, J.~M. Smillie, and B.~R. Webber, ``{Addendum to
  `Distinguishing spins in decay chains at the Large Hadron Collider'},''
  \href{http://arxiv.org/abs/hep-ph/0606212}{{\tt arXiv:hep-ph/0606212}}.

\bibitem{Wang:2006hk}
L.-T. Wang and I.~Yavin, ``{Spin measurements in cascade decays at the LHC},''
  \href{http://dx.doi.org/10.1088/1126-6708/2007/04/032}{{\em JHEP} {\bf 04}
  (2007)  032}, \href{http://arxiv.org/abs/hep-ph/0605296}{{\tt
  arXiv:hep-ph/0605296}}.

\bibitem{Kilic:2007zk}
C.~Kilic, L.-T. Wang, and I.~Yavin, ``{On the existence of angular correlations
  in decays with heavy matter partners},''
  \href{http://dx.doi.org/10.1088/1126-6708/2007/05/052}{{\em JHEP} {\bf 05}
  (2007)  052}, \href{http://arxiv.org/abs/hep-ph/0703085}{{\tt
  arXiv:hep-ph/0703085}}.

\bibitem{Csaki:2007xm}
C.~{Cs\'{a}ki}, J.~Heinonen, and M.~Perelstein, ``{Testing gluino spin with
  three-body decays},''
  \href{http://dx.doi.org/10.1088/1126-6708/2007/10/107}{{\em JHEP} {\bf 10}
  (2007)  107}, \href{http://arxiv.org/abs/0707.0014}{{\tt arXiv:0707.0014
  [hep-ph]}}.

\bibitem{Burns:2008cp}
M.~Burns, K.~Kong, K.~T. Matchev, and M.~Park, ``{A general method for
  model-independent measurements of particle spins, couplings and mixing angles
  in cascade decays with missing energy at hadron colliders},''
  \href{http://dx.doi.org/10.1088/1126-6708/2008/10/081}{{\em JHEP} {\bf 10}
  (2008)  081}, \href{http://arxiv.org/abs/0808.2472}{{\tt arXiv:0808.2472
  [hep-ph]}}.

\bibitem{Gedalia:2009ym}
O.~Gedalia, S.~J. Lee, and G.~Perez, ``{Spin determination via third generation
  cascade decays},'' \href{http://dx.doi.org/10.1103/PhysRevD.80.035012}{{\em
  Phys. Rev.} {\bf D80} (2009)  035012},
  \href{http://arxiv.org/abs/0901.4438}{{\tt arXiv:0901.4438 [hep-ph]}}.

\bibitem{Cheng:2010yy}
H.-C. Cheng, Z.~Han, I.-W. Kim, and L.-T. Wang, ``{Missing momentum
  reconstruction and spin measurements at hadron colliders},''
  \href{http://dx.doi.org/10.1007/JHEP11(2010)122}{{\em JHEP} {\bf 11} (2010)
  122}, \href{http://arxiv.org/abs/1008.0405}{{\tt arXiv:1008.0405 [hep-ph]}}.

\bibitem{MoortgatPick:2011ix}
G.~Moortgat-Pick, K.~Rolbiecki, and J.~Tattersall, ``{Early spin determination
  at the LHC?},'' \href{http://dx.doi.org/10.1016/j.physletb.2011.03.064}{{\em
  Phys.Lett.} {\bf B699} (2011)  158},
  \href{http://arxiv.org/abs/1102.0293}{{\tt arXiv:1102.0293 [hep-ph]}}.

\bibitem{Kats:2009bv}
Y.~Kats and M.~D. Schwartz, ``{Annihilation decays of bound states at the
  LHC},'' \href{http://dx.doi.org/10.1007/JHEP04(2010)016}{{\em JHEP} {\bf 04}
  (2010)  016}, \href{http://arxiv.org/abs/0912.0526}{{\tt arXiv:0912.0526
  [hep-ph]}}.

\bibitem{Carone:2003ms}
C.~D. Carone, J.~M. Conroy, M.~Sher, and I.~Turan, ``{Universal extra
  dimensions and Kaluza-Klein bound states},''
  \href{http://dx.doi.org/10.1103/PhysRevD.69.074018}{{\em Phys. Rev.} {\bf
  D69} (2004)  074018}, \href{http://arxiv.org/abs/hep-ph/0312055}{{\tt
  arXiv:hep-ph/0312055}}.

\bibitem{Fabiano:2008xk}
N.~Fabiano and O.~Panella, ``{Threshold production of metastable bound states
  of Kaluza Klein excitations in universal extra dimensions},''
  \href{http://dx.doi.org/10.1103/PhysRevD.81.115001}{{\em Phys. Rev.} {\bf
  D81} (2010)  115001}, \href{http://arxiv.org/abs/0804.3917}{{\tt
  arXiv:0804.3917 [hep-ph]}}.

\bibitem{Appelquist:2000nn}
T.~Appelquist, H.-C. Cheng, and B.~A. Dobrescu, ``{Bounds on universal extra
  dimensions},'' \href{http://dx.doi.org/10.1103/PhysRevD.64.035002}{{\em Phys.
  Rev.} {\bf D64} (2001)  035002},
  \href{http://arxiv.org/abs/hep-ph/0012100}{{\tt arXiv:hep-ph/0012100}}.

\bibitem{Hooper:2007qk}
D.~Hooper and S.~Profumo, ``{Dark matter and collider phenomenology of
  universal extra dimensions},''
  \href{http://dx.doi.org/10.1016/j.physrep.2007.09.003}{{\em Phys. Rept.} {\bf
  453} (2007)  29}, \href{http://arxiv.org/abs/hep-ph/0701197}{{\tt
  arXiv:hep-ph/0701197}}.

\bibitem{Cheng:2002ab}
H.-C. Cheng, K.~T. Matchev, and M.~Schmaltz, ``{Bosonic supersymmetry? Getting
  fooled at the CERN LHC},''
  \href{http://dx.doi.org/10.1103/PhysRevD.66.056006}{{\em Phys. Rev.} {\bf
  D66} (2002)  056006}, \href{http://arxiv.org/abs/hep-ph/0205314}{{\tt
  arXiv:hep-ph/0205314}}.

\bibitem{Cheng:2002iz}
H.-C. Cheng, K.~T. Matchev, and M.~Schmaltz, ``{Radiative corrections to
  Kaluza-Klein masses},''
  \href{http://dx.doi.org/10.1103/PhysRevD.66.036005}{{\em Phys. Rev.} {\bf
  D66} (2002)  036005}, \href{http://arxiv.org/abs/hep-ph/0204342}{{\tt
  arXiv:hep-ph/0204342}}.

\bibitem{Servant:2002aq}
G.~Servant and T.~M.~P. Tait, ``{Is the lightest Kaluza-Klein particle a viable
  dark matter candidate?},''
  \href{http://dx.doi.org/10.1016/S0550-3213(02)01012-X}{{\em Nucl. Phys.} {\bf
  B650} (2003)  391}, \href{http://arxiv.org/abs/hep-ph/0206071}{{\tt
  arXiv:hep-ph/0206071}}.

\bibitem{Cheng:2002ej}
H.-C. Cheng, J.~L. Feng, and K.~T. Matchev, ``{Kaluza-Klein dark matter},''
  \href{http://dx.doi.org/10.1103/PhysRevLett.89.211301}{{\em Phys. Rev. Lett.}
  {\bf 89} (2002)  211301}, \href{http://arxiv.org/abs/hep-ph/0207125}{{\tt
  arXiv:hep-ph/0207125}}.

\bibitem{Georgi:2000ks}
H.~Georgi, A.~K. Grant, and G.~Hailu, ``{Brane couplings from bulk loops},''
  \href{http://dx.doi.org/10.1016/S0370-2693(01)00408-7}{{\em Phys. Lett.} {\bf
  B506} (2001)  207}, \href{http://arxiv.org/abs/hep-ph/0012379}{{\tt
  arXiv:hep-ph/0012379}}.

\bibitem{Carena:2002me}
M.~S. Carena, T.~M.~P. Tait, and C.~E.~M. Wagner, ``{Branes and orbifolds are
  opaque},'' {\em Acta Phys. Polon.} {\bf B33} (2002)  2355,
  \href{http://arxiv.org/abs/hep-ph/0207056}{{\tt arXiv:hep-ph/0207056}}.

\bibitem{Flacke:2008ne}
T.~Flacke, A.~Menon, and D.~J. Phalen, ``{Non-minimal universal extra
  dimensions},'' \href{http://dx.doi.org/10.1103/PhysRevD.79.056009}{{\em Phys.
  Rev.} {\bf D79} (2009)  056009}, \href{http://arxiv.org/abs/0811.1598}{{\tt
  arXiv:0811.1598 [hep-ph]}}.

\bibitem{Martin:2008sv}
S.~P. Martin, ``{Diphoton decays of stoponium at the Large Hadron Collider},''
  \href{http://dx.doi.org/10.1103/PhysRevD.77.075002}{{\em Phys. Rev.} {\bf
  D77} (2008)  075002}, \href{http://arxiv.org/abs/0801.0237}{{\tt
  arXiv:0801.0237 [hep-ph]}}.

\bibitem{DePree:2005yv}
E.~De~Pree and M.~Sher, ``{Kaluza-Klein mesons in universal extra
  dimensions},'' \href{http://dx.doi.org/10.1103/PhysRevD.72.097701}{{\em Phys.
  Rev.} {\bf D72} (2005)  097701},
  \href{http://arxiv.org/abs/hep-ph/0507313}{{\tt arXiv:hep-ph/0507313}}.

\bibitem{Beneke:2009rj}
M.~Beneke, P.~Falgari, and C.~Schwinn, ``{Soft radiation in heavy-particle pair
  production: all-order colour structure and two-loop anomalous dimension},''
  \href{http://dx.doi.org/10.1016/j.nuclphysb.2009.11.004}{{\em Nucl. Phys.}
  {\bf B828} (2010)  69}, \href{http://arxiv.org/abs/0907.1443}{{\tt
  arXiv:0907.1443 [hep-ph]}}.

\bibitem{Peskin-Schroeder}
M.~E. Peskin and D.~V. Schroeder, {\em An Introduction to Quantum Field
  Theory}.
\newblock Westview Press, 1995.
\newblock See Section 5.3.

\bibitem{Novikov:1977dq}
V.~A. Novikov {\em et al.}, ``{Charmonium and gluons},''
  \href{http://dx.doi.org/10.1016/0370-1573(78)90120-5}{{\em Phys. Rept.} {\bf
  41} (1978)  1}.

\bibitem{Tyutin:1982fx}
I.~V. Tyutin and B.~B. Lokhvitskii, ``{Charge conjugation of non-Abelian gauge
  fields},'' \href{http://dx.doi.org/10.1007/BF00906208}{{\em Sov. Phys. J.}
  {\bf 25} (1982)  346}. [Izv. Vys. Uch. Zav. Fiz. {\bf 4} (1982) 62].

\bibitem{Smolyakov:1980wq}
N.~V. Smolyakov, ``{Furry theorem for non-abelian gauge Lagrangians},''
  \href{http://dx.doi.org/10.1007/BF01016449}{{\em Theor. Math. Phys.} {\bf 50}
  (1982)  225}. [Teor. Mat. Fiz. {\bf 50} (1982) 344].

\bibitem{Burgess-Moore}
C.~P. Burgess and G.~D. Moore, {\em The Standard Model: A Primer}.
\newblock {Cambridge University Press}, 2007.

\bibitem{Goldman:1984mj}
J.~T. Goldman and H.~Haber, ``{Gluinonium: The hydrogen atom of
  supersymmetry},'' \href{http://dx.doi.org/10.1016/0167-2789(85)90161-7}{{\em
  Physica} {\bf 15D} (1985)  181}.

\bibitem{Keung:1983wz}
W.-Y. Keung and A.~Khare, ``{Two-gluino bound states},''
  \href{http://dx.doi.org/10.1103/PhysRevD.29.2657}{{\em Phys. Rev.} {\bf D29}
  (1984)  2657}.

\bibitem{Kauth:2009ud}
M.~R. Kauth, J.~H. {K\"{u}hn}, P.~Marquard, and M.~Steinhauser, ``{Gluinonia:
  Energy levels, production and decay},''
  \href{http://dx.doi.org/10.1016/j.nuclphysb.2010.01.019}{{\em Nucl. Phys.}
  {\bf B831} (2010)  285}, \href{http://arxiv.org/abs/0910.2612}{{\tt
  arXiv:0910.2612 [hep-ph]}}.

\bibitem{Martin:2009iq}
A.~D. Martin, W.~J. Stirling, R.~S. Thorne, and G.~Watt, ``{Parton
  distributions for the LHC},''
  \href{http://dx.doi.org/10.1140/epjc/s10052-009-1072-5}{{\em Eur. Phys. J.}
  {\bf C63} (2009)  189}, \href{http://arxiv.org/abs/0901.0002}{{\tt
  arXiv:0901.0002 [hep-ph]}}.

\bibitem{Kim:2008bx}
C.~Kim and T.~Mehen, ``{Color octet scalar bound states at the LHC},''
  \href{http://dx.doi.org/10.1103/PhysRevD.79.035011}{{\em Phys. Rev.} {\bf
  D79} (2009)  035011}, \href{http://arxiv.org/abs/0812.0307}{{\tt
  arXiv:0812.0307 [hep-ph]}}.

\bibitem{Hagiwara:2009hq}
K.~Hagiwara and H.~Yokoya, ``{Bound-state effects on gluino-pair production at
  hadron colliders},''
  \href{http://dx.doi.org/10.1088/1126-6708/2009/10/049}{{\em JHEP} {\bf 10}
  (2009)  049}, \href{http://arxiv.org/abs/0909.3204}{{\tt arXiv:0909.3204
  [hep-ph]}}.

\bibitem{Barger:1984qg}
V.~D. Barger and A.~D. Martin, ``{Quarkonium production at $p\bar p$
  colliders},'' \href{http://dx.doi.org/10.1103/PhysRevD.31.1051}{{\em Phys.
  Rev.} {\bf D31} (1985)  1051}.

\bibitem{Barger:1987xg}
V.~D. Barger {\em et al.}, ``{Superheavy quarkonium production and decays: A
  new Higgs signal},'' \href{http://dx.doi.org/10.1103/PhysRevD.35.3366}{{\em
  Phys. Rev.} {\bf D35} (1987)  3366}. Erratum \textit{ibid.} \textbf{D38}
  (1988) 1632.

\bibitem{Arik:2002nd}
E.~Arik, O.~\c{C}ak{\i}r, S.~A. \c{C}etin, and S.~Sultansoy, ``{Fourth
  generation pseudoscalar quarkonium production and observability at hadron
  colliders},'' \href{http://dx.doi.org/10.1103/PhysRevD.66.116006}{{\em Phys.
  Rev.} {\bf D66} (2002)  116006},
  \href{http://arxiv.org/abs/hep-ph/0208169}{{\tt arXiv:hep-ph/0208169}}.

\bibitem{Martin:2009dj}
S.~P. Martin and J.~E. Younkin, ``{Radiative corrections to stoponium
  annihilation decays},''
  \href{http://dx.doi.org/10.1103/PhysRevD.80.035026}{{\em Phys. Rev.} {\bf
  D80} (2009)  035026}, \href{http://arxiv.org/abs/0901.4318}{{\tt
  arXiv:0901.4318 [hep-ph]}}.

\bibitem{Younkin:2009zn}
J.~E. Younkin and S.~P. Martin, ``{QCD corrections to stoponium production at
  hadron colliders},'' \href{http://dx.doi.org/10.1103/PhysRevD.81.055006}{{\em
  Phys. Rev.} {\bf D81} (2010)  055006},
  \href{http://arxiv.org/abs/0912.4813}{{\tt arXiv:0912.4813 [hep-ph]}}.

\bibitem{Herrero:1987df}
M.~Herrero, A.~M\'{e}ndez, and T.~Rizzo, ``{Production of heavy squarkonium at
  high energy $pp$ colliders},''
  \href{http://dx.doi.org/10.1016/0370-2693(88)91137-9}{{\em Phys.Lett.} {\bf
  B200} (1988)  205}.

\bibitem{Barger:1988sp}
V.~D. Barger and W.-Y. Keung, ``{Stoponium decays to Higgs bosons},''
  \href{http://dx.doi.org/10.1016/0370-2693(88)90915-X}{{\em Phys.Lett.} {\bf
  B211} (1988)  355}.

\bibitem{Drees:1993uw}
M.~Drees and M.~M. Nojiri, ``{Production and decay of scalar top squarkonium
  bound states},'' \href{http://dx.doi.org/10.1103/PhysRevD.49.4595}{{\em
  Phys.Rev.} {\bf D49} (1994)  4595},
  \href{http://arxiv.org/abs/hep-ph/9312213}{{\tt arXiv:hep-ph/9312213}}.

\bibitem{Borisov:1986ev}
G.~V. Borisov, Y.~F. Pirogov, and K.~R. Rudakov, ``{Pair production of exotic
  particles at$\,^({\bar p}^)p$ colliding beams},''
  \href{http://dx.doi.org/10.1007/BF01579137}{{\em Z. Phys.} {\bf C36} (1987)
  217}.

\bibitem{Chikovani:1996bk}
E.~Chikovani, V.~Kartvelishvili, R.~Shanidze, and G.~Shaw, ``{Bound states of
  two gluinos at the Tevatron and CERN LHC},''
  \href{http://dx.doi.org/10.1103/PhysRevD.53.6653}{{\em Phys. Rev.} {\bf D53}
  (1996)  6653}, \href{http://arxiv.org/abs/hep-ph/9602249}{{\tt
  arXiv:hep-ph/9602249}}.

\bibitem{Cheung:2004ad}
K.~Cheung and W.-Y. Keung, ``{Split supersymmetry, stable gluino, and
  gluinonium},'' \href{http://dx.doi.org/10.1103/PhysRevD.71.015015}{{\em Phys.
  Rev.} {\bf D71} (2005)  015015},
  \href{http://arxiv.org/abs/hep-ph/0408335}{{\tt arXiv:hep-ph/0408335}}.

\bibitem{BouhovaThacker:2004nh}
E.~Bouhova-Thacker, V.~Kartvelishvili, and A.~Small, ``{Search for
  gluino-gluino bound states},''
  \href{http://dx.doi.org/10.1016/j.nuclphysbps.2004.04.148}{{\em Nucl. Phys.
  Proc. Suppl.} {\bf 133} (2004)  122}.

\bibitem{BouhovaThacker:2006pj}
E.~Bouhova-Thacker, V.~Kartvelishvili, and A.~Small, ``{Search for
  gluino-gluino bound states with {\sc ATLAS}},''
  \href{http://dx.doi.org/10.1016/j.nuclphysbps.2005.08.059}{{\em Nucl. Phys.
  Proc. Suppl.} {\bf 152} (2006)  300}.

\bibitem{Drees:1993yr}
M.~Drees and M.~M. Nojiri, ``{Proposed new signal for scalar top-squark
  bound-state production},''
  \href{http://dx.doi.org/10.1103/PhysRevLett.72.2324}{{\em Phys.Rev.Lett.}
  {\bf 72} (1994)  2324}, \href{http://arxiv.org/abs/hep-ph/9310209}{{\tt
  arXiv:hep-ph/9310209}}.

\bibitem{Macesanu:2002db}
C.~Macesanu, C.~D. McMullen, and S.~Nandi, ``{Collider implications of
  universal extra dimensions},''
  \href{http://dx.doi.org/10.1103/PhysRevD.66.015009}{{\em Phys. Rev.} {\bf
  D66} (2002)  015009}, \href{http://arxiv.org/abs/hep-ph/0201300}{{\tt
  arXiv:hep-ph/0201300}}.

\bibitem{Lin:2005ix}
C.~Lin, ``{A search for universal extra dimensions in the multi-lepton channel
  from $p\bar{p}$ collisions at $\sqrt{s} = 1.8$~TeV},''
  FERMILAB-THESIS-2005-69.

\bibitem{Haisch:2007vb}
U.~Haisch and A.~Weiler, ``{Bound on minimal universal extra dimensions from
  $\bar B \to X_s\gamma$},''
  \href{http://dx.doi.org/10.1103/PhysRevD.76.034014}{{\em Phys. Rev.} {\bf
  D76} (2007)  034014}, \href{http://arxiv.org/abs/hep-ph/0703064}{{\tt
  arXiv:hep-ph/0703064}}.

\bibitem{Gogoladze:2006br}
I.~Gogoladze and C.~Macesanu, ``{Precision electroweak constraints on universal
  extra dimensions revisited},''
  \href{http://dx.doi.org/10.1103/PhysRevD.74.093012}{{\em Phys. Rev.} {\bf
  D74} (2006)  093012}, \href{http://arxiv.org/abs/hep-ph/0605207}{{\tt
  arXiv:hep-ph/0605207}}.

\bibitem{Feng:2009te}
J.~L. Feng, J.-F. Grivaz, and J.~Nachtman, ``{Searches for supersymmetry at
  high-energy colliders},''
  \href{http://dx.doi.org/10.1103/RevModPhys.82.699}{{\em Rev. Mod. Phys.} {\bf
  82} (2010)  699}, \href{http://arxiv.org/abs/0903.0046}{{\tt arXiv:0903.0046
  [hep-ex]}}.

\bibitem{Khachatryan:2011tk}
{\bf CMS} Collaboration, ``{Search for supersymmetry in $pp$ collisions at $7$
  TeV in events with jets and missing transverse energy},''
  \href{http://dx.doi.org/10.1016/j.physletb.2011.03.021}{{\em Phys.Lett.} {\bf
  B698} (2011)  196}, \href{http://arxiv.org/abs/1101.1628}{{\tt
  arXiv:1101.1628 [hep-ex]}}.

\bibitem{Collaboration:2011hh}
{\bf ATLAS} Collaboration, ``{Search for supersymmetry using final states with
  one lepton, jets, and missing transverse momentum with the ATLAS detector in
  $\sqrt{s} = 7$ TeV $pp$ collisions},''
\href{http://arxiv.org/abs/1102.2357}{{\tt arXiv:1102.2357 [hep-ex]}}.

\bibitem{daCosta:2011qk}
{\bf ATLAS} Collaboration, ``{Search for squarks and gluinos using final states
  with jets and missing transverse momentum with the ATLAS detector in $\sqrt s
  = 7$ TeV proton-proton collisions},''
  \href{http://arxiv.org/abs/1102.5290}{{\tt arXiv:1102.5290 [hep-ex]}}.

\bibitem{Chatrchyan:2011wc}
{\bf CMS} Collaboration, ``{Search for supersymmetry in $pp$ collisions at
  $\sqrt s = 7$ TeV in events with two photons and missing transverse
  energy},'' \href{http://arxiv.org/abs/1103.0953}{{\tt arXiv:1103.0953
  [hep-ex]}}.

\bibitem{Chatrchyan:2011bz}
{\bf CMS} Collaboration, ``{Search for physics beyond the Standard Model in
  opposite-sign dilepton events at $\sqrt s = 7$ TeV},''
  \href{http://arxiv.org/abs/1103.1348}{{\tt arXiv:1103.1348 [hep-ex]}}.

\bibitem{Aad:2011yf}
{\bf ATLAS} Collaboration, ``{Search for stable hadronising squarks and gluinos
  with the ATLAS experiment at the LHC},''
  \href{http://arxiv.org/abs/1103.1984}{{\tt arXiv:1103.1984 [hep-ex]}}.

\bibitem{Alwall:2008ve}
J.~Alwall, M.-P. Le, M.~Lisanti, and J.~G. Wacker, ``{Searching for directly
  decaying gluinos at the Tevatron},''
  \href{http://dx.doi.org/10.1016/j.physletb.2008.06.065}{{\em Phys. Lett.}
  {\bf B666} (2008)  34}, \href{http://arxiv.org/abs/0803.0019}{{\tt
  arXiv:0803.0019 [hep-ph]}}.

\bibitem{Sjostrand:2006za}
T.~{Sj\"{o}strand}, S.~Mrenna, and P.~Skands, ``{\textsc{Pythia} 6.4 Physics
  and Manual},'' \href{http://dx.doi.org/10.1088/1126-6708/2006/05/026}{{\em
  JHEP} {\bf 05} (2006)  026}, \href{http://arxiv.org/abs/hep-ph/0603175}{{\tt
  arXiv:hep-ph/0603175}}.

\bibitem{Sjostrand:2007gs}
T.~{Sj\"{o}strand}, S.~Mrenna, and P.~Skands, ``{A Brief Introduction to
  \textsc{Pythia} 8.1},''
  \href{http://dx.doi.org/10.1016/j.cpc.2008.01.036}{{\em Comput. Phys.
  Commun.} {\bf 178} (2008)  852}, \href{http://arxiv.org/abs/0710.3820}{{\tt
  arXiv:0710.3820 [hep-ph]}}.
  See~also~\url{http://home.thep.lu.se/~torbjorn/Pythia.html}.

\bibitem{Salam:2007xv}
G.~P. Salam and G.~Soyez, ``{A practical seedless infrared-safe cone jet
  algorithm},'' \href{http://dx.doi.org/10.1088/1126-6708/2007/05/086}{{\em
  JHEP} {\bf 05} (2007)  086}, \href{http://arxiv.org/abs/0704.0292}{{\tt
  arXiv:0704.0292 [hep-ph]}}.
  See~also~\url{http://projects.hepforge.org/siscone/}.

\bibitem{ATL-PHYS-PUB-2009-018}
{\bf ATLAS} Collaboration, G.~Aad {\em et al.}, ``{Performance of the ATLAS
  $b$-tagging algorithms},'' ATL-PHYS-PUB-2009-018 (2009).

\bibitem{CMS-PAS-BTV-09-001}
{\bf CMS} Collaboration, ``{Algorithms for $b$ jet identification in CMS},''
  CMS PAS BTV-09-001 (2009).

\bibitem{CMS-PAS-EXO-10-019}
{\bf CMS} Collaboration, ``{Search for Randall-Sundrum gravitons decaying into
  two photons in $7$~TeV $pp$ collisions with the CMS detector},'' CMS PAS
  EXO-10-019 (2011).

\bibitem{Dicus:2000hm}
D.~A. Dicus, C.~D. McMullen, and S.~Nandi, ``{Collider implications of
  Kaluza-Klein excitations of the gluons},''
  \href{http://dx.doi.org/10.1103/PhysRevD.65.076007}{{\em Phys. Rev.} {\bf
  D65} (2002)  076007}, \href{http://arxiv.org/abs/hep-ph/0012259}{{\tt
  arXiv:hep-ph/0012259}}.

\end{thebibliography}\endgroup

\end{document}